\documentclass[twocolumn,tighten]{aastex62}

\usepackage{apjfonts}
\usepackage{natbib}
\usepackage{amssymb, amsmath, amsbsy, booktabs,subfigure}

\newcommand{\mgal}{\ensuremath{M_\star}}
\newcommand{\bhatobhb}{\ensuremath{\ha^\mathrm{b}/\hb^\mathrm{b}}}

\newcommand{\ledd}{\ensuremath{L\mathrm{_{Edd}}}}
\newcommand{\lratio}{\ensuremath{L/\ledd}}

\newcommand{\msun}{\ensuremath{M_{\odot}}}

\newcommand{\ergs}{\ensuremath{\mathrm{erg~s^{-1}}}}
\newcommand{\kms}{\ensuremath{\mathrm{km~s^{-1}}}}
\newcommand{\mbh}{\ensuremath{M_\mathrm{BH}}}

\newcommand{\chisq}{\ensuremath{\chi^2}}

\newcommand{\ha}{{\rm H\ensuremath{\alpha}}}
\newcommand{\hb}{{\rm H\ensuremath{\beta}}}

\newcommand{\hd}{H\ensuremath{\delta}}

\newcommand{\hii}{H\,{\footnotesize II}}
\newcommand{\nii}{[N\,{\footnotesize II}]}
\newcommand{\sii}{[S\,{\footnotesize II}]}
\newcommand{\oiii}{[O\,{\footnotesize III}]}

\newcommand{\oii}{[O\,{\footnotesize II}] $\lambda$3727}
\newcommand{\oi}{[O\,{\footnotesize I}]}

\newcommand{\caii}{Ca\,{\footnotesize II}}

\newcommand{\mgib}{Mg\,{\footnotesize I}\,$b$}

\newcommand{\dbreak}{\ensuremath{D_{4000}}}
\newcommand{\hda}{H\ensuremath{\delta_{\rm A}}}

\newcommand{\etal}{et~al.}

\newcommand{\sersic}{S\'{e}rsic}

\begin{document}

\title{Local Active Galactic Nuclei with Large Broad-H$\alpha$ Variability Reside in Red Galaxies}

\author{Wen-Juan~Liu}
\affiliation{Yunnan Observatories, Chinese Academy of Sciences, Kunming, Yunnan 650011, China; \mbox{wjliu@ynao.ac.cn}}


\author{Paulina~Lira}
\affiliation{Departamento de Astronom\'ia, Universidad de Chile, Casilla 36-D, Santiago, Chile; \mbox{plira@das.uchile.cl} }

\author{Su~Yao}
\affiliation{Max-Planck Institut f{\"u}r Radioastronomie, Auf dem H{\"u}gel 69, 53121 Bonn, Germany; \mbox{syao@mpifr-bonn.mpg.de}}

\author{Dawei~Xu}
\affiliation{Key Laboratory of Space Astronomy and Technology, National Astronomical Observatories,
      Chinese Academy of Sciences, Beijing 100101, China}
\affiliation{School of Astronomy and Space Science, University of Chinese Academy of Sciences, Beijing, China}

\author{Jing~Wang}
\affiliation{Guangxi Key Laboratory for Relativistic Astrophysics, School of Physical Science and Technology, Guangxi University,
Nanning 530004, China}
\affiliation{Key Laboratory of Space Astronomy and Technology, National Astronomical Observatories,
      Chinese Academy of Sciences, Beijing 100101, China}

\author{Xiao-Bo~Dong}
\affiliation{Yunnan Observatories, Chinese Academy of Sciences, Kunming, Yunnan 650011, China; \mbox{wjliu@ynao.ac.cn}}


\author{Jorge~Mart{\'\i}nez-Palomera}
\affiliation{Bay Area Environmental Research Institute, P.O. Box 25, Moffett Field, CA 94035, USA}
\affiliation{Departamento de Astronom\'ia, Universidad de Chile, Casilla 36-D, Santiago, Chile; \mbox{plira@das.uchile.cl} }


\begin{abstract}
  Inspired by our serendipitous discovery of six active galactic nuclei (AGNs) with varying broad-\ha\ fluxes
	over years from our search for intermediate-mass black holes (IMBHs),
	we conduct a systematic investigation of changing-look (CL) and large-variability AGNs.
	We collect all the CL AGNs at $z<0.15$ and
	the reverberation-mapped AGNs with strongly variable broad \ha,
	and perform careful decomposition fittings to both their images and spectra.
	We find two observational facts:
	(1) The host galaxies of local CL and large-variability AGNs, mainly being Seyferts,
	are in the red (gas-poor) tail of the general Seyfert galaxy population.
	(2) In contrast, there is a significant trend that their more luminous counterparts,
	namely CL and extremely variable quasars (CLQs and EVQs)
	are different: CLQs are generally in blue galaxies;
	in terms of the diagram of star formation rate and \mgal,
	local CL Seyfert galaxies are located in the green valley,
	whereas CLQ hosts are in the star-forming main sequence.
	We propose explanations for those strongly variable Seyferts and quasars, respectively,
	under the assumption that accretion disks broadly depend on nuclear fueling modes.
	Local large-variability and CL Seyferts are in nuclear famine mode,
	where cold-gas clumps can be formed stochastically in the fueling flow,
	and their episodic infall produces sharp peaks in the accretion-rate curve.
	CLQs and EVQs are in feast fueling mode, which
	may account for both their preference for blue galaxies
	and their variability pattern (high-amplitude tail of the continuous distribution).
    Lastly, we propose a new thinking: to search for IMBHs by optical variability in red galaxies.
 \end{abstract}

\keywords{galaxies: active --- galaxies: nuclei --- galaxies: Seyfert --- accretion --- quasars: supermassive black holes}

\section{Introduction \label{sec:intro}}

It has been known for about five decades
that all active galactic nuclei (AGNs, including Seyfert galaxies and more luminous ones --- quasars)
are variable at some level in their optical continua and broad emission lines.
Generally, the optical variability is of small amplitude, 
with the rms fluctuation typically being
within 10--20\% on timescales of weeks to years
\citep[e.g.,][]{2001ApJ...555..775C,Walsh-2009ApJS..185..156W}, 
and behaves in a stochastic (or chaotic) manner
(e.g., \citealt{Kelly-2009ApJ...698..895K}).
For a long time, there were only a few exceptions discovered serendipitously,
varying with much larger amplitudes
 (e.g., by a factor of 5; \citealt{Alloin-1985ApJ...288..205A};
 see \citealt{Antonucci_2018NatAs...2..504A} for a historical commentary).
Yet this field is being renewed, as a welcome to the so-called time-domain astronomy nowadays.
Large or extreme flux changes on timescales of months to years have been observed in
$\gtrsim$\,100 AGNs,
and the number is still increasing owing to the systematic searches
based on large-scale photometric and spectroscopic surveys
\citep[e.g.,][]{2017ApJ...846L...7S,2018ApJ...854..160R,YangQ-WuXB-2018,
  2019ApJ...874....8M, 2019ApJ...885...44D,2020MNRAS.491.4925G}.
The change can be so dramatic that optical broad emission lines can disappear completely
or (re)appear, i.e., the so-called optically changing-look (CL) AGNs \citep[e.g.,][]{LaMassa+2015ApJ}.


It is fair to say that
the studies on the large/extreme variability of either the AGN optical continuum
\citep[e.g.,][]{2018ApJ...854..160R} or broad emission lines
\citep[e.g.,][]{2019ApJ...874....8M,2019ApJ...885...44D} are just developing.
For instance,
there are no unified definitions or criteria yet for
the terms ``large or extreme variability'' and ``optically changing look'' in the literature.
The optical CL AGNs (e.g., \citealt{LaMassa+2015ApJ}; and see \citealt{YangQ-WuXB-2018})
are often referred to as AGNs that experienced the spectral type transitions
between types 1.8--2 (namely having weak or no broad \hb, or even no broad \ha)
and Type 1 (namely having both broad \hb\ and \ha).
However, the detection of broad-\hb\ lines usually depends on the spectral signal-to-noise ratio (S/N) 
level (see \S3.3 of \citealt{2019ApJ...874....8M}; also \S\ref{subsec:magebHa} below).
A consensus has emerges from the recent studies that those optical CL AGNs (including CL quasars) and
extremely variable AGNs (mainly being quasars in the literature so far, usually shortened as EVQs)
generally reflect variable accretion rate rather than variable obscuration
\citep[e.g.,][]{LaMassa+2015ApJ,2017ApJ...846L...7S,2019ApJ...885...44D,
2019ApJ...874....8M,yanfei-2020ApJ...900...25J,2020MNRAS.491.4925G}.
Yet the concrete mechanisms are not clear so far (see, e.g.,
\citealt{2018NatAs...2..102L,2019MNRAS.483L..17D,2020A&A...641A.167S,yanfei-2020ApJ...900...25J}).
There is a trend from the latest large-sample studies
\citep{2018ApJ...854..160R,2019ApJ...874....8M,2019ApJ...885...44D}
that EVQs and optical CL quasars appear to be the high-amplitude tail of the
continuous distribution of quasar variability;
this variability pattern is different from that of their less luminous counterparts,
nearby CL Seyferts such as Fairall 9, Mrk 590 and Mrk 1018 (cf. \S5.6 of \citealt{2019ApJ...885...44D}).
Those local Seyferts instead exhibit a secular pattern.
For example, according to the available data,
Mrk 1018 was a Seyfert 1.9 in the year of 1979 and became Type 1 in 1984;
then, the AGN brightness remained constant roughly (as a Seyfert 1),
with a small rms (0.15 dex) for at least 10 yr (i.e., 2001--2011 with observations available),
and steadily decreased after 2011 (returning to Type 1.9 in 2015)
by 3--4 mag until a minimum around October 2016,
and again gradually got brightened at a low level \citep{2017A&A...607L...9K,2019ApJ...885...44D}.
It appears to keep the slow brightening so far, according to
the latest observation reporting it as a faint Type 1.5 \citep{2020A&A...644L...5H}.



Naturally, one extends the exploration of CL AGNs to their host-galaxy properties.
There have been several studies in the literature.
For instance, two detailed studies on the nuclear cold gas \citep{2016MNRAS.455.2745K} and
nuclear warm gas \citep{2019MNRAS.486..123R} of the prototypal Mrk 590 
reveal complex gas distributions and rich structures in the inner kiloparsec and 100 pc scales.
Those nuclear features are normal in local (invariable) Seyfert galaxies
and indicate secular processes for AGN fueling.
There are also three studies on the host-galaxy properties of small samples.
\citet{2019ApJ...876...75C} analyzed the host galaxies of four faded CL quasars ($z\gtrsim 0.2$)
using Gemini imaging and found that 
their hosts mostly have major-merger features and reside in the ``green valley'' between
blue star-forming galaxies and dead red galaxies;
i.e., the hosts of CL quasars are just ``like the majority of [luminous] AGNs'', as they concluded.
\citet{Yu-Shi-2020MNRAS.498.3985Y} analyzed in detail the host galaxies of five nearby CL AGNs ($z\lesssim 0.05$)
in the MaNGA survey based on their spectra, imaging, and IFU datacube.
Among their results, there is an interesting finding:
in terms of the diagram of global (namely whole-galaxy) star-formation rate (SFR) versus stellar mass ($M_\star$),
the host galaxies of their CL AGNs are located in the star-forming main sequence (SFMS),
just like the hosts of their non-CL (i.e., invariable) broad-line AGNs.
Most recently, \citet{2021ApJ...907L..21D} compiled a list of 17 CL AGNs from the literature
and divided them into two groups: 11 at low redshifts $z<0.15$ and 6 at $0.15<z<0.25$.
They presented a surprising result:
in the global SFR versus \mgal\ diagram, their $z> 0.15$ CL AGN sample
and their reference AGN sample (namely local Seyfert 2 galaxies)
fall in the SFMS, whereas, in stark contrast,
their $z<0.15$ CL AGN sample is located in the green valley (see their Figure~1).
Besides, they found that CL AGNs and highly variable AGNs
have high galaxy \sersic\ indices and high bulge-to-total light ratios,
``implying high stellar density in their cores''
(compared with their reference AGN catalog; see Figures~3 and 4 of \citealt{2021ApJ...907L..21D}).
The above results by different teams are not consistent with each other.
We believe one major reason is their small sample sizes,
which result in their respective biases
(e.g., the high fraction of LINERs
in the $z<0.15$ sample of \citet{2021ApJ...907L..21D},
see \S\ref{subsec:agn-properties} below;
see the fourth conclusion in their \S6).
A second reason lies in the (unwary) use of ready catalogs
of physical quantities derived by mass production; for example,
SFR values differ significantly between different catalogs,
with the difference even being formidable for certain populations
(e.g, more than 1~dex systematically; see \S8 of \citealt{Salim-2016ApJS..227....2S}),
due to their respective systematic errors
(e.g., an obvious yet not the most important one: aperture correction).
This is particularly the case when a galaxy harbors an AGN
and thus its SFR calculation has to account for the AGN contamination
(i.e., subtracting the narrow emission lines of the AGN,
dust emission heated by the AGN,
and to be worse, direct AGN continuum and/or broad emission lines if the AGN is of Type 1).


In this work, we report (1) our discovery of six strongly variable (low-mass) AGNs%
\footnote{\,Following \citet{2007ApJ...670...92G} and \citet{Dong-2012ApJ...755..167D},
	hereinafter we refer to BHs with $\mbh \lesssim 10^6$~\msun\ at the centers of galaxies as ``low-mass'' or
	``intermediate-mass'' BHs (IMBHs); accordingly, for the ease of narration, 
	wherever it is not ambiguous,
	hereinafter we call AGNs hosting low-mass BHs as low-mass AGNs or IMBH AGNs.}
preferentially in red galaxies, and (2) as an extension of the discovery, 
our systematic investigation of the host galaxies of low-$z$ CL AGNs and large-variability AGNs,
particularly as to any connections between host-galaxy properties and
the large/extreme-variability (including CL) phenomenon.
\footnote{\label{ftn:CLandLV}The meaning of ``changing look'' (CL) is explained in the 
  second paragraph above (see also \S\ref{subsec:magebHa}).
  The term ``large/extreme variability'' literally means large or extreme variability
  in AGN continua or emission lines;
  see \S\ref{subsubsec:MagEsample} for a quantitative definition of ``large variability'' used in this study.
  We can sense that the category of CL AGNs generally is included in the category of large-variability AGNs.
  In terms of host-galaxy properties, we will see that
  local CL AGNs and large-variability AGNs have no difference (see \S\ref{subsec:redcolor}).
  Thus, for the ease of narration, wherever it is not ambiguous,
  hereinafter we may use ``large-variability (including CL) AGNs'' or similar,
  or even just ``large-variability AGNs,''
  to represent both categories.} 
The broad-\ha\ fluxes of the six sources
varied by $1.3 - 3.0$ times (namely $0.3 - 1.2$\,mag) on a timescale of years.
Their variability was found unexpectedly during our spectroscopic campaign
initially planned to search for IMBHs in broad-line AGNs,
using the Magellan Echellette Spectrograph (MagE; \citealp{2008SPIE.7014E..54M}) mounted on the 6.5m Magellan
Baade telescope.
Most surprisingly, among our broad-line AGNs identified by MagE (15 broad-line sources in total),
those hosted by blue galaxies generally vary little,
whereas a significant fraction of those hosted by red galaxies exhibit large broad-\ha\ variability,
i.e., the six sources are predominantly in relatively red galaxies
(even redder than the general population of low-$z$ Seyfert 2 galaxies; see \S \ref{subsec:redcolor}).
Their large broad-\ha\ variability is confirmed by our ensuing spectroscopic observations.

In order to avoid the possible bias caused by the small sample size of our broad-line AGNs by MagE,
we collect all the low-$z$ CL AGNs ($z<0.15$) available in the literature
and the strongly variable AGNs observed by reverberation mapping (RM AGNs in short)
with available variability measures.
Those data secured our conclusion:
in terms of all diagnostic diagrams of color versus \mgal,
stellar absorption-line index \hda\ versus 4000\AA\ break (\dbreak),
and SFR versus \mgal,
the large-variability and CL AGNs (mostly being Seyferts not LINERs)
are predominantly in red galaxies
and redder than the general population of low-$z$ Seyfert 2 galaxies.

The structure of the present paper is as follows.
In \S\ref{sec:data}, we describe the data, data reductions and analyses for
the six MagE AGNs, the low-$z$ CL AGNs, and the strongly variable RM AGNs.
In order to make the logical flow of the text friendly to the reader,
we put two parts of the data analyses in Appendix~\ref{sec:appendix},
which are specific to the six variable MagE AGNs and handle their multiband light curves
and X-ray spectra.  
In \S\ref{sec:results}, first we present the results:
the genuineness of the large variability of the six MagE sources (\S\ref{subsec:magebHa});
the BPT and other AGN properties of
the large-variability and CL AGNs (\S\ref{subsec:agn-properties});
our discovery that low-$z$ CL and large-variability Seyferts generally
reside in redder, SFR-deficit host galaxies compared with the control sample
of local Seyfert 2 galaxies,
while CL quasars at relatively higher redshifts tend to be in the opposite,
preferring to blue host galaxies
that are located in the star forming main sequence (\S\ref{subsec:redcolor}).
These discoveries inspired us to think about the idea that
accretion disks broadly depend on nuclear fueling modes;
in \S\ref{subsec:AD-Fuels} we propose theoretical explanations
for those strongly variable Seyferts and quasars, respectively.
Lastly (in \S\ref{subsec:newthinking}),
we return to IMBH research, which was
the initial goal of our observational campaign that branched out in 
the present study, by proposing
a new thinking on optical-variability selection for IMBHs,
i.e., an implication of the discovery of this work.
\S\ref{sec:summary} is the summary.

Throughout the paper, we assume a cosmology with $H_{0} =70$ km~s$^{-1}$~Mpc$^{-1}$,
$\Omega_{\rm m} = 0.3$, and $\Omega_{\Lambda} = 0.7$.

\section{Data and Analysis \label{sec:data}}

\subsection{Samples}

\subsubsection{Brief Description of the MagE Sources \label{subsubsec:MagEsample}}

The six largely variable (low-mass) AGNs are the following:
SDSS J083909.65$+$072431.5 (hereafter J0839$+$0724),
SDSS J111349.83$+$000733.9 (hereafter J1113$+$0007),
SDSS J125710.76$+$272417.6 (hereafter J1257$+$2724),
SDSS J134245.69$+$243524.0 (hereafter J1342$+$2435),
SDSS J141249.63$-$030720.9 (hereafter J1412$-$0307), and
SDSS J144242.63$+$011911.2 (hereafter J1442$+$0119).
They are a part of sources observed by MagE in 2017 March and July,
with the initial goal of identifying IMBHs in broad-line AGNs 
(namely low-mass AGNs).
Throughout the whole MagE run of target selection and observing,
AGN variability did not come to our mind.
From the 2017 observing run, there are 15 broad-line AGNs confirmed in total.
The host-galaxy colors $g-i$ (measured from their Petrosian magnitudes) of the 15 sources
range from 0.43 to 1.14, with eight sources being red with $g-i > 1$.
Details of the MagE observation, data reduction, and all sources
will be presented in another paper (W.-J.~Liu et al. 2021, in preparation).
In this paper hereafter, we will call the six largely variable (low-mass) AGNs the MagE sample.

J0839$+$0724 decreased in broad-\ha\ flux between the two epochs of the SDSS and MagE spectra,
while the other five sources increased.
We took follow-up spectroscopic observations for the five sources if possible,
which confirmed their broad-\ha\ variability (see \S \ref{subsec:magebHa} for details).
\newline

Here, we define the ``large or strong variability'' of the six sources
to be a broad-\ha\ flux change (the ratio of the maximum flux to the minimum flux)
greater than 1.3\,.
This ratio is defined essentially in the same way as the variability measure $R_\mathrm{max}$
commonly used in the RM literature
\citep[e.g.,][]{2004ApJ...613..682P,2015ApJS..217...26B}. 
Another common measure of the overall light-curve variability is the so-called
fractional variability amplitude $F_\mathrm{var}$
(see the Appendix \S\ref{subsec:LCdata} for details),  
which is designed to measure the intrinsic variability amplitude
(i.e., correcting for the effects of measurement errors)
and is thus more robust than $R_\mathrm{max}$ to noises and outlier values
(see \S\ref{subsec:LCdata}; also \citealt{2015ApJS..217...26B}).
Certainly, the two measures are significantly correlated.
$R_\mathrm{max}=1.3$ for \ha\ or \hb\
(i.e., our above threshold value) roughly corresponds to $F_\mathrm{var} = 0.07$
(see Table~3 of \citealt{2015ApJS..217...26B}).
According to \citet{2015ApJS..217...26B}, $F_\mathrm{var} > 0.1$
means strong variability, roughly corresponding to $R_\mathrm{max} \gtrsim 1.55$ (see their Table~3).
This number is higher than the broad-\ha\ flux changes
of three sources in our MagE sample, which are as follows:
J0839$+$0724 (1.30), J1342$+$2435 (1.41), and J1412$-$0307 (1.32).
We keep the three as strongly variable sources
because of the variabilities of their continua in other wavelength bands (see \S\ref{subsec:magebHa}),
particularly because of the behavior of J1412$-$0307.
Although belonging to the two sources least variable in broad \ha,
J1412$-$0307 has $R_\mathrm{max} > 10$ in soft X-ray,
1.75 (0.61mag) in the WISE W2-band light curve (binned every half years),
and 1.51 (0.45 mag) even in the optical $V$-band light curve
(binned every 90 days) without removing the dominant starlight component.
Mentioned in passing,
$R_\mathrm{max} > 1.3$ (i.e., \mbox{0.3\,mag})
is  larger than the general variability amplitude
of low-mass AGNs selected by optical continuum variability
\citep[e.g.,][]{2018ApJ...868..152B,2020ApJ...889..113M,2020ApJ...896...10B,}.

\subsubsection{Collected Low-$z$ CL AGNs \label{subsubsec:clagns}}


Because our MagE sample is small and the emission-line variability of the sources is not extreme,
we decide to extend our exploration to the sources in the extreme and dramatic end.
Thus, we collect all the CL AGNs at $z \leqslant 0.15$ reported in the literature so far.
The redshift cut is set to ensure a better decomposition of the AGN and starlight components of the
SDSS images (see also \citealt{2021ApJ...907L..21D}).
We obtain 31 CL AGNs, constituting the low-$z$ CL sample.
All the sources have SDSS images, and 29 of them (except NGC\,2617 and NGC\,3516) have SDSS spectra.
For the two CL AGNs, we find one optical spectrum publicly available in the 6dF survey
(see also \citealt{2014ApJ...788...48S}) for NGC\,2617
and one nuclear optical spectrum in \citet{1992ApJS...79..255K} for NGC\,3516, respectively.

\subsubsection{RM AGNs with Large Broad-\ha\ Variability}


We also collected RM AGNs with large/strong variability in \ha\ or \hb\ emission lines,
which should belong to the same population as the 6 variable MagE sources.
The criterion is this: RM AGNs with measured
$F_{\rm var} \gtrsim 0.1$ (\citealt{2015ApJS..217...26B}; see our \S\ref{subsubsec:MagEsample})
for \ha\ or \hb\ available in the literature.
We obtain 15 RM AGNs finally.
They turn out to be at very low redshifts ($0.002<z<0.04$),
and from three RM samples \citep{2004ApJ...613..682P,2009ApJ...697..160B,2015ApJS..217...26B}.
\\

 \subsection{Optical Spectra \label{subsec:optiSpecForAllSamples}}

\subsubsection{Observations and Data Reductions for the MagE Sample \label{subsubsec:specfor6agn}}

Brief information on all the spectroscopic observations
for the six variable sources in the aforementioned MagE sample
is listed in Table~\ref{table:lineinfo}.
Below we describe the instrumental and observing matter of every observation
and the corresponding data reduction.

\noindent $\bullet$ SDSS.\\
 \indent  All the six variable sources in the MagE sample
 have archival {\it SDSS} spectra, and J1113$+$0007 was spectroscopically observed twice.
The SDSS is an imaging and spectroscopic survey, using
a dedicated 2.5 m telescope to image one-quarter of the
sky and to perform follow-up spectroscopic observations.
 Fibers that feed the SDSS spectrographs have an aperture of 3\arcsec\ diameter.
 The nominal total exposure time of the survey spectra is 45 minutes,
 which typically yields an S/N of 4.5~pixel$^{-1}$ for sources with a $g$-band magnitude of 20.2.
 The spectra are flux- and wavelength-calibrated by the SDSS pipeline,
 with 4096 pixels from 3800 to 9200~\AA\
 at a resolution $R \equiv \lambda/\Delta \lambda \approx 1800$ (i.e.,
 instrumental dispersion $\sigma_\mathrm{instr} \approx 70$ \kms).

\noindent $\bullet$ MagE.\\
 \indent We conducted spectroscopic observation with MagE aboard the Magellan Baade telescope
 on UT 2017 March 25, using a 1\arcsec-slit, which gives an instrumental
 dispersion $\sigma_\mathrm{instr} \approx 26$ \kms\ as measured from the arc lamp spectra.
 The spectral coverage is approximately 3200-10000\AA\ across 15 echellette orders.
 One-dimensional spectra were extracted and wavelength-calibrated using MASE reduction pipeline
 \citep{2009PASP..121.1409B}. The telluric H$_{2}$O absorption features
 longward of 6800\AA\ were removed by dividing with the normalized spectrum of a standard star.
 More details of the data reduction can be found in W.-J Liu \etal\ (2021, in preparation).

\noindent $\bullet$ Follow-up by P200/DBSP and Xinglong/BFOSC.\\
\indent  To further check and monitor the broad-\ha\ variability for the six sources,
 we carried out several follow-up optical spectroscopic observations, as summarized below.

 For J1257$+$2724, J1412$-$0307, and J1442$+$0119,
 spectra were taken using
 the Double Spectrograph (DBSP) on the Palomar 200 inch Hale telescope (P200) at
 Palomar Observatory on UT 2019 June 24--25.
 The instrument is configured with a D55 dichroic, 1\farcs5 slit, 600/4900 grating
 for the blue side and 600/7600 grating for the red side, which gives a spectral resolution
 of $\sigma_\mathrm{instr} \approx 108$ \kms\ for the blue side,
 and $\sigma_\mathrm{instr} \approx 70$ \kms\ for the red side
 (i.e., almost the same as the SDSS resolution),
 and provides a wavelength coverage of 3400--9300\AA.
 The three sources were observed with two separate 500-900 s exposures.

 For J1257$+$2724, we took a spectrum with the Beijing Faint Object
 Spectrograph and Camera (BFOSC) mounted on the 2.16m telescope at Xinglong Observatory 
 on UT 2020 January 1.
 The grating G4 and a slit of width 2\farcs3 were used.
 This setup gives a spectral resolution of $\sigma_\mathrm{instr} \approx 230$ \kms,
 as measured from the sky emission lines and comparison arcs, and
 provides a wavelength coverage of 3850--8000\AA.
 Two separate 1800s exposures were taken.

 The two-dimensional spectral data of the DBSP and BFOSC observations
 were reduced with the standard routines for longslit spectra in IRAF.
 The telluric H$_{2}$O absorption features longward of 6800\AA\ were removed by dividing with the
 normalized spectrum of a standard star.
 \newline

\subsubsection{Spectral fitting and analysis \label{subsubsec:specfitting}}

For the six variable MagE sources and all the 31 low-$z$ CL AGNs,
as well as several sources as comparison (e.g., the four CL quasars of \citealt{2019ApJ...876...75C}),
we fit their optical spectra, in order to
measure their host-galaxy properties (such as \dbreak\ and \hda) and emission lines.
Based on the measurements of emission lines, we in turn can derive SFR (from \oii) and
AGN parameters such as \mbh\ and \lratio.
The variable MagE sources have multiepoch spectra as described in \S\ref{subsubsec:specfor6agn}.
Among the CL sources, 29 use the SDSS spectra, NGC\,2617 uses the 6dF spectrum,
and NGC\,3516 uses the spectrum observed by Steward Observatory 2.3 m telescope (see \S\ref{subsubsec:clagns})

Regarding the 15 strongly variable RM sources, because their  publicly available optical spectra 
are heterogeneous and incomplete, and generally dominated by the AGN emission,
we give up to fit them.
We do not need to use the spectrum-based quantities for their host-galaxy properties
and instead use the quantities based on imaging decomposition
(such as color, magnitude, and derived stellar mass, after correcting for AGN contamination; see
Table~\ref{table:rmAGN} and \S\ref{subsubsec:whole-color-Mgal}).

Regarding the comparison samples for our low-$z$ CL or large-variability AGNs,
such as the four CL quasars at $z \gtrsim 0.2$ in \citet{2019ApJ...876...75C},
the Seyfert 2 sample selected by \citet{Dong-2010ApJ...721L.143D},
and the non-AGN galaxy sample selected by \citet{Dong-2012ApJ...755..167D},
the aforementioned spectral properties are based on their SDSS spectra;
we either fit their SDSS spectra by ourselves (for the CL quasars)
or obtain the fitting results from the corresponding researchers (for the latter two control samples).
\newline

\noindent $\bullet$ Continuum fitting.\\
\indent
  Prior to fitting, all spectra were de-redshifted to the vacuum rest-frame wavelength and corrected for
  Galactic extinction using the extinction map of \citet{1998ApJ....500..525S} and the reddening curve of
  \citet{1999PASP..111...63F}.
  We use the same procedures as \citet{Dong-2012ApJ...755..167D} for the continuum modeling and emission-line
  profile fitting; we only provide a brief description here.
  We model the continuum of each of the spectra using the starlight templates
  and a single reddened power law to represent the AGN continuum in the spectral fitting.
  For SDSS, DBSP, and BFOSC spectra, we use the starlight templates from \citet{2006AJ....131..790L},
  built from the spectra of simple stellar populations of \citet{2003MNRAS.344.1000B}.
  The spectral resolution of the starlight templates used is 3\,\AA\ across the wavelength range of 3200-9500\AA,
  corresponding to a median resolving power $\lambda/\Delta\lambda \approx 2000$, which is comparable with
  the spectral resolution of SDSS spectra (69\,\kms).
  The DBSP spectra have identical spectral resolution to the SDSS spectra, and the BFOSC spectra have a much lower one.
  The resolution of MagE spectra is about twice as good as that of SDSS spectra; a starlight template built
  from the MagE spectra of four types of stars (B0, A5, K0III, and M4III), which were observed by MagE on the same day
  with the six MagE AGNs, is adopted in continuum fitting. More detailed information on this will be provided in
  W.-J. Liu et al.(2021, in preparation).
  The starlight templates are broadened by convolving with a Gaussian of different widths and shifted to match the
  stellar velocity dispersion, so that the stellar absorption lines could be well subtracted.
  A \chisq\,minimization is performed iteratively over the whole spectroscopic wavelength range,
  except for the regions emission lines (e.g., \ha, \hb, \oiii$\lambda5007$,
  \sii$\lambda\lambda6716,6731$, \oi$\lambda6300$, and \oii).
  In the fitting, we treat each spectrum of the same AGN independently to find the best-fit model for each spectrum.

Mentioned in passing, we also measured the stellar velocity dispersions from the MagE spectra
for the six MagE sources along with the entire MagE sample.
Details are in W.-J. Liu et al. (2021, in preparation).
Briefly, the fitting method is almost the same as
described in \citet{2011ApJ...739...28X}, in which the spectra of velocity
template stars are broadened and fit to the galaxy spectra
locally in a specific spectral region, e.g.,
\mgib\,(5040-5430\AA), or \caii\,triplet (8450-8750\AA);
the spectra of velocity template stars were also observed by MagE on the same day as the MagE AGNs.
We do not use the $\sigma_\star$ data in this work and only list them
in Table~\ref{table:galaxy} for any possible interest of the reader.
\newline

\noindent $\bullet$ Emission-line fitting.\\
\indent
  We then fit the continuum-subtracted spectrum (namely, simultaneously fitting
  \hb\,+\oiii\,+\oi\, +\nii\, +\sii\, and \oii\,), following the methodology of \citet{2008MNRAS.383..581D}.
  Specifically, we assume the broad and narrow components of \hb\, have the same profiles as the
  respective components of \ha.
  The \oiii\,$\lambda\lambda4959,5007$ doublet lines are assumed to have the identical profiles and fixed in
  separation by their laboratory wavelength; the same is applied to \nii$\lambda\lambda6548,6583$
  and to \sii$\lambda\lambda6716,6731$ doublet lines.
  The flux ratio of \oiii$\lambda5007/\lambda4959$ is fixed to the theoretical
  value of 2.98; the flux ratio of \nii$\lambda6583/\lambda6548$ is fixed to the
  theoretical value of 2.96.
  Every narrow or broad component of the emission lines is modeled with Gaussian,
  starting with one Gaussian and adding in more if the fit can be improved significantly
  according to the $F$-test.
\newline

\noindent $\bullet$ Calibrating and scaling multiepoch spectra (MagE sources).\\
\indent
  Due to the difference in adopted apertures/slit widths, seeings, and flux calibrations of the spectra taken
  by different instruments, their nuclei-emission-line flux levels are not exactly the same.
  Therefore, the emission-line-flux calibrations between spectra of different epochs are needed,
  and the \oi$\lambda6300$ is the ideal forbidden line to do this. \oi$\lambda6300$ is mainly produced in the
  ``partially ionized zone'' related primarily to AGN radiation, so it is little affected by star formation in galaxies.
  The disadvantage of using \oi$\lambda6300$ for calibration is that it could be weak in some AGNs, and
  its S/N is poor.
  Another usable emission line is \oiii$\lambda5007$. \oiii$\lambda\lambda4959,5007$ are mainly produced in the AGN
  narrow-line region, but are more affected by star formation in galaxies than \oi$\lambda6300$.
  The advantage is that \oiii$\lambda5007$ is usually very strong in AGN spectra.
  For most AGNs, both \oi$\lambda6300$ and \oiii$\lambda5007$ lines are basically constant.

  Here we used \oi$\lambda6300$ or \oiii$\lambda5007$ to scale the spectra of different epochs of the six MagE AGNs.
  The \oi$\lambda6300$ emission lines in J0839$+$0724, J1257$+$2724, J1412$-$0307, and J1442$+$0119 are weak,
  so their MagE, DBSP, and BFOSC spectra were scaled to the early SDSS epoch assuming the constant
  \oiii\ emission lines.
  For J1342$+$2435, the S/N of continuum-subtracted spectra under \oi$\lambda6300$ is better than that
  of \oiii$\lambda5007$, so we choose to use \oi$\lambda6300$ to scale its MagE and DBSP spectra.
  While for J1113$+$0007, its following spectra were scaled to the SDSS spectrum according to
  the \oi$\lambda6300$ flux, which brings a more reasonable scaling result than the \oiii\ flux.
  We checked its two epochs of SDSS spectra. Without any scaling, their continua show
  almost the same flux level, with a flux difference less than 1\%, and the flux difference between
  two \oi$\lambda6300$ is less than < 1$\sigma$ flux error. However, the second-epoch spectrum has a
  stronger \oiii\ doublet than the first-epoch spectrum, with flux difference larger than the 2$\sigma$ flux error.
  The enhanced trend in the \oiii\ flux is more evident when we compared the MagE and SDSS spectra after
  scaling the spectra by \oi$\lambda6300$ flux.
  Similar \oiii\ behavior is also seen in an ``Turn-on'' AGN SDSS\,J1115$+$0544 \citep{2019ApJ...874...44Y},
  which is explained by the increase of the ionization continuum.
  The fitting results also indicate that the continua of the six AGNs are dominated by starlight.
  Their AGN continuum fluxes at 5100\AA\ account for only 5\%--22\% of the total continuum fluxes.

  We also calculate the AGN spectral types for every spectra of the six MagE AGNs.
  We adopt the definition of spectral subtypes introduced
  by \citet[][see also \citealt{Osterbrock1981ApJ_subtypes} and \citealt{2001A&A...374...92V}]{1992MNRAS.257..677W},
    which uses a combination of the presence or absence of broad \ha\ and broad \hb,
    and the flux ratio of total \hb\ to \oiii.
    We adopt S/N $>3$ as the detection threshold of broad emission lines.
    The final emission-line parameters of every spectra, as well as the spectral subtypes,
    are listed in Table~\ref{table:lineinfo}.
    \newline

\noindent $\bullet$ Calculating spectrum-based quantities for host galaxies.\\
\indent
To quantify the properties of host galaxies,
we calculate three spectrum-based quantities: the 4000\,\AA\ break \dbreak\ \citep{1999ApJ...527...54B}, the Balmer absorption-line index
\hda\ \citep{1997ApJS..111..377W}, and the SFR,
from their SDSS spectra
\footnote{NGC 2617, a source in the low-$z$ CL sample, has no SDSS spectrum,
thus we actually use its available 6dF spectrum publicly. Because \oii\ is not covered by the 6dF spectrum,
we cannot give its SFR in Table~\ref{table:clagn}. NGC 3516 is another low-$z$ CL AGN that has no SDSS spectrum
and image. \citet{2011ApJ...739...57K} observed its $u,g,r,i,z$ image using the Kitt Peak 2.1 m telescope.
In Table~\ref{table:clagn}, we directly use the galactic $g,r,i$ magnitudes, $r$-band PSF fraction, and stellar mass
given by \citet{2011ApJ...739...57K}, 
and use the bulge and disk \sersic\ indexes decomposed from the Hubble Space Telescope (HST) F814W image by \citet{2017ApJS..232...21K}. 
The \dbreak\, and \hda\, are measured from its nuclear optical spectrum observed
using the Steward Observatory 2.3 m telescope during 1989-1991 \citep{1992ApJS...79..255K}.
Because the flux units of the spectrum are counts, so we cannot give its SFR in Table~\ref{table:clagn}.}
for the six MagE sources and for the low-$z$ CL sources except two.
The two exceptional sources have too strong AGN emission in the optical,
with the AGN fraction $>30$\% in the $r$-band images (see Table~\ref{table:clagn})
and larger in the 3\arcsec\ aperture SDSS spectra.
As for the RM sources, only a fraction have spectra in SDSS,
and their SDSS fiber spectra are generally dominated by AGNs (because those sources are very nearby).
Thus we do not calculate or use the three quantities for the RM sources in this work.
Besides, we also calculate the 3 quantities based on their SDSS spectra
for the four CL quasars of \citet{2019ApJ...876...75C}.

\dbreak\ is a good indicator of stellar age,
and its value is small for young stellar populations and is large for old, metal-rich galaxies.
It increases monotonically as the luminosity-weighted mean age of the stellar population increases \citep{2003MNRAS.341...33K}.
\hda\ is defined to measure the strength of stellar \hd\ absorption line
and indicates the burstiness of the star forming activity.
Strong stellar \hd\ absorption (positive large \hda\ values) means that
the galaxies experienced a burst of star formation that ended within 0.1--1\,Gyr ago.
The two indices used in combination can give the mean stellar age and
tell the star formation is bursty or secular.
We calculate the two indices from the stellar continua
decomposed from the SDSS spectra of the MagE and CL samples
and listed them in Tables \ref{table:galaxy} and \ref{table:clagn}.
The SFRs are calculated according to the measured \oii\ flux from their SDSS spectra
using the calibration for the AGN hosts of \citet{2019ApJ...882...89Z},
which removes the \oii\ flux from the AGN and accounts for
the influence of metal abundance on the SFR.
Note that SDSS and BOSS spectra are taken within a fiber aperture
of 3\arcsec\ or 2\arcsec\ diameter,
while the angular sizes of most of the above low-$z$ AGN hosts
(see Tables \ref{table:galaxy} and \ref{table:clagn} for their $r$-band Petrosian radii)
are much larger.
Thus, the three spectrum-based quantities
mainly reflect the host-galaxy properties in the inner regions.
\newline


\subsection{Image Fitting and Analysis\label{subsec:imaging}}

All of the sources in the MagE and low-$z$ CL samples, 
as well as 10 sources in the RM sample, have archival SDSS images. 
For those sources except two,
we fit their SDSS images in order to remove the AGN emission and measure the host-galaxy properties.
The two exceptions are the following:
J111803.22$+$450646.8 in the low-$z$ CL sample, because a bright foreground star sits in its central region;
and NGC 4051 in the RM sample, because it is very nearby and the AGN is bright,
and thus there are too many saturated pixels in the central region of its SDSS images.
As for the five rest RM sources without SDSS images, as well as NGC 4051,
fortunately \citet{2011ApJ...739...57K} have taken deeper optical images and performed two-dimensional fitting;
we simply adopt the host-galaxy magnitudes and colors (with the AGN contamination removed)
as well as the AGN fraction in the $r$ band there.


The SDSS imaging data were collected in a drift-scan mode in five
bandpasses ($u$, $g$, $r$, $i$, and $z$) on nights of pristine conditions,
with a typical seeing of 1.\arcsec5 in the $r$ band.
The images have a total exposure time of 54 s per filter.
The photometric calibration is accurate to 5\%, 3\%, 3\%, 3\%, and 5\%, respectively.

Figure~\ref{fig:sdssimage}
displays the $g$-$r$-$i$ composite images
for all the 6 MagE sources (panel a), 30 low-$z$ CL AGNs that have SDSS images (panel b),
and the aforementioned 10 sources in the RM sample (panel c). 
Almost all of the host galaxies are well resolved,
except several objects in the low-$z$ CL sample
(roughly with Petrosian radius $<4$\arcsec\ in the $r$ rand; see Table \ref{table:galaxy},\ref{table:clagn},\ref{table:rmAGN}):
\footnote{For the broad-line sources of this work, the Petrosian radius values
	given by the SDSS photometric pipeline for their host galaxies
	are affected to some degree by the AGN emission,
yet this problem is not important for our purpose.}:
J0909$+$4747, J1003$+$3525, J1132$+$0357,
J1358$+$4934, J1428$+$1723, and J1545$+$2511.
It is also obvious from Figure~\ref{fig:sdssimage}
that in most of the images the AGN emission does not swallow up the host galaxies.

We perform a two-dimensional (2D) decomposition of AGNs and host galaxies
into their SDSS images in the $g$, $r$ and $i$ bands of every aforementioned source
(44 sources in total) using GALFIT \citep{2002AJ....124..266P,2010AJ....139.2097P}.
Prior to the fitting of every image,
all foreground or background photometric objects identified by the SDSS pipeline
are carefully masked out manually.
In the fitting, the sky background is set to be free.
The AGNs are represented by a PSF component.
The PSF model images
are reconstructed from the psfFiled files provided by SDSS for every given field.
Concerning the host galaxies,
for most sources the galactic morphologies
are basically regular (axisymmetric),
and we adopt the traditional fitting approach implemented in GALFIT.
Yet for about a dozen sources,
their images are rather nonaxisymmetric, e.g.,
with the starlight being dominated by grand-design spirals or bars (e.g., Mrk\,1511,
J1533$+$4432), with merger features and so on,
or have prominent rings,
and we make use of the new machinery implemented in the ``new-generation'' (NG) GALFIT
(namely version 3; \citealp{2010AJ....139.2097P});
hereafter, we call this approach ``NG fitting''
because the new methodology is somehow antitraditional in certain respects,
as can be sensed in our description below.

In traditional fittings of the regular galaxies,
we model the host galaxies with the following three models
(with an increasing number of free parameters):
a single \sersic\ $r^{1/n}$ function \citep{1968adga.book.....S},
or one \sersic\ plus one exponential,
or two \sersic\ functions.
Here the \sersic\ and exponential functions represent
the azimuthally averaged, radial SB profiles,
which describe the intensity falloff away from the peak
(assuming the isophotes being ellipses in shape, i.e., azimuthal symmetry).
The exponential function is the special case of a \sersic\ function 
with the index fixed to be $n=1$,
commonly used to represent galactic disks;
yet nowadays researchers realize that the surface-brightness (SB) profiles of
(the outer) disks are not necessarily exponential,
and \sersic\ functions with $n<1$ are also common (see also \citealp{2010AJ....139.2097P}).
Following the common practice of this kind of imaging fitting,
we begin the fitting of every image
with the simplest scheme (``PSF $+$ S\'{e}rsic'') allowing the scaling factor of
the PSF (AGN) and all parameters of the \sersic\ (host galaxy) to vary.
Then, we try schemes with the other two advanced models for host galaxies,
if the fit can be improved significantly
in terms of \chisq\ (in the spirit of the $F$-test)
and the analysis of residual images; see \citet{Dong-2007ApJ...657..700D} for details.
In the case of two-component models for the host galaxies, following the common practice
hereinafter we interpret
the inner \sersic\ as the bulge or pseudo-bulge,
and the outer one as the disk.

In NG fittings,
any galactic component is still represented by the same basic SB profile models
used in the traditional fittings,
but their azimuthal shapes can be modified from the fundamental ellipse shape
by four novel devices:
bending, Fourier, coordinate rotation, and truncation modes.
Those devices, or called shape operators in the mathematical physics fashion,
are carefully devised so that the traditional profile parameters almost keep
their original, intuitive meaning intact;
in fact, they are merely higher-order modifying functions of the fundamental shape
from the perspective of mathematics.
In terms of the practical effects, however, those shape operators are magic
(particularly when used in combination):
for instance, they can metamorphose a \sersic\ component into
almost any arbitrary shape,
thus representing a galactic component/substructure of any realistic morphology
(e.g., a spiral with any number of arms, a spiral with a bar, or a tidal tail,
let alone a bulge, bar, or their subcomponents; see \S7 of \citealp{2010AJ....139.2097P}).
This power naturally results in a new methodology,
in contrast to the traditional wisdom of decomposing galactic images.
Specifically, for the fitting of the galactic disks we concerned with in this work,
the traditional wisdom is this: use a global \sersic\ or exponential model to
represent the (presumably) main body of the disk,
then add localized subcomponents onto it (e.g., a bar).
This tradition is actually not a choice, limited by previously available tools
(see \citealt{2010AJ....139.2097P}).
And worse, it has reinforced the misimpression that
there should be a dominant disk component as the main body
and the other (sub)components such as spirals and bars be secondary,
which certainly is not true in many galaxies (see \S7 of \citealt{2010AJ....139.2097P}).
On the contrary, the NG fitting may divide a galactic ``disk''
into several (somehow localized) pieces,
without a backbone disk component;
e.g., the galactic disk of NGC 289 is decomposed into
three spiral components each modeled with a morphed \sersic\ each
by \citealt[][see their Figure\,22]{2010AJ....139.2097P};
i.e., the sum of the three spiral components
is the commonly called ``disk'' component.
Following the advice of \citet{2010AJ....139.2097P},
in order to get accurate bulge/disk decomposition,
particularly the accurate index values ($n$) for the inner \sersic\ components
(see the thorough analyses and remarks in \S7 of \citealp{2010AJ....139.2097P}),
for about a dozen galaxies with complex morphologies
(such as NGC 2611, Mrk 1511, J1533$+$4432 and J1605$+$4526 in the CL and RM samples,
as well as two mild ones in the MagE sample --- J1113$+$0007 and J1343$+$2435;
see their demonstrations in Figure~\ref{fig:galfit}),
we adopt the NG fitting approach.
The fitting procedure is the same as described in detail in \S7 of \citet{2010AJ....139.2097P}.
All the inner components (roughly corresponding to bulges or pseudo-bulges) of the galaxies
can be well modeled with a \sersic\ model (morphed or not),
and the outer components (the commonly called ``galactic disks'') usually
need two morphed \sersic\ models.
To the purpose of this work, we do not list all the fitting results here.
Instead, because for their outer components
we only need the global/averaged \sersic\ indexes (see Tables~\ref{table:galaxy} and \ref{table:clagn}),
again we run a second fitting for every one of those galaxies
with two or more \sersic\ models for the outer components in their aforementioned NG fittings.
In the second fit of an image, we fix the inner component to
the best fit of its first NG fitting
and use just one (morphed) \sersic\ model to fit the outer.
Then we adopt the best-fit \sersic\ indexes of the second fittings as the final ones
for the outer components.

The final $g$, $r$ and $i$ magnitudes of the host-galaxy starlight (i.e., with AGN light removed),
as well the \sersic\ indexes of the inner and outer (if present) components,
are summarized in Tables~\ref{table:galaxy} and \ref{table:clagn}.
The listed magnitudes have been corrected for Galactic extinction using the dust map of \citet{1998ApJ....500..525S}
and the reddening curve of \citet{1999PASP..111...63F};
we list the magnitudes without k-corrections on purpose to avoid the uncertainty caused by k-correction.
The global (namely whole-galaxy) $g-i$ colors of the host galaxies
are also listed in Table~\ref{table:galaxy}, with the k-corrections applied with
the public code $kcor$ provided by \citet{2010MNRAS.405.1409C}.
Six galaxies in the CL sample are of too small size (Petrosian radius $<4$\arcsec\ in the $r$ band)
to get reliable indexes for their inner \sersic\ components
(e.g., J1358$+$4934 particularly, which is small, and to be worse has a bright AGN and a relatively high redshift);
for them we mark their inner \sersic\ indexes with a superscript `$^?$' in the table.
For several additional galaxies that are of relatively small size
or somehow with a bright AGN, or at a little bit high redshift,
it is not certain whether their outer \sersic\ component is present or not,
and we denote it with a `$?$' symbol in the Outer \sersic\ Index column.
If our best-fit model for a galaxy does not require an outer \sersic\ component,
we denote with a `--' symbol in that column.
Mentioned in passing, J0915$+$4814 (in the CL sample) has a very small outer \sersic\ index of $\approx$0.1,
which actually indicates that there exists an outer ring with a radius of $\approx$8\arcsec.



Regarding the \mgal\ estimation for the host galaxies,
our strategy is as follows.
According to Tables~\ref{table:galaxy} and \ref{table:clagn},
all the 6 MagE sources and most of the low-$z$ CL sources have a small fraction of AGN emission
to the total (AGN $+$ starlight) light ($<10$\% in the $r$ band).
Thus, we uniformly estimate \mgal\ for the sources of the two samples
based on their NIR magnitudes and do not perform AGN--starlight decomposition into their NIR images.
\footnote{For several sources in the low-$z$ CL sample, the AGN fraction in the $r$ band exceeds 10\%,
thus our estimation based on NIR magnitudes
would overestimate the \mgal\ values to some degree
(albeit less significantly than in the $r$ band). But this overestimation only weakens
our related conclusion in \S\ref{subsec:redcolor}.}
We use the mass-to-light ratio in the $K$ band ($M/L_{\rm K}$),
which is relatively insensitive to either dust absorption or stellar population age.
The $M/L_{\rm K}$ formula was calibrated by \citet{2013MNRAS.430.2715I}
and involved the $K_{\rm S}$ magnitude and $g-i$ color,
and \mbox{log\,$M/L_{\rm Ks} =$} $0.794(g - i) \,-0.997$ (with a scatter of $\pm$0.1 dex; see their Table 3).
All of the above sources were detected in the Two Micron All-Sky Survey
(2MASS, \citealp{2006AJ....131.1163S}),
and most are extended sources.
we prefer the photometric data in the 2MASS Extended Source Catalog (XSC),
for the emission from the whole galaxies;
as for some sources not in XSC, we then use their 2MASS Point Source Catalog (PSC) data.
Galactic extinction corrections and the k-corrections were performed
on the $g$, $i$, and $K_{\rm s}$ magnitudes for the \mgal\ estimation.
As for the sources in the RM sample, we just use the \mgal\ data available
in the two references---\citet{2018ApJ...864..146B} and \citet{2011ApJ...739...57K},
both with AGN contamination removed; see Table~\ref{table:rmAGN}
(except one source, IC 4218, for which no ready \mgal\ is available, and we estimate it
in the same way as for the MagE and low-$z$ CL sources).
For sources listed in \citet{2018ApJ...864..146B},
we adopt their \mgal\ values estimated using the \citet{2013MNRAS.430.2715I} formula,
consistent with our above estimation.
If a source has \mgal\ in both references,
we prefer \citet{2018ApJ...864..146B}, because their AGN--starlight decomposition
was based on NIR images observed with HST.
\newline

\section{Results and Discussion\label{sec:results}}


\subsection{Large Variability of MagE Sources\label{subsec:magebHa}}

\noindent $\bullet$ Broad-\ha\ and continuum variability.\\
 \indent The broad-\ha\ fluxes of the six variable MagE sources
change by  $1.3 - 3.0$ times (namely $0.3 - 1.2$\,mag) on a timescale of years,
exceeding the common variability amplitudes observed in reverberation mapping experiments
(\citealt{Kelly-2009ApJ...698..895K}; \citealt{Walsh-2009ApJS..185..156W}).
First of all, we need to verify that
the broad-\ha\ changes of the six sources are real,
not due to instrumental effect.
Because the spectral resolution of our MagE observations is
more than twice that of the SDSS (as well as better observing condition),
there are possibilities that the broad-\ha\ fluxes of the followup MagE spectra
are different from the SDSS ones, either increasing
(e.g., recovering the broad-\ha\ wing to a deeper flux density level)
or decreasing (e.g., extracting a finer or even recognizing a false profile).
Thus we conducted ensuing spectroscopic observations
instead with similar (or somehow inferior)
spectral resolution and observing conditions to the SDSS,
for as many sources as possible
by P200/DBSP and Xinglong 2.16m/BFOSC (see \S\ref{subsubsec:specfor6agn}).
For all the four MagE sources with ensuing lower-resolution spectroscopy,
their DBSP or BFOSC observations confirmed the presence of broad-\ha\ lines,
with S/N $> 10$ (see Table~\ref{table:lineinfo});
moreover, the measured broad-\ha\ fluxes
are still higher than their SDSS ones,
consistent with the increase of their MagE fluxes over the SDSS.
Hence their broad-\ha\ flux changes are secured.
For the remaining two sources (with RA smaller than 12 hr),
we did not find opportunities to take decent spectra
with spectral resolution smaller than the MagE ones.
Fortunately for J1113$+$0007, there are two SDSS spectra
taken with a time separation of about one year;
the later one gives a larger broad-\ha\ flux larger by 1.6 times
than the former one (albeit with a 2$\sigma$ significance only,
according to the flux difference and errors
listed in Table~\ref{table:lineinfo});
this increasing trend is consistent with
the MagE observation taken six additional years later.

The genuineness of their broad-\ha\ flux changes,
as we mentioned in the definition of this large-variability MagE sample (\S\ref{subsubsec:MagEsample}),
is supported by their continuum variability in the optical, MIR, and/or X-ray.
Actually, the optical and MIR light curves displayed in Figure~\ref{fig:light_curve}
are seriously diluted by host-galaxy starlight,
and any dramatic variability is not expected.
But we are still able to observe significant variability of several sources.
First,  
as listed in Table~\ref{table:lightcurve},
all of the light curves by ASAS-SN and WISE for the six MagE AGNs
satisfy $\sigma^2_{\rm rms}-err(\sigma^2_{\rm rms}) >0$,
indicative of continuum variability (see \S\ref{subsec:LCdata}).
Furthermore, in terms of the threshold for ``strong variability'' by \citet{2015ApJS..217...26B},
$F_{\rm var} > 0.1$ (see also \S\ref{subsubsec:MagEsample}),
three sources (J0839$+$0724, J1113$+$0007 and J1412$-$0307) satisfy it
according to their ASAS-SN light curves;
this is remarkable because those optical light curves are actually dominated by starlight.
Third, 
according to the peak-to-valley changes
measured from the binned light curves
(corresponding to $R_\mathrm{max}$ in Table~\ref{table:lightcurve}),
there are considerable optical or MIR continuum variabilities as follows
(only listing those changing by $>0.2$\,mag peak to valley):
J0839$+$0724, 0.3\,mag in the $W2$ band;
J1113$+$0007, 0.4\,mag in the $W2$ band;
and the most dramatic one, J1412$-$0307,
which is in fact the delimiter of this MagE sample
(i.e., its broad-\ha\ $R_\mathrm{max} \approx 1.3$),
0.45\,mag in the ASAS-SN $V$ band, and
0.61\,mag in the $W2$ band
(see also \S\ref{subsubsec:MagEsample}).
Such a large change in the WISE $W2$ band
reaches the same variability degree as the CL AGNs in the literature
\citep[e.g.,][]{2017ApJ...846L...7S}.
Lastly,
among the three sources having two- or multiepoch X-ray observations,
the two-epoch fluxes of J1442$+$0119 change by 2.74 times,
and J1257$+$2724 and J1412$-$0307 exhibit peak-to-valley changes by $>20$ times.
We should note that
strong/extreme variability in the X-ray usually has different origins from
the strong variability or CL events in the optical of the same AGNs,
e.g., due to variable obscuration by dust-free gas (see, e.g., \citealt{Risaliti+2009});
even in the cases without variable X-ray obscuration,
the connection may be rather complicated between X-ray and optical (CL) variabilities
(see, e.g., \citealt{Ricci+2021}).
Thus, the observational support from the strong variability in the X-ray of the three MagE sources
is not conclusive. But for the three sources at least their X-ray spectra exhibit
little gas absorption (see below).

\noindent $\bullet$ Evidence against variable obscuration.\\
 \indent The variability of the MagE sources is not likely
 due to variable dust obscuration;
this is supported by several lines of evidence.
The direct evidence
comes from the constancy of the broad-line Balmer decrements (\bhatobhb)
of their optical spectra taken at different epochs.
As \citet{2008MNRAS.383..581D} demonstrated,
the intrinsic (unreddened) ratios of Balmer recombination lines from the BLR
have a very small scatter in the normal radio-quiet AGN population,
and thus the observed broad-line Balmer decrements can be used as
an indicator of dust obscuration in the broad emission lines.
If the dust obscuration of a source is time variable, then the measured Balmer decrements
are variable accordingly; equivalently, the contrapositive holds.
J1257$+$2724, J1342$+$2435, and J1442$+$0119 have reliable broad-\hb\ measurements
in their multiepoch spectra (see Table~\ref{table:lineinfo}).
For J1257$+$2724, the \bhatobhb\ ratios of the four spectra
(ordered by increasing observing Date) are
3.58$\pm$0.63, 3.52$\pm$0.18, 3.71$\pm$0.17, and 3.79$\pm$0.34, respectively.
For J1342$+$2435, the ratios of its three spectra are
3.40$\pm$0.51, 3.42$\pm$0.36, and 2.93$\pm$0.34.
For J1442$+$0119, the ratios are 6.23$\pm1.79$, 5.88$\pm$0.44, and 6.78$\pm$1.01.
All the three sources have insignificant variation in \bhatobhb,
well within 1$\sigma$ uncertainty, during those years.
Besides, for every object,
the consistency of the \mbh\ values estimated from the multiepoch spectra
is also against the scenario of variable obscuration
(see \S\ref{subsec:agn-properties} below).

Another line of evidence, for the three sources with X-ray observations,
comes from their X-ray spectral fitting:
the model with free intrinsic absorption is not favored by the data
of any spectra (either in high- or low-flux states),
and even if the free intrinsic absorption is added into the model,
the best-fit $N_H$ has a very small impact on the fitting
(see \S\ref{subsec:XraySpecFitting} for the detail).

It is worth noting that the X-ray spectral shape of J1257$+$2724 is soft
($\Gamma = 2.3$) in the high-flux state
and gets hard ($\Gamma = 1.1$) in the low-flux state.
This behavior may be similar to the soft-to-hard state transition
at $\lratio \approx$ a few $\times 10^{-2}$
found in Galactic X-ray binaries \citep{2003MNRAS.345L..19M};
if so, the variability may be due to the change of accretion state.

\noindent $\bullet$ Spectral type transitions.\\
 \indent  
 As stated in \S\ref{subsubsec:specfitting} (see also Table~\ref{table:lineinfo}),
 we classify every spectra of the 6 sources into spectral subtypes
 according to the traditional definition
 \citep{Osterbrock1981ApJ_subtypes,1992MNRAS.257..677W,2001A&A...374...92V}.
 The spectral type transitions of the six sources
 are not as ``dramatic'' as prototypical CL AGNs;
 i.e., not between type~1 and type~1.8--2 with the dramatic appearance or disappearance
 of a strong \hb\ component \citep{LaMassa+2015ApJ}.
J0839$+$0724 (type~1.9), J1113$+$0007 (type~1.9 or 2),
J1342$+$2435 (type~1.2) and J1442$+$0119 (type~1.5)
do not show significant type transitions.
J1257$+$2724 varied between type~1.5 and 1.2\,.
J1412$-$0307 transited from type~1.9 to 1.5 during a timescale of 15 years,
and then returned to type~1.9 recently.
In its type~1.5 state, the broad-\hb\ component is measured with S/N $> 10$.
As described above, J1412$-$0307 also showed significant continuum variabilities
in the soft X-ray, optical, and MIR;
in particular, the variability measures of its WISE $W2$ light curve
are similar to those of typical CL AGNs \citep[see, e.g.,][]{2017ApJ...846L...7S}.
Hence we regard it as a \textit{bona fide} CL AGN.
\footnote{We also tried the spectral subtype definition of
\citet[][see their \S4.1]{YangQ-WuXB-2018},
which is based on the S/N (or significance levels) of broad \ha\ and \hb.
It gave that J1412$-$0307 as transiting from type 1.8 to 1.0 and then back to 1.8,
safely being a CL AGN.}

\subsection{AGN Properties \label{subsec:agn-properties}}


For the six variable MagE AGNs (\S\ref{subsubsec:MagEsample}) and
the 31 low-$z$ CL AGNs (\S\ref{subsubsec:clagns}),
we begin with exploring their distribution
in the diagnostic diagrams of narrow-line ratios
(Figure~\ref{fig:bpt}),
the so-called BPT diagrams \citep{1981PASP...93....5B,2001ApJ...556..121K,2006MNRAS.372..961K,Kau-bpt_2003MNRAS.346.1055K},
which are a powerful tool to separate
Seyfert galaxies, low-ionization nuclear emission-line region
sources (LINERs; \citealt{1980A&A....87..152H}), and \hii\ galaxies.
Following \citet{Kau-bpt_2003MNRAS.346.1055K},
in the \oiii/\hb\ versus \nii/\ha\ diagram
we classify the sources below
the empirical demarcation line of
\citet[][the dashed line in panel a]{Kau-bpt_2003MNRAS.346.1055K}
as \hii\ galaxies,
sources between the \citet{Kau-bpt_2003MNRAS.346.1055K} line and
the theoretical maximum starburst line of \citet[][the dotted line]{2001ApJ...556..121K}
as transition objects,
and those above the maximum starburst line as pure Seyferts.
Commonly the transition objects are in fact Seyferts
with considerable contribution in the narrow emission lines
from star formation in the host galaxies.
Following \citet{2006MNRAS.372..961K}, we use the empirical line
in terms of \sii/\ha\ versus \oiii/\hb\
(the green dashed line in panel b)
to separate Seyferts and LINERs.
In the six MagE sources,
J0839$+$0724, J1113$+$0007, J1257$+$2724, and J1412$-$0307 are Seyferts;
J1342$+$2435 is a LINER;
concerning J1442$+$0119,
its two observations are in the LINER region
and the third observation in the Seyfert region,
and thus we deem it a 50\% probability LINER and 50\% probability Seyfert.
Thus, among the MagE sources
the LINER fraction (LINER/total) is 1.5/6 (25.0\%).
In the 31 sources of the low-$z$ CL sample,
22 are pure Seyferts, 1 transition object, 7 pure LINERs,
and the remaining 1 sits just on the boundary line between Seyferts and LINERs;
thus, the LINER fraction is 7.5/31 (24.2\%).

All of the 11 low-$z$ CL AGNs ($z<0.15$)
of \citet{2021ApJ...907L..21D} are included
in our low-$z$ CL sample.
Their BPT classifications are as follows:
four Seyferts, one transition objects, and six LINERs.
Thus, the LINER fraction of the low-$z$ CL sample of \citet{2021ApJ...907L..21D}
is 6/11 (54.5\%), much higher than our low-$z$ CL sample.

We also plot the four CL quasars of \citet{2019ApJ...876...75C} on the BPT diagrams.
It turns out to be two Seyferts and two transition objects.

The sources in our low-$z$ CL sample have \mbh\ estimates in the literature (see the \mbh\ values
and references in Table~\ref{table:clagn}),
and there are estimations and discussions of their accretion rates (or \lratio)
and accretion-rate changes (or state changes) also in the literature.
Regarding the sources in the RM sample,
there are RM-based \mbh\ measurements and corresponding \lratio\ estimates
(see the references in Table~\ref{table:rmAGN}).
Thus here we only describe the estimation for the MagE sources.

We follow the common practice to estimate the virial masses
based on the measured luminosity and line width of the broad-\ha\ emission line.
The mass formalism is given by \citet[their Equation\,(6)]{2011ApJ...739...28X},
which is based on \citet{2005ApJ...630..122G} but incorporates the updated relation
between broad-line region size and AGN luminosity calibrated by \citet{2009ApJ...697..160B}.
The \mbh\ values of the six MagE AGNs are between $10^{6}-10^{7.3}$\msun\,(see Table~\ref{table:lineinfo}).
We find that for any source the derived \mbh\ values from the spectra of different epochs
are consistent with each other, with difference $\lesssim 0.3$\,dex,
well within the measurement uncertainty of the viral mass estimation method.

To further estimate their \lratio,
we use the conversion,
$\log L_{\rm bol} = 0.91 \log \lambda L_\lambda(5100\AA) + \log(0.75) + 4.89$
\citep[their Equations~11 and 13]{2012MNRAS.422..478R} to get the bolometric luminosity ($L_\mathrm{bol}$),
where $\lambda L_\lambda(5100\AA)$ is derived from the \ha\ luminosity
\citep[their Equation~1]{2005ApJ...630..122G}.
The \lratio\ values of the six sources are between 0.0075 to 0.038 (see Table~\ref{table:lineinfo}).

Besides, for the three sources with X-ray measurements (\S\ref{subsec:XraySpecFitting}),
J1257$+$2724, J1412$-$0307, and J1442$+$0119,
we also use their $L_{X}$(2--10\,keV) to estimate $L_\mathrm{bol}$ and \lratio.
The bolometric correction relation of \citet{2004MNRAS.351..169M} is used, which
are $\approx 10$ for the $L_{X}$(2--10\,keV) of the three sources.
We use the mean value of different-epoch \mbh\ values of each source
to calculate their Eddington luminosity (\ledd).
For J1257$+$2724, the \lratio\ of the high- and low-flux states is 0.023 and 0.012, respectively;
for J1412$-$0307, it is 0.016 and 0.0029.
For J1442$+$0119, there is only one measurement, and its \lratio\ is 0.0072.
Those X-ray based \lratio\ estimates
are quite consistent with the broad-\ha\ based ones.

The variable MagE sources stay in the relatively low-accretion regime,
with $0.01 \lesssim \lratio \lesssim 0.04$,
i.e., definitely not in or close to the Eddington accretion.
We note that a similar low-accretion trend
was found in EVQs and CL quasars.
\citet{2018ApJ...854..160R} found that EVQs with $\Delta g > 1$~mag
have lower \lratio\ than the control sample of quasars matched in redshift and luminosity.
\citet{2019ApJ...874....8M} reported that CL quasars
with a large optical variability ($|\Delta g| >1$~mag and $|\Delta r| > 0.5$~mag) have
lower \lratio\ compared with the overall quasar population.
Such a trend of large- and extreme-variability AGNs in a relatively low-accretion state
may shed light on the underlying accretion-flow physics.
\newline



\subsection{Hosts Redder than General Seyfert 2 Galaxies \label{subsec:redcolor}}



Broadly viewed in Figure \ref{fig:sdssimage},
almost all of the host galaxies of the large-variability AGNs in the three samples
are in secular evolution, without ongoing major-merger activity (except two in the RM sample).
It is easy to judge that they are not luminous elliptical galaxies;
many of them have more or less signatures of galactic disks, such as bars and spirals,
while some may be (low-luminosity) S0 or spheroidal galaxies.
These characteristics are consistent with the general trend of local AGNs
and also the general trend of local galaxies \citep{2013ARA&A..51..511K}.
Beyond the 0th-order general similarity, however,
there appears a perceivable discrepancy from our general impression of the host galaxies of local AGNs:
the images in Figure~\ref{fig:sdssimage} are seldom blue, but slant seriously toward red;
this is particularly true when looking at their inner regions.

Based on the derived quantities in \S\ref{subsec:imaging} (magnitudes, colors, inner and/or outer \sersic\ indexes,
\mgal, \dbreak, \hda\ and SFR; corrected for AGN contamination),
below we investigate the host-galaxy properties and
explore any possible clues of these quantities against or for the large variability of the central AGN.
We first utilize the diagnostic tools based on the global (namely whole-galaxy) quantities (\S\ref{subsubsec:whole-color-Mgal}),
and then invoke the tools based on the quantities of the inner regions (i.e.,
quantities derived from the fiber-aperture spectra,
properties of the inner \sersic\ imaging components, etc.; see \S\ref{subsubsec:inner-region-diagrams}).

\subsubsection{In Terms of Whole-galaxy Properties \label{subsubsec:whole-color-Mgal}}

A traditional tool to diagnose galaxies is the diagram of global color versus \mgal\
\citep[e.g.,][]{2003MNRAS.341...54K,2007ApJS..173..512S}.
Here we use rest-frame $g-i$ to represent the color,
just following \citet{2019ApJ...876...75C} in their study of the host galaxies of CL quasars;
this is because the SDSS $g$, $r$ and $i$ images have similar quality
(better than $u$ and $z$) and
the wavelength span between $g$ and $i$ is the largest among all combinations.
To represent the general color$-M_{\star}$ parameter space populated by local Seyfert hosts,
we use the $\sim$26000 low-$z$ Seyfert 2 galaxies
homogeneously selected from SDSS DR4 by \citet{Dong-2010ApJ...721L.143D}.
The $g-i$ values of those Seyfert 2 galaxies are calculated from
the SDSS $g$ and $i$ Petrosian magnitudes, accounting for both Galactic extinction and k-correction.
The stellar masses of the Seyfert 2 galaxies are basically derived
in the same way as described in \S\ref{subsec:imaging}.
A fraction of Seyfert2 galaxies have no $2MASS$ $K_{\rm S}$ data,
and we then use
their WISE $W1$ 3.4\micron\ luminosity to estimate \mgal,
with the mass-to-light ratio, $\log\,M_{\star}/\nu L_{\nu}(3.4\micron)$ calibrated
by \citet{2013MNRAS.433.2946W}.

In addition, we retrieve the data of the 4 CL quasars at $z \gtrsim 0.2$ \citep{2019ApJ...876...75C},
and the 52 Seyfert 1 hosts at $0.145 < z < 0.3$ in the SDSS Stripe 82 region
provided by \citet{2015MNRAS.454.4103B} for comparison.

Figure \ref{fig:C-M-diagram} shows the distributions on the $g-i$ versus $M_{\star}$ diagram.
Seyfert 2 galaxies, being the reference sample,
are plotted as background (gray dots).
We plot the contours of the Seyfert 2 distribution, representing
90\% (the outermost), 50\%, and 20\%, respectively, the number fraction of enclosed sources.
We also group the Seyfert 2s into nine \mgal\ bins,
and plot the median $g-i$ colors of every bins and corresponding standard deviation
(black solid circles with $\pm1\sigma$ error bars).

The majority of the large-variability AGNs in any samples
are redder than the median-color line of general Seyfert 2s (black dashed line):
3 of the 5 MagE non-CL variable sources (60\%),
22 of the 31 $z<0.15$ CL AGNs
(including 30 $z<0.15$ CL AGNs in Table~\ref{table:clagn} and
the MagE CL AGN J1412$-$0307; 73.3\%)
\footnote{Here we formally take the single MagE CL AGN from
the category of MagE large-variability (non-CL) sources
and bring it into the CL category.
But just as we point out in footnote~\ref{ftn:CLandLV},
we actually need not to elaborate such strict separation between the two categories
in this study concerning their host-galaxy properties.
Anyway, this formal operation involves only one source (J1412$-$0307),
and does not impact the statistics of
either the MagE sample or low-$z$ CL sample.}, 
and 10 of the 15 RM sources (66.7\%).
We perform 2-dimensional Kolmogorov-Smirnov (KS) tests between the variable-AGN samples
and the reference Seyfert 2 sample.
The $p$-value (change probability)
between all the three variable-AGN samples (51 sources in total)
and the Seyfert 2 sample is 0.0021, that
between the the above 31 CL sources
and the Seyfert 2,
0.0035\,. 
That is, there is a statistically significant difference in the color--\mgal\ diagram
between large-variability AGN hosts and general Seyfert 2 galaxies.
\newline

Regarding the galactic structure of these host galaxies,
according to our 2D imaging fittings
it is clear that they are predominantly disk galaxies with a bulge or pseudo-bulge.
Among the 24 galaxies with both reliable inner \sersic\ and outer \sersic\ indexes
in the MagE and CL samples (see Tables~\ref{table:galaxy} and \ref{table:clagn}),
there are merely about four or even fewer spheroidal galaxies.
\footnote{If low-luminosity (dwarf) galaxies are well fitted by
	a single \sersic\ with a small index (e.g., $n=1.32$ for J1636$+$4102),
then it is somehow arbitrary to classify them as pure disk or spheroidal galaxies.}
In light of the limited resolution and depth of the SDSS images,
as well as the sample incompleteness induced,
we refrain from advanced explorations of
the structural and morphological properties of the host galaxies.

Just as we stated in the beginning of this \S\ref{subsec:redcolor},
although there are a few galaxies in the three samples showing somehow blue colors
(say, $g-i<1$; see also the galaxies below the line of the median colors
of Seyfert 2 galaxies in the $g-i$ versus \mgal\ diagram),
their central regions are almost all red.
For instance,
the two MagE sources with the bluest global color
in the sample (J0839$+$0724 and J1113$+$0007)
have $g-i <1$ (see Figure~\ref{fig:C-M-diagram});
however, if we use the SDSS fiber magnitudes to derive their colors,
then their $g-i$ values (after Galactic reddening correction and k-correction)
would be 1.17 and 1.28, respectively,
and well above the median-color line of Seyfert 2 galaxies in Figure~\ref{fig:C-M-diagram}.
As it is not easy to perform AGN--starlight decomposition to the fiber magnitudes,
we do not delve into the use of fiber magnitudes 
but use instead the quantities based on fiber spectra
and the properties of the inner \sersic\ imaging components in the following.

\subsubsection{In Terms of Inner-region Properties \label{subsubsec:inner-region-diagrams}}


For the six MagE sources and all except two sources in the CL sample,
\footnote{Concerning the two CL sources with the AGN fraction in the $r$ band
	$>30$\% (see Table~\ref{table:clagn} and \S\ref{subsubsec:specfitting}),
	their SDSS spectra have significant AGN contribution
	and cannot give accurate \dbreak\ and \hda\ of the host galaxies.}
we measured the \dbreak\ and \hda\ values of the host galaxies on the basis of their SDSS spectra.
The SDSS fiber aperture is 1\farcs5 in radius, corresponding to 1.47 kpc
at the median redshift ($z=0.05$) of the 35 MagE and CL sources.  

As for the large-variability RM sources,
because of the lack of useful spectra
as stated in the last part of \S\ref{subsubsec:specfitting},
we did not use them in the investigation concerning the inner regions of the host galaxies.

First, we employ the traditional diagnostic diagram of \dbreak\ and \hda.
Figure \ref{fig:d4000-hda} shows the distributions
of the MagE and low-$z$ CL AGNs,
together with non-AGN galaxies and Seyfert 2 galaxies as comparison.
The distribution of Seyfert 2 galaxies is peaked at \dbreak\
$\approx 1.3$ and \hda\ $\approx 1.7$\AA.
All but one MagE source are outside the densest 40\% contour of Seyfert\,2s,
with larger \dbreak\ and/or smaller \hda;
this trend is similarly followed by the low-$z$ CL sources.
We perform a two-dimensional K-S test to the distributions in the \hda\ versus \dbreak\ diagram.
The $p$-value (change probability)
between the low-$z$ CL plus MagE AGNs and Seyfert 2 galaxies are 0.0007.
If we apply to the Seyfert 2 sample the same redshift cut $z<0.15$
as the definition of the low-$z$ CL sample,
then the $p$-value is 0.0025.
Thus the difference between large-variability AGNs and general Seyfert 2s
is significant in terms of the \hda\ versus \dbreak\ diagram.
The difference is more significant than in the color--\mgal\ diagram, just as we expected.

It is easy to see the advantage of using inner-region properties over whole-galaxy properties.
For instance, among the five bluest galaxies in the low-$z$ CL sample,
with global color $0.81 \leqslant  g-i \leqslant 0.95$ (see Table~\ref{table:clagn}),
four (except J1126$+$5134) are bluer than the median-color curve of Seyfert 2 galaxies
in the diagram of global color and \mgal.
In contrast, now in the diagram of \hda\ and \dbreak,
four (except J1316$+$3015) have $\dbreak >1.5$ indicating the mean stellar age
older than \mbox{1\,Gyr} (for the inner-region stars),
and four (except J1126$+$5134) are not above the median-\hda\ curve of Seyfert galaxies.
By the way,
the latest IFU observation by MUSE/VLT of Mrk\,590
demonstrated that most of the stars
within the central 10$\arcsec$ (4.9\,kpc)
have older ages $>$5~Gyr \citep{2019MNRAS.486..123R}.
\newline


Because those old so-called Lick indexes such as \dbreak\ and \hda\
do not directly denote the physical properties of galaxies,
and because the measurement errors of \dbreak\ is
considerable --- particularly considering the small dynamic range of \dbreak\
for the entire galaxy population,
below we employ the relatively new diagnostic tool: the diagram of SFR versus \mgal,
which is favored by recent studies (e.g., \citealt{Yu-Shi-2020MNRAS.498.3985Y},
\citealt{2021ApJ...907L..21D}).

In Figure~\ref{fig:sfr-mgal}, the blue dashed line represents the so-called SFMS
with the slope and intercept values calibrated by
\citet[][see their Eq.~4]{Chang-SFR-2015ApJS..219....8C},
which are almost the same as those given by \citet{Jin-2016MNRAS.463..913J}.
The two green dotted lines mark the boundary of the green valley,
which are given by \citet{Jin-2016MNRAS.463..913J} and \citet{ChenYM-2016NatCo...713269C},
as follows:
the upper boundary line,
\mbox{$\log \mathrm{SFR} / \msun\,\mathrm{yr}^{-1}$}
$= 0.86\,\log \mgal / \msun\, -\, 9.29 \,$,
is just the 1-$\sigma$ (0.5~dex) downward of their SFMS line;
the lower one,
\mbox{$\log \mathrm{SFR} / \msun\,\mathrm{yr}^{-1}$}
$= \log \mgal / \msun\, -\, 14.65 \,$.
The red dashed line represents the quiescent sequence
given by \citet[][see their Figure~2]{ChenYM-2016NatCo...713269C},
\mbox{$\log \mathrm{SFR} / \msun\,\mathrm{yr}^{-1}$}
$= \log \mgal / \msun\, -\, 15 \,$.
It is clear that
Seyfert 2 galaxies are basically located in the green valley,
or in other words, are below the SFMS line predominantly (by $>90$\% in number).
All six MagE sources, in SFR, are below the median SFRs of Seyfert 2 galaxies.
All except two of the low-$z$ CL sources are below the median SFR line of Seyfert 2s.
Thus, it is unquestionable that large-variability (including CL) AGNs have smaller SFRs than
Seyfert 2 galaxies of similar stellar masses.
Our large-variability AGNs are mostly Seyfert 1 galaxies
(not biased to being LINERs)
according to the BPT diagram (see \S\ref{subsec:agn-properties}).
On the other hand, Seyfert 1s are the same as Seyfert 2s
in terms of the cold-gas content and star formation activity of their host galaxies
(e.g., \citealt{Zou+2019ApJ...878...11Z};
even for the high-luminosity counterparts namely quasars,
\citealt{ShangguanHo2019ApJ...873...90S}).
Thus, we conclude that the host galaxies of large-variability and CL Seyferts
are in the red tail (i.e., the gas-poor, SFR-deficit tail)
of the general Seyfert galaxy population.

We performed various checks and tests to our above conclusion.
First of all, because the SFR values we use are based on
the calibration of \citet{2019ApJ...882...89Z} with AGN-emitted \oii\ flux being removed,
we replot the SFR--\mgal\ diagram using SFR values from the [O\,{\footnotesize II}] based estimator
without accounting for such nonstellar contamination;
the result is displayed in Figure~\ref{fig:sfr-mgal} (left panel),
and clearly the above trend changes little.
We tried several other SFR estimators, and the conclusion is the same.
We also tested by using the 22 pure Seyferts in the low-$z$ CL AGN sample,
and again the conclusion remains intact.
Recently, as described in \S\ref{sec:intro},
\citet{2021ApJ...907L..21D} presented a surprising result:
local Seyfert 2 galaxies fall on the SFMS, whereas their $z<0.15$ CL AGNs
are on the green valley of the SFR--\mgal\ diagram.
Because their $z<0.15$ CL AGN sample is small (11 sources in total)
and seriously dominated by LINERs (6 LINERs, accounting for 54.5\%; see \S\ref{subsec:agn-properties}),
we believe that their host-galaxy difference is mainly between LINERs and Seyferts,
and thus we do not make further comparison.

We noticed that there were attempts to the link central concentration of host galaxies to
CL variability \citep[e.g.,][]{2021ApJ...907L..21D}.
We plot such a diagram of inner \sersic\ index and \mbh\ (see Figure~\ref{fig:sersic_mbh})
using our carefully fitted \sersic\ indexes for the MagE and low-$z$ CL sources (\S\ref{subsec:imaging})
as well as the data for the four CL quasars at $z\gtrsim 0.2$ presented by \citet{2019ApJ...876...75C}.
Inspecting both our Figure~\ref{fig:sersic_mbh} here and Figure~3 of \citet{2021ApJ...907L..21D},
we can see that neither our low-$z$ sources nor the 
CL quasars of \citet{2019ApJ...876...75C} are different from
the reference AGN sample (namely Seyfert 2s) in this respect.
\footnote{Please see Figure~3 of \citet{2021ApJ...907L..21D} for the distribution of
the reference Seyfert 2 sample. Unlike Type 1 AGNs,
the \sersic\ indexes of those Seyfert 2s
(taken from the catalog of \citealt{2011ApJS..196...11S}) did not suffer from
the contamination of AGN emission.
}
That is, the claim is not supported for a connection between high stellar density in the core region
(high central concentration) and CL AGN phenomenon.
Again, just as we warn in Section \ref{sec:intro} against the unconscious use of the SFR values
in the ready-made catalogs, there is a similar caveat here:
the galactic bulge and disk parameters
listed in the ready, mass-produced catalogs \citep[e.g.,][]{2011ApJS..196...11S}
were fitted with a general scheme aimed at normal galaxies,
and thus special treatments are usually required for specific galaxies,
particularly when the galaxies of interest are
significantly nonaxisymmetric and/or have a Type 1 AGN
(see \S\ref{subsec:imaging}).
In particular, the abnormally large \sersic\ indexes ($n>4$) are generally due to
poor fitting (e.g., when a galaxy is not well resolved).
\newline

{\it Difference between CL quasars and local CL Seyferts.}
More interestingly, we note that CL quasars (namely high-luminosity AGNs)
seem to be different from low-$z$ CL/large-variability Seyferts in host-galaxy properties.
In terms of the diagrams of color versus \mgal, \hda\ versus \dbreak,
and SFR versus \mgal\ (see Figures~\ref{fig:C-M-diagram}, \ref{fig:d4000-hda} and \ref{fig:sfr-mgal}),
the host galaxies of the four CL quasars of \citet{2019ApJ...876...75C}
are relatively young, above the upper SFR boundary of the green valley (consistent with the SFMS);
the host galaxies of the six CL AGNs at $z>0.15$ of \citet{2021ApJ...907L..21D}
hold similar properties.
There is a trivial factor contributing to this difference
between CL quasars at relatively high redshifts versus CL Seyferts at low redshifts:
the aperture effect, wherein the spectroscopic aperture takes more outer-region starlight
for high-$z$ AGNs than for low-$z$ AGNs, and the outer regions of galaxies are generally
bluer and younger and have more SFRs than their inner regions.
But this factor seems unlikely to explain here,
because even in terms of whole-galaxy properties
(global color, \citealt{2019ApJ...876...75C}; global SFR, \citealt{2021ApJ...907L..21D})
see also \S\ref{subsubsec:whole-color-Mgal})
the difference between CL quasars and local CL Seyferts is evident already.
It seems that there is no difference in global colors and SFR
between CL quasars and normal quasars. 

Thus we tend to believe that the host-galaxy difference is real and physical
(albeit the evidence comes only from the above 10 CL quasars in total):
such a difference corresponds to the difference in variability pattern, namely
secular pattern (low-$z$ Seyferts) versus high-amplitude tail (EVQs and CL quasars) as stated in
the second paragraph of \S\ref{sec:intro}.
If this were the case, there should be a deep physical link from nuclear fueling flows to
the structure of accretion disks; we will follow this line of thought
in \S\ref{subsec:AD-Fuels},
in detail for the part of the local large-variability (CL) Seyferts 
and briefly for CL quasars and EVQs.
\newline

\subsection{Dependence of Accretion Disks on Nuclear Fuels \label{subsec:AD-Fuels}}

Why do the accretion rates of these large-variability (CL) AGNs, indicated in the continuum and emission-line
light curves, change so significantly on timescales of years?
This is the current hotly debated theoretical question
concerning the physics and structure of accretion disks
\citep[e.g.,][]{2018MNRAS.480.4468R,2019MNRAS.483L..17D,yanfei-2020ApJ...900...25J},
which is beyond the scope of this work.
Rather than exploring accretion disks that directly produce the optical continuum emission,
here, following the discussions on Mrk 1018 and Mrk 590
\citep[e.g.,][]{2014ApJ...796..134D,2016A&A...593L...9H,2019MNRAS.486..123R},
we would like to discuss broadly from the perspective of nuclear fuel supply.
This is inspired by two factors:
the preference of large-variability (including CL) Seyferts for red, SFR-deficit (gas-poor) galaxies
as discovered in this work
and the curve of fueling rate as a function of time produced by numerical simulations of
nuclear fuel supply from the $\sim$1pc scale down toward central massive black holes
\citep[e.g.,][]{2006MNRAS.366..358C,2008MNRAS.383..458C,2018MNRAS.478.3544R}.
Also, this is along the line of thought suggested by the host-galaxy difference
between CL quasars and CL Seyferts
as discussed in \S\ref{subsubsec:inner-region-diagrams}:
a deep physical link from nuclear fueling flows (or fueling modes)
to accretion disks.
At the end of this subsection,
we discuss to some extent
the difference between CL quasars and local CL Seyferts.
\newline

\subsubsection{Scenario for Local CL/Large-variability
Seyferts (Nuclear Famine Fueling):
Cold-clump Formation and Episodic Accretion}

Low-$z$ Seyferts are generally triggered and fueled by various secular processes
(internal or environmental; \citealt{2004ARA&A..42..603K}),  
and their AGN activity does not depend on the host-galaxy properties on larger scales
than, e.g., nuclear star clusters or even smaller structures
\citep{2008ARA&A..46..475H,2013ARA&A..51..511K}.
\citet{2009MNRAS.397..135K} proposed two distinct regimes of AGN fueling in nearby galaxies.
One is the ``feast mode'' in blue galaxies with huge deposit of cold gas
ready on $\gtrsim$1\,pc scales, which could sustain a steady fueling flow inward.
The other one is the so-called ``famine mode'' in red galaxies with old stellar populations,
where the SMBH fuel supply is mainly from slow stellar winds produced by evolved stars
\citep[e.g.,][]{2007ApJ...671.1388D},
or from external cold gas
via minor mergers or via accretion of cold-gas clumps in the intergalactic medium
particularly when the gas-poor galaxies are in small galaxy groups
as realized later
(e.g., \citealt[their \S8.1, \S8.2.2 and \S8.2.3]{Davies+2014}; \citealt{Davies+2017}).
In this context, the 51 large-variability (including CL) AGNs in the 3 samples
of the present study should be in the ``famine mode,'' being in old red galaxies.

We must point out that
the theoretical scenario we propose here is
actually a modified and narrowed version of
the ``feast vs. famine'' notion of \citet{2009MNRAS.397..135K}:
while \citet{2009MNRAS.397..135K} took into account the amount of
cold gas in galactic bulges (i.e., on kiloparsec scales),
we adopt the inference of \citet{King&Pringle07Fuelling} that
only the cold gas in the nuclear region (on $\lesssim$ 1 pc scales)
can affect BH accretion.
\footnote{Although the smallest spatial scale
observationally probed in the present work
is still on the bulge or pseudo-bulge scale (the spectra
as well as the decomposed inner \sersic\ imaging component;
\S\ref{subsubsec:inner-region-diagrams}),
we are convinced by the arguments
of \citet{King&Pringle07Fuelling} for local Seyferts
and believe that the related cold-gas supply
is on the $\lesssim$ 1pc scale. See also \citet{2008ARA&A..46..475H}
and \citet{2013ARA&A..51..511K}.
} 
Thus we deliberately use the terms ``nuclear feast'' and ``nuclear famine.''
In the nuclear famine mode, whether from slow stellar winds
or externally from the galactic environment,
the information on the cold-gas origin is still retained in the nuclear region
(see, e.g., \citealt{Hobbs+2011MN});
in other words, the BH accretion process does not need larger-scale information than
the nuclear cold gas can provide.
In the nuclear feast mode, the cold-gas supply on a $\approx$1 pc scale is
much more than required to feed the central BH,
and the accretion-disk properties do not care
what is the origin of the nuclear cold gas,
whether via wet major mergers or secularly from the galactic disks of late-type galaxies
(see, e.g., \citealt{Davies+2014}).

However, presently there is no observational census for cold gas on $1$\,pc or smaller scales of AGNs,
and the fueling passage from the \mbox{1\,pc} scale (i.e., AGN tori, if present)
down to the outer boundary of accretion disks is beyond the capacity of current observing facilities.
It is just now that even ALMA has been launching observations for nearby AGNs
and can merely resolve the molecular tori of only a few parsecs \citep[e.g.,][]{2019A&A...623A..79C},
let alone the subparsec fuel flows.

Fortunately, we can get insights from the three-dimensional simulations of the fueling passage
on these scales in the literature, although those simulations are
for the fueling of Sgr A$^{*}$ (as a dim AGN) in the Galactic center
where the fuel is fast stellar winds from young stars
\citep{2006MNRAS.366..358C,2008MNRAS.383..458C,2018MNRAS.478.3544R}.
According to \citet[see their Fig.~10]{2018MNRAS.478.3544R},
the accretion rate measured at $1.5 \times 10^{-4}$ pc
(namely 370 times the Schwarzschild radius of the Sgr A$^*$ BH)
can vary by a factor of 9 within tens of years.
This is consistent with the simulation of
\citet[see their Figures 1--3 where the time sampling is every 30 years]{2008MNRAS.383..458C}.
\citet{2008MNRAS.383..458C} clearly demonstrated that
while the accretion-rate curve of hot gas is smooth and has only small-amplitude fluctuation,
the episodic infall of cold-gas clumps produces {\em sharp peaks} in the accretion rate!
If the time resolution gets higher (e.g., without averaging of snapshots; see \citealt{2006MNRAS.366..358C}),
the increasing of accretion rate by cold clumps would get more steep and abrupt,
and the duration of the peaks could be shorter (e.g., $\lesssim$ a few years).

We can imagine that such sharp accretion-rate peaks
\footnote{To put more precisely, as described above, the fueling rate peaks.}
should be more common in red galaxies,
which do not have young stars but AGB and red giant stars to produce slow stellar winds
(accordingly, the AGN is not as dim as Sgr A$^{*}$);  
see, e.g., \citet{2006MNRAS.366..358C} and \citet{2014ApJ...782..103S} for detailed discussions.
Certainly, if we make an analogy of AGN fueling by those simulations of Sgr A$^*$,
those sharp peaks actually reflect merely reflect the variability in fueling rate
into the outer boundary of accretion disks
and will be modulated (or smoothed?) by accretion disks later.
At present it is unclear what the signal of the fueling-rate peaks would look like
in the optical continuum emission of accretion disks.
On the other hand, in fact,
the notion of episodic fueling and thus episodic accretion
was perceived and formulated from different perspectives
(\mbox{R.\,Davies} 2021, private communication).
Initially in the study of cooling flows in galaxy clusters,
there are theoretical models, numerical simulations and observations
that cold-gas clumps
(or called blobs, streams, filaments or alike in the literature)
can condensate out of the hot intergalactic gas,
and several researchers have further argued that
some of the cold clumps can sink toward the central BHs of galaxies
(e.g., \citealt{Pizzolato&soker2005ApJ}; \citealt{Gaspari+13CCA}).
In particular, \citet{Gaspari+13CCA} coined the notion of ``chaotic cold accretion,''
in which the large-scale (namely intergalactic and galaxy-scale)
cold clouds (clumps) can lose angular momentum
and some of them finally ``rain'' into the central BH
via recurrent collisions among clouds \emph{and} and between clouds and the clumpy AGN torus;
the raining is the actual accretion process.
They also proposed that chaotic cold accretion seems
to be an excellent model to explain AGN variability
(see their \S7.3).
\citet{King&Pringle06Chaotic}
put forward and derived a ``chaotic accretion'' mechanism,
characteristic of a series of small-scale, randomly oriented accretion events to feed BHs,
which was specifically applied to (and analyzed in)
the case of nearby Seyferts \citep{King&Pringle07Fuelling};
later on, following this line of thought,
\citet{Hobbs+2011MN}
put forward a concrete scenario emphasizing the role of turbulence.
Turbulence is also important in the simulations of subparsec fueling of
the aforementioned \citet{2006MNRAS.366..358C,2008MNRAS.383..458C}
and \citet{2018MNRAS.478.3544R}.  
After their long-term observational studies on nuclear fueling of AGNs,
\citet{Davies+2014,Davies+2017} proposed that
there are two modes of inflows feeding low-redshift AGNs,
one being quasi-continuous
with a plentiful internal supply of gas (e.g., in gas-rich spiral galaxies),
the other being stochastic events accreted from the galactic environments
(e.g., important for gas-poor galaxies in moderately dense galaxy groups).
\footnote{Davies \etal\ based and focused their explorations on their observations
and thus put emphasis on external accretion events (listing it as
the only case in their second fueling mode), which is the the main fueling process
for their target galaxies, namely S0 galaxies in small galaxy groups with 5--20 members.}
We can see that, besides our approach on AGN variability,
the notion of two (nuclear) fueling modes
has been reached from various perspectives in recent years.
Again, just as we point out in the second paragraph of this scenario,
we would like to stress that for local CL and large-variability AGNs
only the nuclear fuel (on $\lesssim 1$ pc) scale matters.

A last point we would like to mention:
in red galaxies it is probable that
even the broad-line region \textit{per se}\/
is episodic \citep[see][]{2014ApJ...796..134D}.
When the aforementioned cold-gas clumps fall into the right radii,
they get ionized accordingly by the AGN continuum
and become effective in producing certain optical broad lines.
This is an interesting picture, and can be modeled
by taking the ``locally optimally emitting clouds''
(LOC; \citealt{1995ApJ...455L.119B}) approach,
just as, e.g., \citet{2004ApJ...606..749K} did.
Without data from either observations or modelings so far,
we however refrain from a full discussion on episodic 
broad-line regions in this paper.
\newline

\subsubsection{Speculation for CL Quasars (Nuclear Feast Fueling)}
\indent
More excitedly, we cannot help thinking about the trend wherein
the host galaxies of CL quasars and EVQs appear quite different from that of 
local CL Seyfert galaxies.
Although the number of such quasars with host-galaxy properties analyzed
is small (10 sources at most; \citealt{2019ApJ...876...75C}, \citealt{2021ApJ...907L..21D}),
there are reasons to believe this trend is real (\S\ref{subsec:redcolor}).
We think this trend is consistent with the difference between local CL/large-variability Seyferts
and EVQs in variability pattern
(secular variation versus high-amplitude tail),
both differences being suggestive of
a deep link between nuclear fuel supply and the structure of accretion disks.
We speculate that the ``feast mode'' of nuclear fueling
may account for both the preference of CL quasars and EVQs for blue galaxies
and the corresponding accretion-disk structure required by CL quasars and EVQs
(see, e.g., \citealt{yanfei-2020ApJ...900...25J}).
With more than enough cold gas available on $\approx$1 pc scale
(recalling AGN tori as the fuel reservoir),
surely the structure of the accretion disks in the feast fueling mode
is starkly different from that of the accretion flows for local large-variability (CL) Seyferts described in the above.
We have to admit this idea for CL quasars and EVQs is quite speculative,
without solid supports from either observations or simulations so far,
and thus we defer the exploration for the future.


\subsection{New Thinking on the Variability Selection for IMBHs\label{subsec:newthinking}}

The initial goal of the project branching out from this work
is to search for IMBHs.
We now return to the IMBH topic
with the implication of the finding of this work to
the IMBH searching by optical variability.
In the flowering time-domain astronomy era, variability selection for low-mass AGNs is promising
(see \S4.5 of \citealt{Greene+2020araa} for a brief review).
This field is just developing, and there are only a few such searches based on optical variability so far
\citep{2016PASJ...68...40M,2018ApJ...868..152B,2020ApJ...896...10B,2020arXiv200310457G,2020ApJ...889..113M}.

A general trend in those studies is that the optical continuum variability is generally of low level,
less than \mbox{0.1\,mag}, i.e., with the peak-to-valley flux ratio being $<1.1$ \citep[see][]{2020ApJ...889..113M}.
Note that the ``nucleus magnitudes'' (or fluxes)
reported in \citet{2018ApJ...868..152B,2020ApJ...896...10B} and \citet{2020ApJ...889..113M}
are measured through a small aperture (matching to the seeing in order to collect AGN flux),
either by adding together the fluxes of the difference image and template image for each data point
\citep{2018ApJ...868..152B,2020ApJ...896...10B},
or by directly performing aperture photometry on the source images \citep{2020ApJ...889..113M};
this is consistent with the data points for the  nuclear magnitudes/fluxes displayed in most light curves,
such as those used in this work.
To clarify possible confusion, we would like to mention in passing the following point.
\citet{2016PASJ...68...40M} used a different convention to report the data points of their light curves
(see the ``Flux (sub)'' and ``Mag (sub)'' columns in their Table 1):
they adopted the same difference imaging methodology as in \citet{2018ApJ...868..152B},
but they presented the fluxes and magnitudes measured from the difference images
(i.e., with the template subtracted, which is actually more physically motivated).
Thus \citet{2016PASJ...68...40M} reported a large peak-to-valley magnitude change $\Delta g = 0.42$
(i.e., a flux change by 1.5 times),
but this change is with respect to the flux scale of 1\,$\mu$Jy;
in fact the peak-to-valley flux change in physical units is tiny, 1.2\,$\mu$Jy, un-surprisingly.
Recently, \citet{2020arXiv200310457G} reported a low-mass AGN at  $z = 0.823$,
identified from the Dark Energy Survey (DES) Supernova field by optical variability.
Reading from the light curve (their Figure 1), the peak-to-valley magnitude change is really large,
$\Delta g \approx0.5$\,mag (varying around $g\approx22.0$,
PSF magnitude measured from the source images).

Besides, in the literature optical-variability-selected low-mass AGNs as a sample
were reported to tend to be in galaxies bluer than
the general galaxy population \citep{2018ApJ...868..152B},
which agreed with the conclusion of \citet{2016ApJ...826...62H} on the host-galaxy
colors of their variability-selected AGNs.
Moreover, variability-selected low-mass AGNs of \citet{2020ApJ...896...10B}
tended to be even bluer (in $g-r$ color) than the low-mass AGNs selected
in terms of optical narrow lines by \citet{2013ApJ...775..116R},
and mostly resided in the star-forming region of the BPT diagram.
All the above reports on variability-selected low-mass AGNs
are inconsistent with the discovery of our present work.
We are not aware of the reason of the inconsistency at this point.  

Instead, we would like to note that
the present work reminds us not to ignore red galaxies.
It is rewarding to search for low-mass AGNs in red galaxies by optical variability.
First of all, as discovered in the present study,
low-mass AGNs in red galaxies would have continuum variability of
larger amplitude than those in blue galaxies;
or in other words, red galaxies have a larger fraction hosting variable low-mass AGNs
than blue galaxies.  
Additionally, in stark contrast to blue galaxies, red galaxies have little
star-formation dilution of the AGN emission.
We could think a little bit further about designing this kind of variability search.
Galaxy groups and even clusters would be ideal target fields,
where red galaxies such as S0 and spheroidal galaxies
are plentiful, and thus the survey efficiency and productivity would be high.
Besides, if the above interpretation invoking episodic fueling (\S\ref{subsec:AD-Fuels}) is correct,
the environmental secular processes
red galaxies suffering in galaxy groups and clusters \citep{2004ARA&A..42..603K}
can enhance the intermittency in fuel supply \citep{Davies+2017}.
\newline

\section{Summary \label{sec:summary}}

During our spectroscopic MagE campaign initially planned to search
for intermediate-mass black holes (IMBHs) in nearby broad-line AGNs,
we unexpectedly found six variable AGNs with relatively small black hole masses
($\mbh \approx 10^{6-7}$\msun),
their broad-\ha\ fluxes varying by $1.3 - 3.0$ times (namely $0.3 - 1.2$\,mag) on timescale of years.
Most surprisingly, among our broad-line AGNs identified by MagE (15 in total),
those hosted by blue galaxies generally vary little,
whereas a significant fraction of those hosted by red galaxies exhibit large broad-\ha\ variability;
in other words, the six variable sources are predominantly in red galaxies.

The above unexpected ``bonus'' motivated us to explore AGNs with large variability
(including CL AGNs), aiming at their host-galaxy properties
(particularly as to any connections between those properties and
the CL and large-variability AGN phenomenon)
in a systematic way.
We collected all the low-$z$ changing-look AGNs available in the literature
and RM AGNs with broad-\ha\ $F_\mathrm{var} >0.1$,
and performed careful imaging and spectral fittings.
From our observational investigations,
we draw the following two main conclusions about the connection between host-galaxy properties
and the CL and large-variability AGN phenomenon:
\begin{itemize}
	\item Local CL and large-variability AGNs (mainly Seyferts)
	reside in redder, more SFR-deficit (presumably gas-poor) galaxies
	than the control sample of local Seyfert 2 galaxies.
	That is, the host galaxies of those strongly variable Seyferts are
	in the red tail (i.e., the gas-poor, SFR-deficit tail)
	of the general Seyfert galaxy population.
	
  \item In contrast, there is a significant trend that their more luminous counterparts,
    namely CL quasars (CLQs) and extremely variable quasars (EVQs),
    are different from local CL Seyferts
	in host-galaxy properties.
	For instance, in terms of the diagnostic diagram of global color and \mgal,
	the host galaxies of CLQs are generally blue (see also \citealt{2019ApJ...876...75C});
	in terms of the diagram of SFR and \mgal\
	(of the inner regions, Figure~\ref{fig:sfr-mgal}; see also \citealt{2021ApJ...907L..21D}),
	local CL Seyfert galaxies are located in the green valley,
	whereas CLQ hosts are in the so-called SFMS.	
\end{itemize}

These two discoveries inspired our theoretical thinkings
about the physical link between nuclear fuel supply and accretion-disk structure,
and about the implication to the field of IMBH research in turn.
We proposed a physical scenario for local CL and large-variability Seyferts,
a speculation for CLQs and EVQs,
and an implication for IMBH searches based on optical variability, as follows:
\begin{itemize}
	\item We presented an explanation for the preference of
     local CL and large-variability Seyferts
	to old red host galaxies from the perspective of the nuclear fueling mode,
	which is a modified and narrowed concept of ``famine mode'' proposed by \citet{2009MNRAS.397..135K}.
    The concrete mechanism may be revealed by three-dimensional simulations of the fueling passage
    from the 1 pc scale down to the outer boundary of accretion disks.
    In similar existing simulations, cold-gas clumps can be formed stochastically
    in the fueling flow on $\lesssim 1$ pc scales. 
    While the accretion-rate curve of hot gas is smooth and has only small-amplitude fluctuation,
    the episodic infall of the cold-gas clumps produces sharp peaks
    in the accretion rate (measured at the outer boundary of the accretion disk).
	We discussed the feasibility of this scenario, namely the time scales of the rising, lasting, and falling of
    this kind of cold-clump accretion activity (\S\ref{subsec:AD-Fuels}).

	\item We speculated that
	the ``nuclear feast mode''
	may account for both the preference of CL quasars and EVQs to blue galaxies
	and their variability pattern (high-amplitude tail of the continuous distribution)
	that is different from the secular variation of local CL Seyferts.
	With more than enough cold gas piled up on $\approx$1 pc scale
	(say, AGN tori as the reservoir),
    surely the structure of the accretion disks in the feast fueling mode
    (see, e.g., \citealt{yanfei-2020ApJ...900...25J})
    is starkly different from that of accretion flows for local CL Seyferts.
    We defer the exploration along this line of thought for the future.

	\item We proposed a new thinking on the design of optical-variability selection for IMBHs:
	to launch variability searches in red galaxies.
	This strategy would be more efficient than usual blind surveys.
	And it can be regarded as a kind of deliberate debiasing and reminder,
	because the variability-selected low-mass AGNs so far
	tend to be in blue galaxies.
\end{itemize}

We really became excited by the observational discoveries about the connection between
host-galaxy properties and the large-variability (including CL) AGN phenomenon,
triggered by the unexpected ``bonus''  (the six variable low-mass AGNs) out of our
observing campaign actually aimed at IMBHs;
also excited ourselves by the ensuing theoretical thinkings listed above.
There appear to be a lot of lines of fruitful work for the future.
From an observational standpoint, the direct measurement of
cold-gas content of the host galaxies---particularly on $\lesssim 1$ pc
scale---of such strongly variable Seyferts and quasars is in demand;
comparison studies of considerably large samples, i.e., being statistically robust,
are necessary.
Through numerical experiments,
it would be instructive to carry out three-dimensional simulations
for different nuclear fueling modes,
simulating the fueling passage from $\approx$1 pc scale
down toward (the outer boundary of) the accretion disks,
or even handling both the nuclear fueling flows
and some part of accretion flows simultaneously.
And finally, theoretical insights (see \citealt{Antonucci_2018NatAs...2..504A})
are always needed.
\newline

We thank the anonymous referee for a helpful report that significantly improved this paper.
We also thank Richard Davies for helpful comments, particularly on the two modes of AGN fueling;
C. Martin Gaskell and Ski (Robert Antonucci)
for valuable comments and discussions, particularly on accretion
and on early discoveries and facts long ignored (ahead of their time),
Ning Jiang for his guidance on the 2D-decomposition of SDSS images with GALFIT; and
Fuguo Xie for advice and discussions during the course of this work.
This work is supported by the Natural Science Foundation of China grants (NSFC 11703079, 11873083)
and the "Light of West China" Program of Chinese Academy of Sciences (CAS).
Su Yao acknowledges support by an Alexander von Humboldt Foundation Fellowship.
D.W.X. and J.W. are supported by the Natural Science Foundation of China grant 11773036.
This work has made use of the spectra obtained with the Magellan Baade Telescope/MagE, Hale Telescope/DBSP,
and Xinglong 2.16m telescope/BFOSC. 
The Hale Telescope/DBSP data were obtained through the Telescope Access Program (TAP),
which has been funded by the National Astronomical Observatories, Chinese Academy of Sciences,
and the Special Fund for Astronomy from the Ministry of Finance.
Observations obtained with the Hale Telescope at the Palomar Observatory were obtained as part of an
agreement between the National Astronomical Observatories, Chinese Academy of Sciences,
and the California Institute of Technology. This project is supported by the National Natural Science
Foundation of China (NSFC-11421303, NSFC-11603021, and NSFC-11833007).
We also acknowledge the support of the staff of the Xinglong 2.16m telescope, which was partially supported by
the Open Project Program of the Key Laboratory of Optical Astronomy, National Astronomical Observatories,
Chinese Academy of Sciences.

\facilities{Magellan:Baade (MagE), Hale (DBSP), Beijing:2.16m (BFOSC)}

\appendix

\section{Variabilities of the Six MagE AGNs \label{sec:appendix}}

In this Appendix, we present two parts of the data analyses for the six variable MagE sources:
analysis of the multiband light curves and X-ray spectra fitting.

\subsection{Multiband Light Curves of MagE sources \label{subsec:LCdata} }

We construct the multiband light curves for the six MagE sources using
all of the photometric data publicly available (see Figure~\ref{fig:light_curve}).

The optical light curves were constructed using
the data from Catalina Real-Time Transient Survey (CRTS) \citep{2009ApJ...696..870D} and
the $V$-band and $g$-band magnitudes from  
ASAS-SN \citep{2014ApJ...788...48S,2017PASP..129j4502K}.
CRTS is one of the largest time-domain optical surveys currently operating,
which is performed using unfiltered light
and nominally transformed to the $V$-band zero point.
ASAS-SN is a long-term project designed to monitor the entire sky on a rapid cadence to
find nearby supernova and other bright transients,
providing $V$-band and $g$-band photometric data.
We first removed the data points with large uncertainties
and then binned the data with a bin size of 90 days,
which roughly correspond to the natural observing gaps (see Figure~\ref{fig:light_curve}).
In every panel, every original observed data point
is plotted together with their $\pm1\sigma$ error bars.
For each binned data point, we report the median value within a bin,
and the error is the sum of two parts (in quadratic form):
(1) the random error of the mean,
calculated according to the error propagation formula
from the quoted statistical errors of every measured data points in the bin,
and (2) the standard error of the mean, i.e.,
the standard deviation (namely difference) of the measured data points divided by $\sqrt{N}$
($N$ being the number of the data points in a bin).

The MIR light curves are constructed using the $W1$ and $W2$ of
the WISE \citep{2010AJ....140.1868W} and
NEOWISE-R \citep{2014ApJ...792...30M} surveys.
Following \citet{2017ApJ...846L...7S}, we removed bad data points
with poor image quality (``qi\_fact'' < 1) or
with flagged moon masking (``moon mask'' = 1).
We binned the data points every half a year.
The average values and their errors in every bin are calculated
in the same way as the above optical light curves.

The X-ray data are obtained from the literature and archives,
by XMM-Newton, Chandra, and ROSAT.
Three sources, J1257$+$2724, J1412$-$0307, and J1442$+$0119,
have multiple observations over the past decades.
For the purpose of variability, we simply use the X-ray fluxes
retrieved from 4XMM-DR9 catalog \citep{2020A&A...641A.136W},
Chandra source catalog \citep[CSC,][]{2010ApJS..189...37E},
and ROSAT Catalogs
(\citealp{1994AAS...185.4111W} and \citealp{2018MNRAS.473.4937S} for J1257$+$2724;
\citealp{2003A&A...399..351P} for J1412$-$0307;
\citealp{2007AJ....133..313A} for J1442$+$0119).
The {\it XMM-Newton} and {\it Chandra} observations
cover the energy range 0.2--10 keV, while ROSAT
only covers 0.1--2.4 keV.
To build the X-ray light curves,
we adopt the 0.2--2\,keV X-ray fluxes for XMM-Newton and Chandra observations
and convert the ROSAT fluxes into the same energy range
using the measured spectral slope of each source
(or the mean slope if two or more slopes are measured; see \S\ref{subsec:XraySpecFitting}).
The XMM-Newton 0.2--2 keV fluxes are calculated
by summing up the fluxes of
EPIC band 1 (0.2--0.5\,keV), band 2 (0.5--1.0\,keV), and band 3(1.0--2.0\,keV) in the 4XMM-DR9 catalog.
The Chandra 0.2--2.0keV X-ray fluxes are calculated
by summing up the fluxes of ``U\_Flux\_Ap'', ``S\_Flux\_Ap'', and ``M\_Flux\_Ap'' in CSC.
\newline

To characterize the variability of those multiband light curves,
we calculate a set of variability statistics,
which (except for $R_\mathrm{max}$) are otherwise not applicable
when the data points are few (such as the multiepoch broad-\ha\ flux data).
Besides the maximum variability $R_\mathrm{max}$ and fractional variability amplitude $F_{\rm var}$
mentioned in \S\ref{subsubsec:MagEsample},
we use a third measure, normalized excess variance ($\sigma^2_{\rm rms}$),
and its error (see, e.g., \citealt{2020ApJ...889..113M}).
Below we give their functional definitions.

$F_{\rm var}$, called fractional variability amplitude historically in the literature
(see \citealt{Edelson+2002}),
is defined to be the squared root of the excess variance of the total light curve
(the excess is presumably the intrinsic,
with the variance of random measurement errors subtracted),
 which is then divided by the mean of the light curve,
 \begin{equation}\label{eq:fvar}
    F_\mathrm{var} = \frac{\sqrt{\sigma^2 -\overline{\delta^2}}}{\overline{x}} ~.
 \end{equation}
Here the quantity $\sigma^2$ is the variance of the light curve
(i.e., of all the observational data points),
$\overline{x}$ is the mean,
and $\overline{\delta^2}$ is the mean of the squared errors
($\delta_i$) associated with the observations ($x_i$), i.e.,
 \begin{equation}
   \overline{\delta^2} = \frac{1}{N}\sum_{i=1}^{N}\delta^2_i ~,
 \end{equation}
where $N$ is the number of the observational data points
(fluxes or magnitudes) of the light curve.
$F_{\rm var}$ is commonly used in the literature of both X-ray time-series analysis
and optical reverberation mapping.
Yet it has two limitations: (1) by definition, $F_{\rm var}$ cannot apply when
$\sigma^2 < \overline{\delta^2}$ (see Table~\ref{table:lightcurve});
(2) because both $\sigma^2$ and $\overline{\delta^2}$
involve summing over all the data points of the light curve,
it does not appear necessary (and even not good in principle)
to do the summing twice separately.

Thus we introduce also $\sigma^2_{\rm rms}$, called the normalized excess variance,
which is similar to $F_{\rm var}$ being overall measures of
the intrinsic variability of light curves that correct for
measurement errors from photon counting and detector read noise,
but is free from the above two limitations.
The definition is straightforwardly the calculation of
the excess (presumably intrinsic) variance,
normalized by the square of the mean ($\overline{x}$), as follows:
\begin{equation}
  \sigma^2_{\rm rms}=\frac{1}{N\,\overline{x}^2}
  \sum_{i=1}^{N}[(x_{i}-\overline{x})^2-\delta_i^2]  ~.
\end{equation}
The error of $\sigma^2_{\rm rms}$, due to Poisson noise, is as follows (\citealt{2020ApJ...889..113M}),
\begin{equation}
  err(\sigma^2_{\rm rms}) = \frac{S_{D}}{\overline{x}^2 \, \sqrt{N} } ~,
\end{equation}
 \begin{equation}
 S_{D}=\frac{1}{N\,\overline{x}^2} \sum_{i=1}^{N}
      \left\{[(x_{i}-\overline{x})^2 -\delta_i^2] - \sigma_{\rm rms}^2 \,\overline{x}^2 \right\}^2 ~.
 \end{equation}
Light curves with $(\sigma^2_{\rm rms}- err(\sigma^2_{\rm rms})\,)>0$ can be regarded
to be intrinsically variable (\citealt{2020ApJ...889..113M}).


All the above measures are calculated for every light curves, and listed in Table~\ref{table:lightcurve}.
$R_\mathrm{max}$, being the simplest and not accounting for measurement errors,
 is calculated based on the binned light curves (see the above);
 the other are are calculated based on the original data points and their associated measurement errors.

\subsection{X-ray Spectra of MagE sources\label{subsec:XraySpecFitting}}


As described above, J1257$+$2724, J1412$-$0307, and J1442$+$0119 were observed by XMM-Newton or Chandra.
This subsection analyzes their X-ray spectra to check their flux changes and to investigate whether there are any
the possible absorption features.

As demonstrated in Figure \ref{fig:light_curve}, both J1257$+$2724 and J1412$-$0307 have a dozen X-ray observations and
have shown violent variability in X-ray during the past 30 yr.
For J1257$-$2724, we select XMM-Newton observation runs of 2006 June 14 and 2010 December 5 to extract the X-ray spectra,
which have a high enough S/N and represent the low- and high-flux states, respectively.
For J1412$-$0307, in a similar way, we extract the the XMM-Newton spectrum from the observation on 2008 July 27 as
its high-flux state. Its lowest-flux states appear in the observations between 2001 and 2004,
which however have too few statistics to provide a meaningful constraint on the absorption.
Thus we select the observation on 2015 as a test of the absorption in the low-flux state.
The third source, J1442$+$0119, has only one observation run, by Chandra.
For the XMM-Newton data, the spectra are extracted preferentially from PN for its large effective area.
When PN data are not available, we combine spectra from two MOS CCD arrays and response files.
The XMM-Newton spectra are re-binned so that each bin achieves S/N $\geqslant 3$.
For the Chandra data, the spectra are extracted following standard procedures.
The Chandra spectra are rebinned so that each bin contains at least 25 counts.

We started with a single power-law model with hydrogen absorption fixed at the Galactic value for each source,
which gives a well fit to all the spectra.
  Then we added into the model an intrinsic hydrogen absorption component with the column density ($N_{\rm H}$) being a
  free parameter. But this model does not improve the fit in all cases (the five spectra), and the best-fit
$N_{\rm H}$ values are all small, as follows.
For each source, the best-fitting models give upper limits of $N_{\rm H}$ at 90\% confidence
level:
  $< 4\times10^{20}$ and $< 2.8\times10^{21}$cm$^{-2}$
  for the high-state and low-state spectra of J1257$+$2727, respectively;
$< 7\times10^{19}$ and $< 3.6\times10^{20}$cm$^{-2}$ for the high and low states of J1412$-$0307, respectively;
$< 1.1\times10^{21}$cm$^{-2}$ for J1442$+$0119.
Thus we suggest that the intrinsic absorption is insignificant.


We also tried adding a thermal component ({\tt{bbody}} in {\sc{Xspec}}) to the
model for J1257$+$2727 and J1412$-$0307.
It turns out that only the fitting to the high-state spectrum of J1412$-$0307
can be improved in a statistical sense, with an decrease $\Delta\chi^{2}=9$ 
after the adding of two more free parameters
(i.e., the decrease of the degree of freedom $\Delta\,\mathrm{dof} =2$);
the $F$-test is marginally significant, with the $p$-value being 0.017.
This improvement may indicate a possibility
that the X-ray variability can be explained by the variation of a thermal component or similar.
Yet at present only a single spectrum (the high-flux one) can
marginally constrain this additional component,
and thus we cannot say anything about the variation of this component.
Therefore, we leave the advanced investigation of X-ray properties for a future paper,
and in this work adopt the single power-law fitting as the final results.

Figure~\ref{fig:xrayspec} shows the X-ray spectra and their best fits by the single power-law model,
as well as the respective residuals.
For J1257$+$2724, the best-fit photon indexes $\Gamma$ of its high- and low-flux states are
2.3 and 1.1, respectively.
Its highest and lowest \mbox{2--10} keV luminosities are 4$\times 10^{41}$ \ergs and 2.2$\times 10^{41}$ \ergs.
For J1412$-$0307, the best-fit photon indexes keep at 1.8 in both its high and low states, and its
highest and lowest \mbox{2--10} keV luminosities are 2.4$\times 10^{42}$ \ergs and 4.5$\times 10^{41}$ \ergs.
The photon index of J1442$+$0119 is 1.64, with the \mbox{2--10} keV luminosity of 3.92$\times10^{41}$ \ergs.
\newline

\clearpage


%
\begin{deluxetable}{cccccCCCCCCCCCCCCCCCCC}
	\rotate
	\tablewidth{0pt}
	\tabletypesize{\tiny}
	\tablecaption{Observations and Measurements of the Six Large-variability MagE Sources \label{table:lineinfo}}
	\tablehead{
		\colhead{Name} & \colhead{$z$} & \colhead{Obs. Date} & \colhead{Instru.} & \colhead{Exp.} &
		\multicolumn{10}{c}{Flux} & \colhead{FWHM\,\tablenotemark{$+$}} &
		\colhead{log $M_{\rm BH}$} & \colhead{$L_{\rm bol}/L_{\rm Edd}$} & \colhead{Type\tablenotemark{$\vartriangle$}} & \colhead{CL?\tablenotemark{$\triangledown$}}\\
		\colhead{} & \colhead{}  & \colhead{} & \colhead{(UT date)} & \colhead{} & \colhead{(s)} &
		\multicolumn{10}{c}{(10$^{-17}$~erg~s$^{-1}$~cm$^{-2}$)} & \colhead{(\kms)} &
		\colhead{(\msun)} & \colhead{} & \colhead{}\\
		\cline{7-16}
		\colhead{} & \colhead{} & \colhead{} & \colhead{} & \colhead{}  &
		\colhead{[O\,{\tiny III}]$\lambda5007$} & \colhead{[O\,{\tiny I}]$\lambda6300$} &\colhead{[O\,{\tiny II}]$\lambda3727$} &
		\colhead{[N\,{\tiny II}]$\lambda6583$} & \colhead{[S\,{\tiny II}]$\lambda6716$} & \colhead{[S\,{\tiny II}]$\lambda6731$} &
		\colhead{\hb$^{\rm n}$} & \colhead{\hb$^{\rm b}$} &\colhead{\ha$^{\rm n}$} &\colhead{\ha$^{\rm b}$} &
		\colhead{\ha$^{\rm b}$} &   \colhead{} & \colhead{}  & \colhead{} & \colhead{} & \colhead{}
	}
	\startdata
	J083909.65+072431.5 & 0.0465 & 2004-03-17 & SDSS & 2040 & 237\pm7& 13\pm3 &72\pm13 &113\pm4 &32\pm4 &32\pm4 &26\pm4 & 37\pm13 & 94\pm5 & 179\pm45 & 1964 & 6.09 & 0.0382 & 1.9 & no\\
	                    &        & 2017-03-25 & MagE & 2400 & 237\pm3& 13\pm1 &37\pm3  &67\pm2  &21\pm1 &22\pm1 & 3\pm1 & 13\pm8 & 17\pm2 & 138\pm21 & 2878 & 6.38 & 0.0129 & 1.9 & \\
	\hline
	                    &   & 2000-02-28 & SDSS & 2700 & 238\pm7& 25\pm4 & 144\pm37 & 168\pm6 & 66\pm10& 47\pm5 & 50\pm5 & 0.3\pm31  &245\pm7 & 160\pm61 & 6217 & 7.31  & 0.0075 & 2 & \\
	J111349.83+000733.9\tablenotemark{*}& 0.0787& 2001-03-16 & SDSS & 3601 & 314\pm7& 25\pm3 & 107\pm7 & 174\pm5 & 79\pm7 & 59\pm4 & 60\pm4 & 3\pm19 &221\pm5 & 257\pm28 & 4711 & 7.15 & 0.0123 &1.9 & no\\
	                    &  & 2017-03-25 & MagE & 1500 & 466\pm6& 25\pm2 & 105\pm13 & 166\pm5 & 82\pm5 & 66\pm4 & 31\pm3 & 19\pm9 &144\pm5 & 485\pm25 & 4332 & 7.20  & 0.0136  & 1.9 & \\
	\hline
	                &        & 2006-03-26 & SDSS & 3900 & 181\pm8 & 3\pm4 & \nodata & 160\pm6 & 39\pm5 & 31\pm5 & 28\pm4 & 143\pm46 & 137\pm6 & 433\pm33 & 2997 & 6.31 & 0.0112 & 1.5 & \\
	J125710.76+272417.6 & 0.0207 & 2017-03-25 & MagE & 2100 & 181\pm5 & 13\pm43 & 46\pm5  & 155\pm5 & 22\pm3 & 15\pm3 & 28\pm3 & 335\pm16 & 121\pm5 & 1216\pm64 & 2492 &  6.35  & 0.0202  & 1.2 & no\\
	                &        & 2019-06-24 & DBSP & 1800 & 181\pm5 & 7\pm3 3 & 49\pm28 & 151\pm5 & 30\pm3 & 18\pm3 & 22\pm3 & 372\pm38 & 96\pm5  & 1179\pm38 & 2094 & 6.19 & 0.0282 & 1.2 & \\
		 &        & 2020-01-01 & BFOSC& 3600 & 181\pm6 & 15\pm33 & \nodata & 53\pm9 & 32\pm4 & \nodata & 36\pm6 & 294\pm30 & 74\pm12 & 946\pm34  & 2590 &  6.33 & 0.0169 & 1.5 & \\
	\hline
	        &       &2007-05-07 &SDSS &5302 &92\pm6 &99\pm9 & 219\pm19 & 69\pm10&89\pm7 &77\pm7 & 30\pm7 & 225\pm28 &123\pm12 & 766\pm63 &2173 &6.24 & 0.0285 & 1.2 & \\
	J134245.69+243524.0\tablenotemark{*}& 0.0267&2017-03-25 &MagE &2100 &85\pm4 &99\pm7 & 216\pm17 & 71\pm5 &93\pm6 &81\pm5 & 32\pm5 & 317\pm30 &122\pm6 &1083\pm47 & 2246 & 6.34 & 0.0289 & 1.2 & no\\
	        &       &2019-06-24 &DBSP &1800 &92\pm4 &99\pm8 & 234\pm26 & 70\pm7 &95\pm5 &85\pm5 & 34\pm4 & 294\pm30 & 122\pm8 & 861\pm49 & 2210 & 6.28 & 0.0282 & 1.2 & \\
	\hline
                          &        & 2002-04-14 & SDSS & 2646 & 199\pm5 & 21\pm3 & 96\pm11 &87\pm4 & 41\pm4 & 37\pm5 & 25\pm4 & 27\pm21 & 95\pm5 & 250\pm27 & 4677  & 7.12  & 0.0095 & 1.9 & \\
      J141249.63-030720.9 & 0.0751 & 2017-03-25 & MagE & 2400 & 199\pm2 & 12\pm1 & 62\pm5  &67\pm2 & 33\pm2 & 27\pm2 & 24\pm2 & 122\pm9  & 87\pm3 & 299\pm24 & 3659  & 6.94  & 0.0158 & 1.5 & yes\\
                          &        & 2019-06-25 & DBSP & 1200 & 199\pm6 & 19\pm3 & 68\pm11 &82\pm4 & 30\pm3 & 30\pm4 & 20\pm2 & 21\pm18 & 90\pm4 & 330\pm30 & 4589  & 7.16  & 0.0101  & 1.9 & \\
	\hline
	&         & 2001-04-28 & SDSS & 3123 & 180\pm16 & 49\pm6 & 258\pm46 & 339\pm 12 & 120\pm8 & 122\pm8 & 90\pm6 & 88\pm28  & 364\pm13& 548\pm68 & 2456 & 6.37  & 0.0308 & 1.5 & \\
	J144242.63+011911.2 & 0.0337  & 2017-03-25 & MagE & 2400 & 180\pm5  & 42\pm3 & 136\pm5 & 331\pm4  & 136\pm2 & 127\pm2 & 71\pm3 & 138\pm14 & 323\pm4 & 814\pm26 & 2780 & 6.56  & 0.0237&1.5 & no\\
	&         & 2019-06-25 & DBSP & 1000 & 180\pm11 & 56\pm7 & 112\pm32 & 188\pm6   & 75\pm5  & 64\pm4  & 49\pm4 & 125\pm21 & 232\pm8 & 845\pm61 & 3784 & 6.85  & 0.0118 & 1.5 &
	\enddata
	{\textbf Notes:}
	\tablenotetext{+}{\scriptsize Corrected for instrumental broadening.}
	\tablenotetext{\vartriangle}{\scriptsize Spectral type classification
        according to the traditional definition (see \S\ref{subsubsec:specfitting}). }
	\tablenotetext{\triangledown}{\scriptsize Label for CL AGNs. J1412$-$0307 is identified as a CL AGN according
	to its spectral type transition and
    the variability measures of its multiband light curves (see \S\ref{subsec:magebHa}).}
	\tablenotetext{*}{\scriptsize For the two sources, J1113$+$0007 and J1342$+$2435,
		the spectral flux levels of their other spectra are scaled to their first SDSS
		spectra in terms of the [O\,{\tiny I}]$\lambda6300$ flux; for the rest sources,
        in terms of the [O\,{\tiny III}] $\lambda$5007 flux.}
\end{deluxetable}

\clearpage

\begin{deluxetable}{ccCCCCCCRCCCCRCC}
	\rotate
	\tablewidth{0pt}
	\movetabledown=0mm
	\tabletypesize{\scriptsize}
	\tablecaption{Galaxy Properties of the Six Large-variability MagE AGNs \label{table:galaxy}}
	\tablehead{
		\colhead{ID} &\colhead{Name} & \colhead{$g$} & \colhead{$r$} & \colhead{$i$} & \colhead{$g-i$} &
		\colhead{$f_\mathrm{PSF}$} & \colhead{$K_\mathrm{S}$} & \colhead{log $M_{\star}$} & \colhead{$R_\mathrm{petrosian}$} &
		\colhead{$\sigma_{\star}$} & \colhead{SFR} & \colhead{\dbreak} & \colhead{\hda} &
		\colhead{$n_1$} & \colhead{$n_2$}\\
		\colhead{}  & \colhead{}     & \colhead{(mag)} & \colhead{(mag)} & \colhead{(mag)} & \colhead{(mag)} & \colhead{(\%)} &
		\colhead{(mag)} &  \colhead{(\msun)} & \colhead{(arcsec)} &\colhead{(\kms)} & \colhead{(\msun\,yr$^{-1}$)} &
		\colhead{} & \colhead{(\AA)} & \colhead{Inner \sersic} & \colhead{Outer \sersic} \\
		\colhead{(1)}  & \colhead{(2)}     & \colhead{(3)} & \colhead{(4)} & \colhead{(5)} & \colhead{(6)} & \colhead{(7)} &
		\colhead{(8)} &  \colhead{(9)} & \colhead{(10)} &\colhead{(11)} & \colhead{(12)} & \colhead{(13)} & \colhead{(14)} & \colhead{(15)} & \colhead{(16)}
	    }
	\startdata
	1& J083909.65+072431.5 & 15.78\pm0.02& 15.25\pm0.03 & 14.95\pm0.01& 0.78\pm0.02 & 2.68& 12.98\pm0.08 & 10.34 & 12.00 &109\pm5 & 0.04 &1.58&-1.18 & 2.27\pm0.12 & 0.99\pm0.01 \\
	2& J111349.83+000733.9 & 16.72\pm0.01& 16.13\pm0.01 & 15.79\pm0.01& 0.82\pm0.01 & 5.54& 13.73\pm0.18 & 10.52 &  8.05 & 96\pm9 & 0.24 &1.42&0.99  & 2.15\pm0.08 & 0.40\pm0.02 \\
	3& J125710.76+272417.6 & 15.58\pm0.03& 14.89\pm0.02 & 14.53\pm0.01& 1.01\pm0.03 & 5.27& 12.35\pm0.07 & 10.10 & 11.18 & 81\pm4 &0.008 &1.40&0.61  & 2.77\pm0.13 & 1.04\pm0.02 \\
	4& J134245.69+243524.0 & 15.57\pm0.05& 14.91\pm0.03 & 14.51\pm0.02& 1.04\pm0.05 & 8.40& 12.74\pm0.08 & 10.16 &  9.31 &144\pm6 & 0.02 &1.47&-0.91 & 3.13\pm0.08 & 1.69\pm0.10 \\
	5& J141249.63-030720.9 & 17.67\pm0.09& 16.94\pm0.03 & 16.53\pm0.05& 1.02\pm0.10 & 2.54& 14.73\pm0.13 & 10.24 &  4.64 &120\pm7 & 0.11 &1.53&-1.54 & 1.51\pm0.06 & ? \\
	6& J144242.63+011911.2 & 15.90\pm0.02& 15.09\pm0.03 & 14.71\pm0.02& 1.14\pm0.03 & 1.50& 12.22\pm0.08 & 10.67 &  8.68 &155\pm6 & 0.09 &1.51&-0.98 & 2.06\pm0.06 & 1.04\pm0.02 \\
	\enddata
	\tablecomments{Col.(1) Identification number assigned in this paper.
 	               Col.(2) Target name.
	               Cols.(3)-(5) Host-galaxy magnitude from the GALFIT fittings (AGN contamination removed), with Galactic extinction corrected.
	               Col.(6) $g-i$ color of the host galaxies. Calculated from the GALFIT fittings, with Galactic extinction and k-corrections performed.
	               Col.(7) Fraction of AGN light to the total in the $r$-band images, according to our GALFIT fittings.
	               Col.(8) $K_\mathrm{S}$ magnitude from 2MASS, with Galactic extinction corrected.
	               Col.(9) Stellar mass of the host galaxies estimated from $K_\mathrm{S}$ luminosity using the calibration of \citet{2013MNRAS.430.2715I}.
	               Col.(10) Petrosian radius in the $r$ band given by the SDSS photometric pipeline.
		       Col.(11) $\sigma_{\star}$ measured from the MagE spectra.
	               Col.(12) Star formation rate of the host galaxies, estimated from the [O\,{\tiny II}]$\lambda3727$ line in the SDSS spectra using the calibration
	               by \citet{2019ApJ...882...89Z}.
	               Col.(13) 4000\AA\, break.
	               Col.(14) Stellar absorption-line index \hda.
		       Cols.(15)-(16) The inner and outer \sersic\ indexes from the GALFIT fittings. The symbol ``?'' in Col.(16) denotes that
		     it is not sure whether the outer \sersic\ component is present or not.}
\end{deluxetable}

\clearpage

\begin{deluxetable}{ccCCCCCCCCCCRCCRCC}
	\rotate
	\tablewidth{0pt}
	\tabletypesize{\tiny}
	\tablecaption{Properties of the Changing-look AGNs\label{table:clagn}}
	\tablehead{
		\colhead{No.} &\colhead{Name} & \colhead{Ref.\tablenotemark{a}}& \colhead{$z$} &\colhead{$M_{\rm BH}$} &
		\colhead{$g$} & \colhead{$r$} &\colhead{$i$} & \colhead{$g-i$} & \colhead{$f_\mathrm{PSF}$} &
		\colhead{$K_{\rm s}$} & \colhead{log $M_{\star}$} & \colhead{$R_\mathrm{petrosian}$} &
		\colhead{SFR} &\colhead{\dbreak} & \colhead{\hda} &
		\colhead{$n_1$} & \colhead{$n_2$} \\
		\colhead{}   & \colhead{}    & \colhead{} & \colhead{}  & \colhead{\msun} &
		\colhead{(mag)} & \colhead{(mag)} & \colhead{(mag)} &   \colhead{(mag)} & \colhead{(\%)} &
		\colhead{(mag)} & \colhead{(\msun)} & \colhead{(arcsec)} &\colhead{(\msun\,yr$^{-1}$)} & \colhead{} & \colhead{(\AA)}&
		\colhead{Inner \sersic} & \colhead{Outer \sersic} \\
		\colhead{(1)}   & \colhead{(2)}    & \colhead{(3)} & \colhead{(4)}  & \colhead{(5)} &
		\colhead{(6)} & \colhead{(7)} & \colhead{(8)} &   \colhead{(9)} & \colhead{(10)} &
		\colhead{(11)} & \colhead{(12)} & \colhead{(13)} &\colhead{(14)} & \colhead{(15)} & \colhead{(16)} &
		\colhead{(17)} & \colhead{(18)}
	      }
	\startdata
1&Mrk 1018             & (1)& 0.0350 &7.95 &14.32\pm0.04 &13.47\pm0.02 &13.13\pm0.09                    &1.15\pm0.10 &11.7    &10.65\pm0.04&11.32 &10.6 & 0.42 & 1.27 &  3.60 & 5.22\pm0.06 & 0.76\pm0.01 \\
2&Mrk 590              & (5)& 0.0264 &7.68\tablenotemark{*} &13.26\pm0.01 &12.55\pm0.01 &12.12\pm0.01   &1.12\pm0.01 &1.3     & 9.65\pm0.03&11.46 &28.7&0.004 & 2.05 & -4.98 & 2.57\pm0.02 & 0.88\pm0.01 \\
3&J030510.60-010431.6  & (9)& 0.0451 &7.35\tablenotemark{*} &15.26\pm0.01 &14.59\pm0.01 &14.21\pm0.01   &0.99\pm0.01 &1.8     &12.07\pm0.07&10.83 &23.3& 0.21 & 2.01 & -3.45 & 1.03\pm0.05 & 0.62\pm0.05 \\
4&J080020.98+263648.8  & (9)& 0.0267 &7.25 &13.85\pm0.01 &13.24\pm0.01 &12.76\pm0.02                    &1.07\pm0.02 &1.2     &10.57\pm0.03&11.05 &15.5 & 0.32 & 1.33 &  3.19 & 1.33\pm0.01 & 0.55\pm0.01 \\
5&J081319.34+460839.5  & (1)& 0.0540 &7.29 &16.41\pm0.02 &15.62\pm0.02 &15.09\pm0.03                    &1.25\pm0.04 &11.7    &12.78\pm0.06&10.92 & 6.3 & 0.12 & 1.53 & -2.01 & 2.48\pm0.05 & -           \\
6&J081726.42+101210.1  & (4)& 0.0458 &7.34 &16.63\pm0.06 &15.92\pm0.07 &15.51\pm0.01                    &1.06\pm0.06 &7.0     &13.32\pm0.08&10.42 & 5.2 & 0.09 & 1.34 &  3.04 & 2.43\pm0.04 & ?           \\
7&J082842.73+454433.2  & (8)& 0.0491 &7.22 &16.49\pm0.01 &15.83\pm0.01 &15.43\pm0.01                    &0.99\pm0.01 &6.2     &13.07\pm0.09&10.54 & 8.6 & 0.12 & 1.22 &  1.75 & 1.92\pm0.03 & -           \\
8&NGC 2617              & (6)& 0.0140 &7.60\tablenotemark{*} &13.40\pm0.01 &12.71\pm0.01 &12.47\pm0.01   &0.92\pm0.01 &1.7     &10.17\pm0.05&10.53 &21.5&\nodata& 1.56 & -0.63 &1.31\pm0.01 & 0.88\pm0.01 \\
9&J090902.35+133019.4  & (1)& 0.0500 &7.30 &16.06\pm0.02 &15.36\pm0.02 &14.95\pm0.03                    &1.04\pm0.04 &7.2     &12.53\pm0.07&10.79                &11.8& 0.01  & 1.34 & -0.73 &1.46\pm0.06 & 0.45\pm0.01 \\
10&J090932.02+474730.6 & (2)& 0.1171 &7.68 &18.52\pm0.05 &17.47\pm0.04 &17.12\pm0.03                    &1.17\pm0.06 &1.7     &14.86\pm0.14&10.70 & 2.9 & 0.21 & 1.42 &  1.96         & 6.16\pm0.36\,^{?} & ?           \\
11&J091531.04+481407.7 & (4)& 0.1005 &7.80\tablenotemark{*} &17.48\pm0.03 &16.63\pm0.01 &16.24\pm0.02   &1.07\pm0.03 &7.5     &13.91\pm0.09&10.87 &9.5& 0.18& 1.55 & -2.05 & 1.51\pm0.08 & 0.09\pm0.01 \\
12&J100323.46+352503.8 & (2)& 0.1187 &8.11 &18.66\pm0.02 &17.74\pm0.02 &17.31\pm0.01                    &1.13\pm0.02 &5.9     &14.43\pm0.07&10.97 & 2.4 & 0.38 & 1.62 & -0.41         & 4.82\pm0.47\,^{?} & ?           \\
13&NGC 3516             &(10)& 0.0090 &7.67\tablenotemark{*} &12.22        &11.39        &10.93                           &1.29        &10.3    &8.64\pm0.02 &10.46 &\nodata & \nodata & 1.35 & -0.01 &1.15\pm0.57& 0.5    \\
14&J111536.57+054449.7 & (2)& 0.0900 &\nodata&18.29\pm0.03 &17.23\pm0.02 &16.81\pm0.03                  &1.31\pm0.04 &16.5    &14.36\pm0.11&10.81 & 4.1 & 0.09 & 1.51 &  0.75       & 1.58\pm0.11       & ?           \\
15&J111803.22+450646.8 & (8)& 0.1072 &8.64 & \nodata     & \nodata     & \nodata                        & \nodata    &\nodata &12.31\pm0.05&11.69 & 6.2 & 0.   & 1.42 &  2.78 &    \nodata  & \nodata        \\
16&J112637.73+513423.0 & (8)& 0.0264 &6.17 &15.88\pm0.03 &15.27\pm0.03 &14.96\pm0.03                    &0.90\pm0.04 &0.8     &12.90\pm0.07&10.02 & 9.0 & 0.03 & 1.70 &  0.04 & 1.13\pm0.03 & 0.55\pm0.01 \\
17&J113229.14+035729.0 & (2)& 0.0910 &7.44 &17.48\pm0.04 &16.63\pm0.01 &16.24\pm0.02                    &1.09\pm0.04 &4.1     &13.86\pm0.14&10.83 & 3.0 & 0.30 & 1.53 & -0.60         & 3.23\pm0.22\,^{?} & ?           \\
18&J113355.93+670107.0 & (4)& 0.0397 &7.83 &15.42\pm0.09 &16.65\pm0.03 &14.22\pm0.03                    &1.14\pm0.09 &15.9    &12.02\pm0.07&10.88 &15.3 & 0.05 & 1.45 &  0.23 & 1.05\pm0.08 & 1.29\pm0.02 \\
19&J125258.72+591832.7 & (3)& 0.1240 &8.02 &17.68\pm0.02 &16.89\pm0.02 &16.47\pm0.02                    &1.02\pm0.03 &33.0    &13.77\pm0.10&11.08 &16.5 & 0.25 & \nodata &\nodata & 1.67\pm0.15 & 0.95\pm0.02  \\
20&J125403.78+491452.8 & (4)& 0.0670 &7.60\tablenotemark{*} &16.35\pm0.04 &15.57\pm0.06 &15.27\pm0.06   &0.98\pm0.07 &8.8     &13.08\pm0.07&10.78 & 6.4& 0.14& 1.69 & -2.40 & 4.61\pm0.14 & - \\
21&J130716.99+450645.3 & (3)& 0.0840 &6.48 &18.71\pm0.03 &17.95\pm0.02 &17.51\pm0.03                    &1.07\pm0.04 &22.9    &15.13\pm0.13&10.24 & 4.2 & 0.15 & 1.23 &  0.78         & 1.16\pm0.09       & ?           \\
22&J131615.95+301552.2 & (9)& 0.0490 &6.93 &14.77\pm0.01 &14.33\pm0.01 &13.93\pm0.01                    &0.81\pm0.01 &3.9     &11.71\pm0.05&10.92 & 8.9 & 0.23 & 1.28 &  3.11         & 2.03\pm0.03       & 0.12\pm0.01 \\
23&J135855.82+493414.1 & (2)& 0.1159 &7.31 &19.22\pm0.05 &18.64\pm0.02 &17.92\pm0.12                    &1.25\pm0.13 &31.8    &15.20\pm0.13&10.66 & 3.9 & 0.16 &\nodata  & \nodata    & 1.13\pm0.06\,^{?} & ?           \\
24&J142846.71+172353.0 & (3)& 0.1040 &7.97 &18.13\pm0.03 &17.45\pm0.01 &17.04\pm0.02                    &0.95\pm0.04 &4.6     &14.65\pm0.11&10.53 & 2.5 & 1.07 & 1.18 &  4.04         & 5.24\pm0.37\,^{?} & ?           \\
25&J153308.01+443208.2 & (4)& 0.0367 &7.6-8.0\tablenotemark{*}&14.68\pm0.03 &14.18\pm0.02 &13.72\pm0.02 &0.94\pm0.03 &0.3     &11.57\pm0.07&10.82 &15.7& 0.06& 1.54 & 0.09  & 5.67\pm0.37 & 0.78\pm0.01 \\
26&J153355.99+011029.7 & (2)& 0.1426 &7.68 &18.16\pm0.03 &17.15\pm0.05 &16.94\pm0.03                    &0.97\pm0.04 &5.9     &15.09\pm0.15&10.61 & 5.0 & 0.23 & 1.53 &  0.20         & 2.58\pm0.17 & ?           \\
27&J154529.64+251127.9 & (2)& 0.1171 &6.22 &17.75\pm0.03 &16.92\pm0.02 &16.50\pm0.02                    &1.05\pm0.04 &12.0    &13.81\pm0.10&11.02 & 3.7 & 0.55 & 1.28 &  2.81         & 4.48\pm0.12\,^{?} & ?   \\
28&J155258.30+273728.4 & (7)& 0.0865 &8.32\tablenotemark{*} &17.95\pm0.05 &17.23\pm0.02 &16.78\pm0.02   &1.06\pm0.05 &4.1     &13.80\pm0.11&10.80 & 6.6& 0.19& 1.46 & -0.45 & 2.28\pm0.12 & 0.24\pm0.02\\
29&J160505.14+452634.7 & (8)& 0.0433 &7.77 &15.04\pm0.03 &14.43\pm0.01 &14.10\pm0.01                    &0.89\pm0.03 &2.4     &11.72\pm0.07&10.87 &14.7 & 0.13 & 1.58 & -1.98 & 1.08\pm0.02 & 0.47\pm0.01 \\
30&J162501.43+241547.3 & (9)& 0.0503 &6.97 &16.44\pm0.02 &15.79\pm0.01 &15.35\pm0.04                    &1.03\pm0.04 &2.8     &13.00\pm0.07&10.60 & 4.2 & 0.87 & 1.18 &  2.89 & 2.07\pm0.14 & 1.25\pm0.1  \\
31&J163629.66+410222.4 & (9)& 0.0474 &7.36 &16.62\pm0.01 &15.88\pm0.01 &15.52\pm0.01                    &1.03\pm0.01 &7.0     &13.03\pm0.07&10.53 & 5.0 & 0.07 & 1.55 & -0.88 & 1.32\pm0.02 & -           \\
	\hline
	Q1 &J012648.08-083948.0 &(11) & 0.1980 &7.86&     &       &                          &0.75\pm0.02 &        &    & 10.54 &2.6 &1.12 &1.12 & 5.38 & 3.60\pm0.01 &         \\
	Q2 &J015957.64+003310.5 &(11) & 0.3120 &8.00&     &       &                          &1.10\pm0.02 &        &    & 10.67 &1.5 &2.63 &1.22 & 4.36 & 3.44\pm0.02 &         \\
	Q3 &J101152.90+544206.4 &(11) & 0.2460 &7.82&     &       &                          &0.45\pm0.02 &        &    & 10.00 &1.8 &0.84 &1.10 & 5.65 & 3.38\pm0.02 &         \\
	Q4 &J233602.98+001728.7 &(11) & 0.2430 &8.08&     &       &                          &1.53\pm0.02 &        &    & 10.26 &2.5 &0.50 &1.36 & 1.41 & 1.87\pm0.01 &         \\
	\enddata

	\tablecomments{Col.(1) Identification number assigned in this paper.
	   Col.(2) Target name.
	   Col.(3) References for each CL AGN.
	   Col.(4) Redshift measured by the SDSS pipeline.
	   Col.(5) \mbh\ data with an asterisk are from the corresponding papers denoted in the reference column, otherwise from \citet{2019ApJS..243...21L}.
	   Cols.(6)-(8) Host-galaxy magnitude from the GALFIT fittings (AGN contamination removed), with Galactic extinction corrected.
	   Col.(9) $g-i$ color of the host galaxies. Calculated from the GALFIT fittings, with Galactic extinction and k-corrections performed.
	   Col.(10) Fraction of AGN light to the total in the $r$-band images, according to our GALFIT fittings.
	   Col.(11) $K_\mathrm{S}$ magnitude from 2MASS, with Galactic extinction corrected.
	   Col.(12) Stellar mass of the host galaxies estimated from $K_\mathrm{S}$ luminosity using the calibration of \citet{2013MNRAS.430.2715I}.
	   Col.(13) Petrosian radius in the $r$ band given by the SDSS photometric pipeline.
	   Col.(14) Star formation rate of the host galaxies, estimated from the [O\,{\tiny II}]$\lambda3727$ line in the SDSS spectra using the calibration
	   by \citet{2019ApJ...882...89Z}, for both the low-$z$ CL AGNs and the 4 CL quasars.
	   Col.(15) 4000\AA\, break.
	   Col.(16) Stellar absorption-line index \hda.
	   Cols.(17)-(18) The inner and outer \sersic\ indexes of the CL AGN hosts from the GALFIT fittings. In Col.(17), the script ``?''
	   denotes CL AGN hosts of small size (Petrosian radius <4\arcsec\ in the $r$ band). In Col.(18), the symbol ``?'' denotes that
	   it is not sure whether the outer \sersic\ component is present or not.
           For NGC\,3516, its host-galaxy magnitudes($g,r,i$), $f_\mathrm{PSF}$, and log $M_{\star}$ are from \citet{2011ApJ...739...57K}.
           Its inner and outer \sersic\ indexes are the bulge and disk \sersic\ indexes decomposed from HST F814W image by \citet{2017ApJS..232...21K}.
           Objects Q1-Q4 are Four faded CL quasars at $z\gtrsim0.2$. The numbers of their $g-i$ color,
	   \mgal\ and $n_1$ (the host galaxies being fitted with a single \sersic) are taken from the reference \citet{2019ApJ...876...75C}.}
	   \tablenotetext{a}{Reference. (1) \citet{2017ApJ...846L...7S}; (2) \citet{YangQ-WuXB-2018}; (3) \citet{2020ApJ...889...46S};
		(4) \citet{2019ApJ...883...31F}; (5) \citet{2014ApJ...796..134D}; (6) \citet{2014ApJ...788...48S};
		(7) \citet{2020ApJ...890L..29A}; (8) \citet{2020MNRAS.497..192H}; (9) \citet{Yu-Shi-2020MNRAS.498.3985Y};
                (10) \citet{2019MNRAS.485.4790S}; (11) \citet{2019ApJ...876...75C}}

\end{deluxetable}
\clearpage

\begin{deluxetable}{ccCCCCCCCCCC}
	\tablewidth{0pt}
	\movetabledown=0mm
	\tabletypesize{\scriptsize}
	\tablecaption{Properties of the Large-variability RM AGNs\label{table:rmAGN}}
	\tablehead{
		\colhead{No.} & \colhead{Name} & \colhead{Reference}\tablenotemark{$a$} & \colhead{$z$} & \colhead{$M_{\rm BH}$} & \colhead{$F_{var}$} &
		\colhead{$g$} & \colhead{$r$} & \colhead{$i$} & \colhead{$g-i$} & \colhead{$f_\mathrm{PSF}$}  & \colhead{log $M_{\star}$} \\
		\colhead{}   & \colhead{}    & \colhead{} & \colhead{}  & \colhead{\msun} & \colhead{} &
		\colhead{(mag)} & \colhead{(mag)} & \colhead{(mag)} & \colhead{(mag)} &\colhead{(\%)} & \colhead{(\msun)} \\
		\colhead{(1)}   & \colhead{(2)}    & \colhead{(3)} & \colhead{(4)}  & \colhead{(5)} & \colhead{(6)} &
		\colhead{(7)} & \colhead{(8)} & \colhead{(9)} & \colhead{(10)} &\colhead{(11)} & \colhead{(12)}
	      }
	\startdata
	1 & 3C 120         & (1) & 0.0330 & 7.745 & 0.095& 14.10       &13.55        & 13.19        & 0.88        & 48.3 & 10.54\tablenotemark{+}  \\
	2 & NGC 3227       & (1) & 0.0039 & 6.775 & 0.133& 11.6        &10.76        & 10.18        & 1.42        &  7.1 & 10.78\tablenotemark{+}  \\
	3 & NGC 3516       & (1) & 0.0088 & 7.395 & 0.110& 12.22       &11.39        & 10.93        & 1.29        & 10.3 & 10.08\tablenotemark{+}  \\
	4 & SBS 1116+583A  & (2) & 0.0211 & 6.558 & 0.102& 15.92\pm0.04&15.18\pm0.01 & 14.86\pm0.02 & 1.04\pm0.04 &  8.6 & 10.05\tablenotemark{+}  \\
	5 & Mrk 40         & (3) & 0.0211 & 6.670 & 0.200& 15.84\pm0.01&15.15\pm0.01 & 14.72\pm0.01 & 1.11\pm0.01 & 14.4 & 10.19\tablenotemark{+}  \\
	6 & Mrk 1310       & (2) & 0.0194 & 6.212 & 0.108& 15.09\pm0.01&14.54\pm0.01 & 14.20\pm0.01 & 0.88\pm0.01 & 22.4 &  9.53\tablenotemark{+}  \\
	7 & NGC 4051       & (1) & 0.0023 & 6.130 & 0.096& 10.78       &10.26        & 10.00        & 0.78        &  4.1 &  9.56\tablenotemark{+}  \\
	8 & Mrk 50         & (3) & 0.0234 & 7.442 & 0.20 & 15.23\pm0.05&14.52\pm0.04 & 14.16\pm0.02 & 1.05\pm0.05 &  8.5 &  9.90\tablenotemark{*} \\
	9 & NGC 4593       & (3) & 0.0090 & 6.882 & 0.23 & 11.48\pm0.01&10.74\pm0.01 & 10.24\pm0.01 & 1.24\pm0.01 &  2.4 & 10.40\tablenotemark{+}  \\
	10& IC 4218        & (2) & 0.0193 & 6.808 & 0.159& 14.21\pm0.01&13.56\pm0.01 & 13.23\pm0.02 & 0.97\pm0.02 &  5.3 & 10.65  \\
	11& Mrk 279        & (1) & 0.0305 & 7.435 & 0.138& 14.48       &13.74        & 13.35        & 1.10        & 26.7 & 10.86\tablenotemark{+}  \\
	12& NGC 5548       & (1) & 0.0172 & 7.718 & 0.284& 12.96\pm0.01&12.31\pm0.01 & 12.03\pm0.01 & 0.92\pm0.01 &  4.7 & 10.46\tablenotemark{*}  \\
	13& Mrk 817        & (1) & 0.0315 & 7.586 & 0.097& 14.52\pm0.02&13.77\pm0.03 & 13.49\pm0.01 & 0.99\pm0.02 & 25.9 & 10.63\tablenotemark{+}  \\
	14& Mrk 1511       & (3) & 0.0399 &\nodata& 0.12 & 13.94\pm0.10&13.42\pm0.09 & 13.08\pm0.01 & 0.82\pm0.10 &  8.9 & 10.47\tablenotemark{*} \\
	15& NGC 6814       & (2) & 0.0052 & 7.038 & 0.093& 11.05       &10.41        & 10.01        & 1.04        &  3.1 &  9.85\tablenotemark{+}  \\
	\enddata

	\tablecomments{Col.(1) Identification number assigned in this paper.
	   Col.(2) Target name.
	   Col.(3) References for each RM AGN.
	   Col.(4) Redshift.
	   Col.(5) BH masses by reverberation mapping (except IC 4218),
         taken from \citet{2015PASP..127...67B} and \citet{2018ApJ...864..146B}.
         For IC 4218, whose reverberation mapped \mbh\ is not available,
         we measured its \mbh\ based on its SDSS spectrum in the same way as the MagE sources (see \S\ref{subsec:agn-properties}).
	   Col.(6) Fractional variability amplitude of broad \hb,
         characteristic of the overall variability of broad-\hb\ light curves,
         taken from the papers noted in the reference column.
	   Note that here broad \hb\ is used because its $F_\mathrm{var}$ is more commonly available than broad \ha\ in the literature.
	   Cols.(7)--(9) Host-galaxy magnitude, calculated with AGN contamination removed and Galactic extinction corrected.
	   The data with $\pm$1$\sigma$ errors are from our GALFIT fittings, and the others are taken from \citet{2011ApJ...739...57K}.
	   Col.(10) $g-i$ color of the host galaxies, with Galactic extinction and k-corrections performed.
	   Col.(11) Fraction of AGN light to the total in the $r$-band images, either from our GALFIT fittings or from \citealt{2011ApJ...739...57K} (see Cols.(7)-(9)).
	   Col.(12) Stellar mass, estimated with AGN contamination removed. The data marked with a \tablenotemark{$+$} are taken from \citet{2018ApJ...864..146B},
	 those with an \tablenotemark{$*$} are from \citet{2011ApJ...739...57K},
     and the remaining one is our estimate in the same way as for the variable MagE sources.
          }

		  \tablenotetext{a}{References. (1) \citet{2004ApJ...613..682P}; (2) \citet{2009ApJ...697..160B}; (3) \citet{2015ApJS..217...26B}}
\end{deluxetable}
\clearpage

\renewcommand{\thetable}{A\arabic{table}}
\setcounter{table}{0}

\begin{deluxetable}{cccCCcCC}
	\tablewidth{0pt}
	\movetabledown=0mm
	\tabletypesize{\scriptsize}

	\tablecaption{Light-curve Variability Statistics of the Six MagE AGNs \label{table:lightcurve}}

	\tablehead{
	  \colhead{ID} &\colhead{Name} & \colhead{Band} & \colhead{$\sigma^2_{\rm rms}$} & \colhead{$err(\sigma^2_{\rm rms})$} &
	  \colhead{Variable?} & \colhead{$F_{\rm var}$} & \colhead{$R_{\rm max}$} \\
	  \colhead{(1)} &\colhead{(2)} & \colhead{(3)} & \colhead{(4)} & \colhead{(5)} &
	  \colhead{(6)} & \colhead{(7)} & \colhead{8}
	}
	\startdata
	1 & J0839+0724 & CRST V    & 2.05\times10^{-7}  & 2.99\times10^{-9}  &  yes& 0.009  & 1.06\\
  	  &            & ASAS-SN V & 4.35\times10^{-5}  & 3.20\times10^{-7}  &  yes& 0.103  & 1.03\\
	  &            & ASAS-SN g & 3.64\times10^{-5}  & 2.40\times10^{-7}  &  yes& 0.091  & 1.01\\
	  &            & WISE W1   & 4.96\times10^{-5}  & 1.03\times10^{-7}  &  yes& 0.082  & 1.43\\
	  &            & WISE W2   & 3.50\times10^{-5}  & 2.10\times10^{-5}  &  yes& 0.069  & 1.28\\
	2 & J1113+0007 & CRST V    & -6.49\times10^{-7} & 9.25\times10^{-11} &  no & -      & 1.03 \\
	  &            & ASAS-SN V & 6.12\times10^{-5}  & 7.42\times10^{-7}  &  yes& 0.134  & 1.16 \\
	  &            & ASAS-SN g & 5.36\times10^{-5}  & 8.72\times10^{-7}  &  yes& 0.007  & 1.02\\
	  &            & WISE W1   & 1.15\times10^{-5}  & 2.51\times10^{-8}  &  yes& 0.043  & 1.14\\
	  &            & WISE W2   & 5.38\times10^{-5}  & 3.64\times10^{-7}  &  yes& 0.094  & 1.44\\
        3 & J1257+2724 & CRST V    & -1.41\times10^{-5} & 2.46\times10^{-10} & no  & 0.069  & 1.01   \\
	  &            & ASAS-SN V & 1.94\times10^{-5}  & 2.23\times10^{-7}  & yes & 0.073  & 1.07  \\
	  &            & ASAS-SN g & -3.17\times10^{-6} & 4.11\times10^{-6}  & no  & -      & 1.01   \\
	  &            & WISE W1   & 2.03\times10^{-5}  & 3.50\times10^{-8}  & yes & 0.052  & 1.11  \\
	  &            & WISE W2   & 8.79\times10^{-6}  & 2.44\times10^{-8}  & yes & 0.034  & 1.15  \\
	  &            & X-ray(0.2-2keV) & 6.61\times10^{-1} & 1.72\times^{-10} & yes & 0.839 & 25.36  \\
	  4 & J1342+2435 & CRST V  & -6.10\times10^{-6} & 7.93\times10^{-8} & no  & 0.040  & 1.03   \\
	  &            & ASAS-SN V & 3.89\times10^{-5}  & 1.42\times10^{-6} & yes & 0.087  & 1.03  \\
	  &            & ASAS-SN g & 2.48\times10^{-5}  & 5.62\times10^{-7} & yes & 0.097  & 1.02 \\
	  &            & WISE W1   & 9.08\times10^{-6}  & 1.26\times10^{-8} & yes & 0.036  & 1.07  \\
	  &            & WISE W2   & 2.76\times10^{-5}  & 1.85\times10^{-7} & yes & 0.058  & 1.09 \\
	5 & J1412-0307 & CRST V    &  9.18\times10^{-6} & 1.16\times10^{-8} & yes & 0.054 & 1.17    \\
	  &            & ASAS-SN V &  2.48\times10^{-5} & 3.28\times10^{-7} & yes & 0.097 & 1.51   \\
	  &            & ASAS-SN g &  6.28\times10^{-4} & 3.26\times10^{-4} & yes & 0.142 & 1.50   \\
	  &            & WISE W1   &  4.16\times10^{-5} & 1.12\times10^{-7} & yes & 0.080 & 1.32   \\
	  &            & WISE W2   &  8.72\times10^{-5} & 3.28\times10^{-6} & yes & 0.145 & 1.80   \\
	  &            & X-ray (0.2-2keV) & 1.02        & 1.58\times10^{-17}  & yes & 1.061 & 29.25   \\
        6 & J1442+0119 & CRST V    & -1.36\times10^{-5} & 2.79\times10^{-10} & no  & -      & 1.01   \\
 	  &            & ASAS-SN V & 1.89\times10^{-5}  & 7.93\times10^{-8}  & yes & 0.063  & 1.08   \\
 	  &            & ASAS-SN g & 2.10\times10^{-6}  & 2.98\times10^{-7}  & yes & 0.049  & 1.11  \\
 	  &            & WISE W1   & 1.09\times10^{-5}  & 8.28\times10^{-9}  & yes & 0.045  & 1.10    \\
 	  &            & WISE W2   & 1.84\times10^{-5}  & 3.38\times10^{-8}  & yes & 0.047  & 1.10   \\
	  &            & X-ray (0.2-2keV) &             &                    &     &        & 2.74    \\
	\enddata
	\tablecomments{Col.(1) Identification number assigned in this paper.
	   Col.(2) Target name.
	   Col.(3) Wavelength bands of the light curves.
	   Col.(4) Normalized excess variance, calculated from the magnitude data of the light curves \citep[see][]{2020ApJ...889..113M}.
	   Col.(5) The uncertainty of $\sigma^2_{\rm rms}$ due to Poisson noise \citep[see][]{2020ApJ...889..113M}.
	   Col.(6) Light curves with $(\sigma^2_{\rm rms}-err(\sigma^2_{\rm rms})>0)$ can be regarded to
          be intrinsically variable \citep[see][]{2020ApJ...889..113M}.
	   Col.(7) Fractional variability amplitude $F_{\rm var}$, calculated from the flux data of light curves
          \citep[see][]{2015ApJS..217...26B}.
        Light curves with $\sigma^2 -\overline{\delta^2} <0$ (see Equation~\ref{eq:fvar}) are marked with `$-$',
        indicating that $F_{\rm var}$ is not applicable.
	   Col.(8) Maximum variability amplitude of light curves.
	   The X-ray data of J1442$+$0119 are of two epochs only and thus not possible to calculate the statistics in Cols.(4--7).
	 }
\end{deluxetable}

\clearpage


  \begin{figure}[htbp]
      \centering
      \includegraphics[width=6.3in]{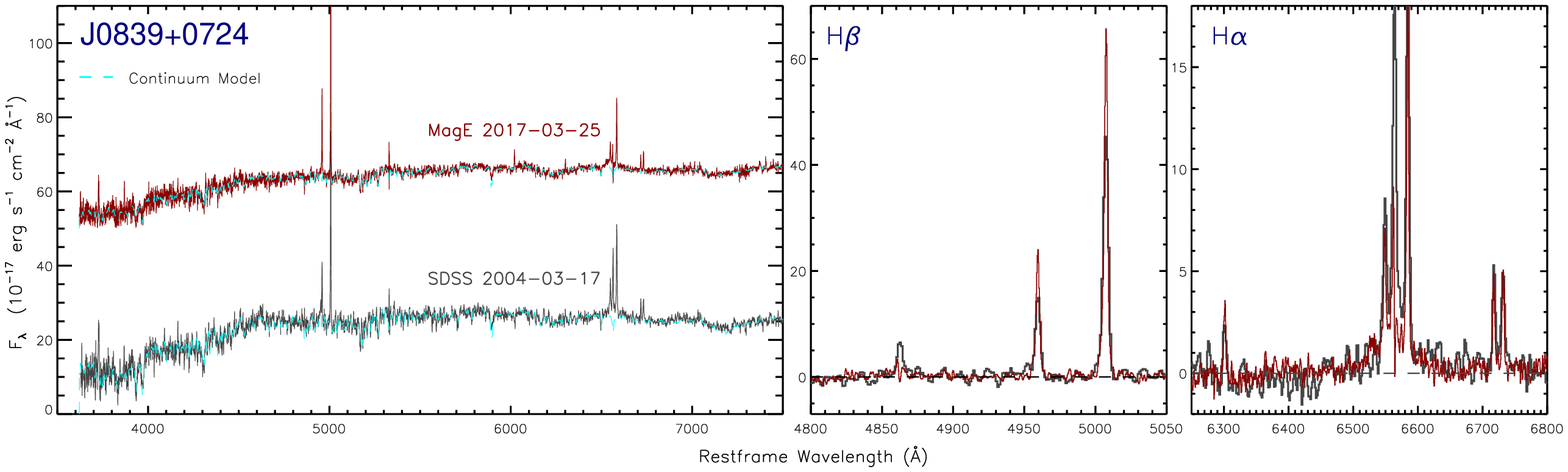}
      \includegraphics[width=6.3in]{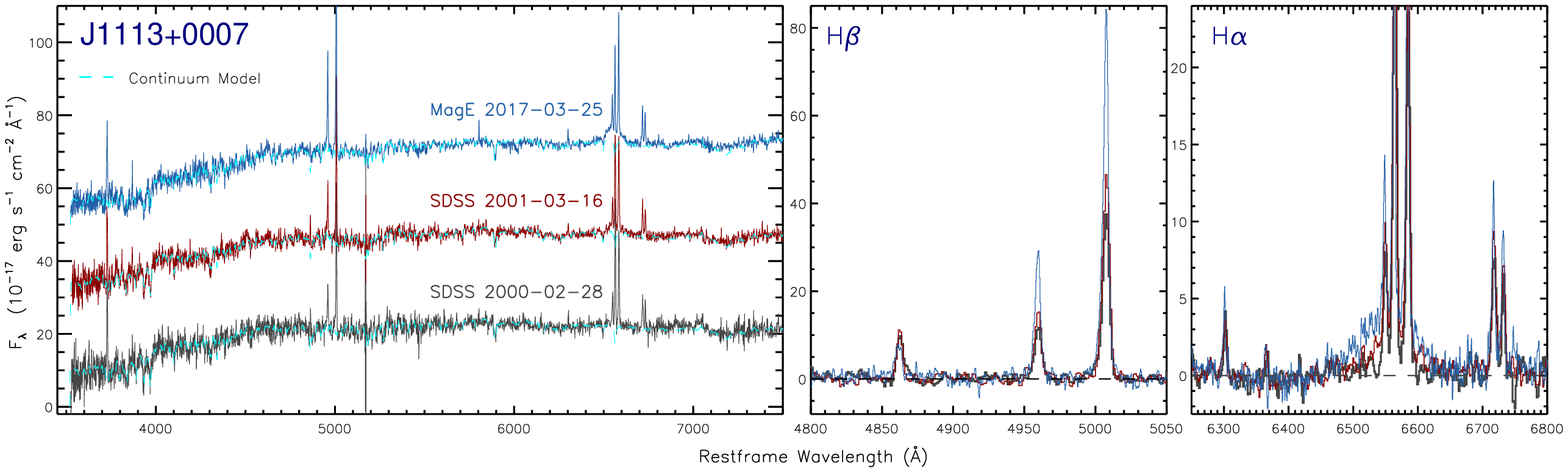}
      \includegraphics[width=6.3in]{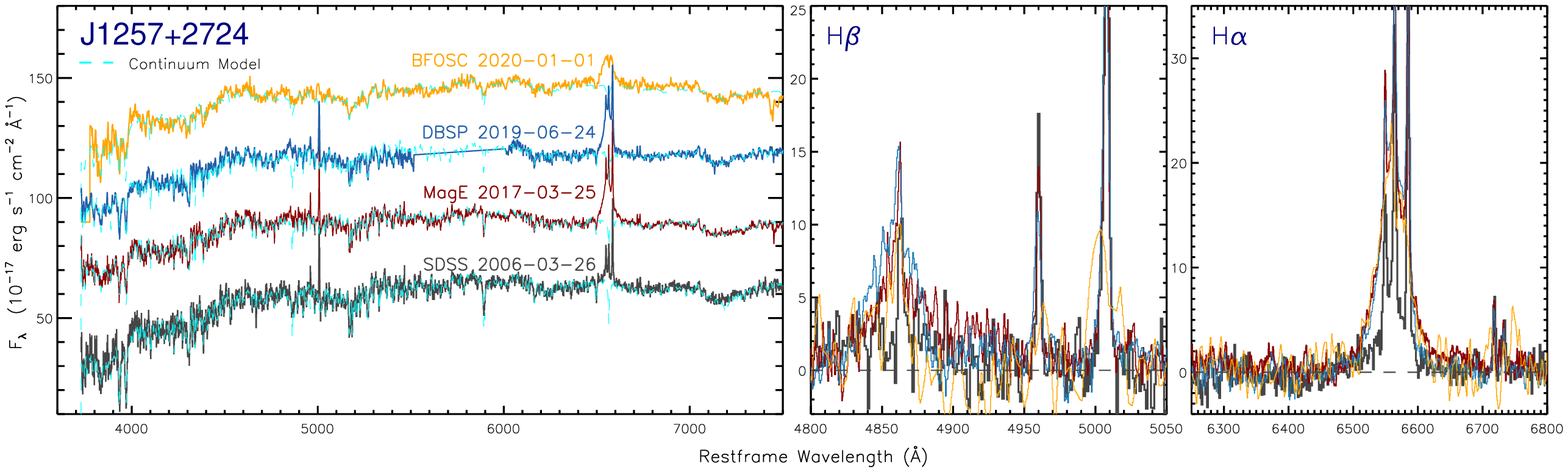}
      \includegraphics[width=6.3in]{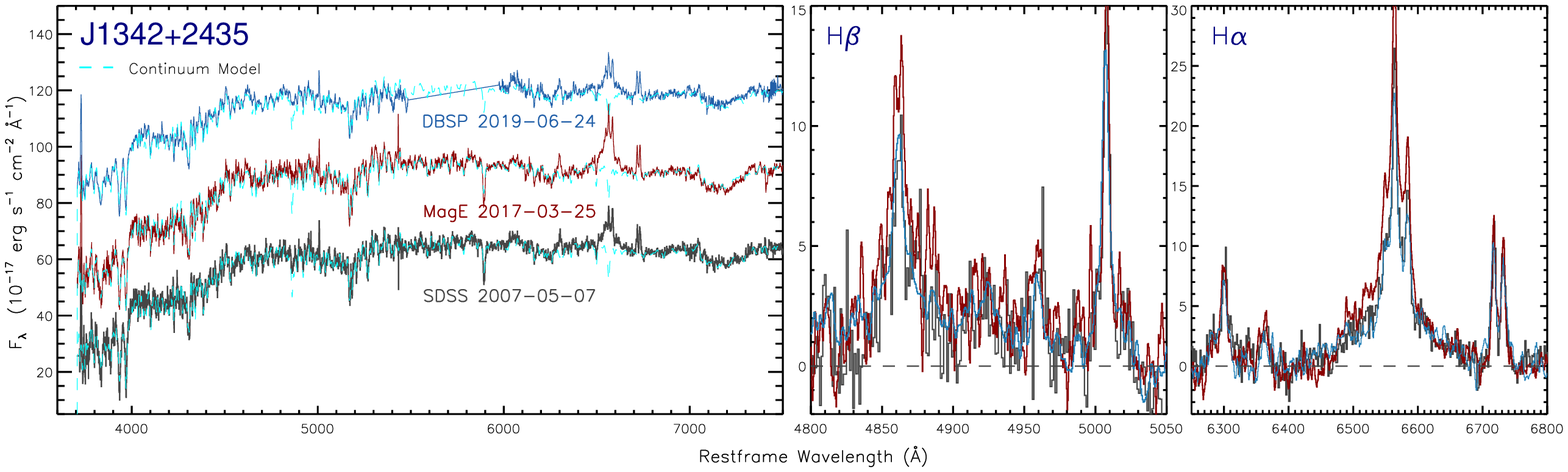}
      \caption{\footnotesize Multi-epoch spectra of the six variable MagE sources.
	{\it Left:} The multiepoch observed spectra are plotted in different colors
	(with arbitrary vertical offsets for clarification),
	together with the best-fits of their continua (cyan dashed lines).
	For every sources, the followup spectra are scaled with respect to
	the earliest SDSS spectrum (the black solid line)
	in terms of [O\,{\scriptsize I}]$\lambda6300$ flux (for J1113$+$0007 and J1342$+$2435)
	or [O\,{\scriptsize III}]$\lambda5007$ (for the rest).
    {\it Center and Right: } Close-up of the continuum-subtracted, emission-line spectrum of the \hb\ and \ha\
        regions, with the same coloring of the left panels. \label{fig:optspec} }
  \end{figure}

\addtocounter{figure}{-1}

  \begin{figure}[htbp]
      \centering
      \includegraphics[width=6.3in]{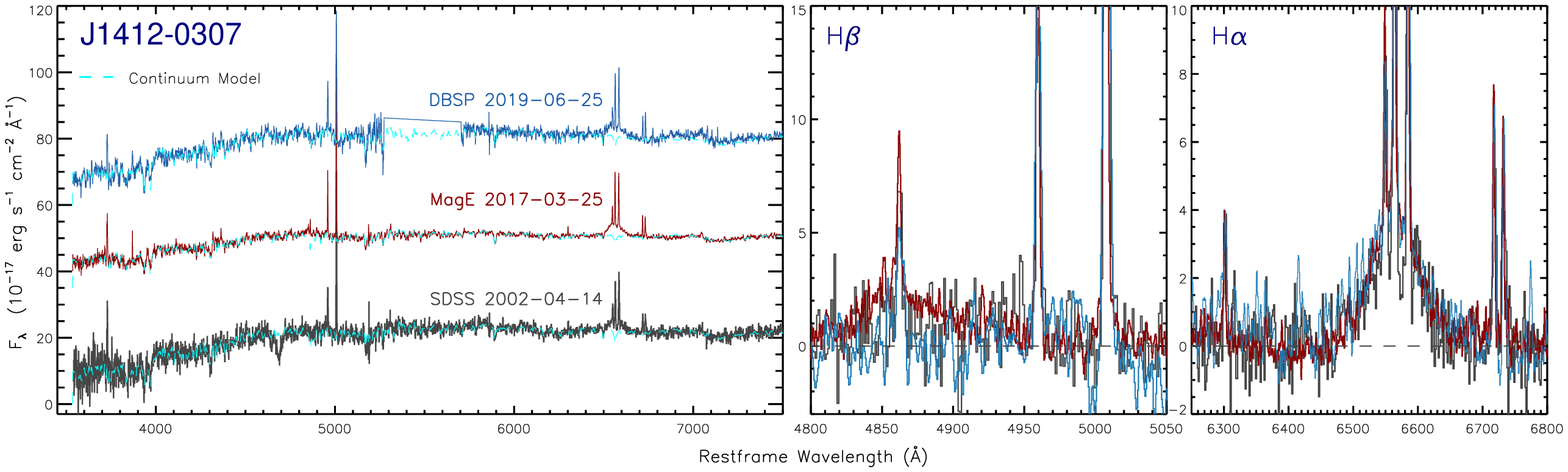}
      \includegraphics[width=6.3in]{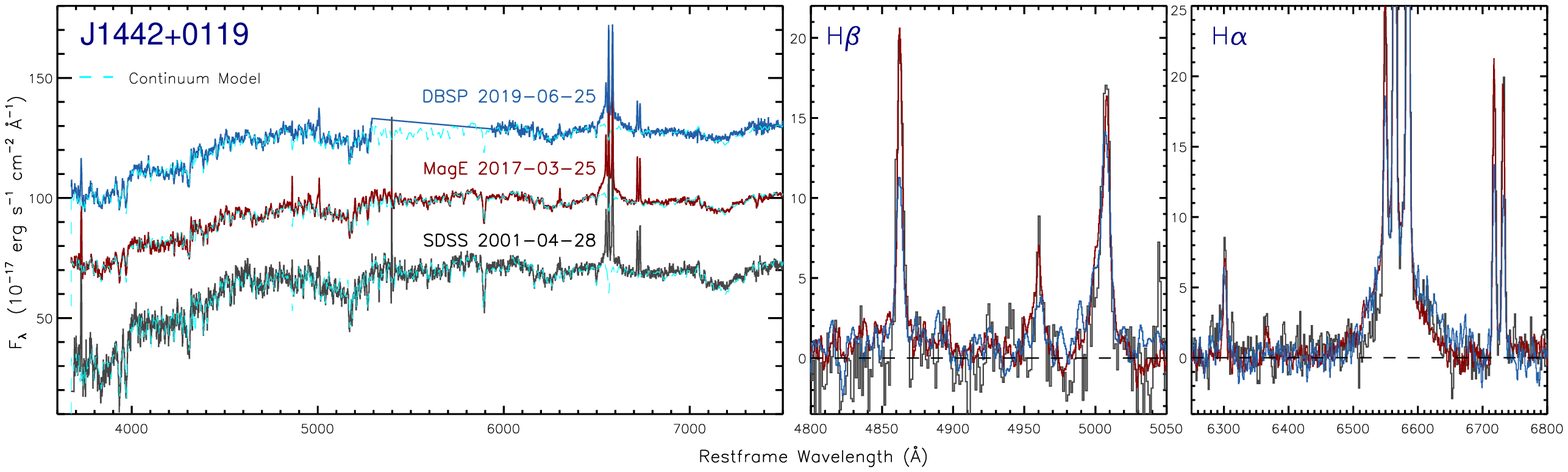}
      \caption{\footnotesize Continued. }
  \end{figure}

\begin{figure}[htbp]
	\centering
	\subfigure[SDSS $g,r,i$ composite images of the large-variability MagE sources. \label{fig:sdssimage_mage}]{
	  \includegraphics[width=0.7\textwidth]{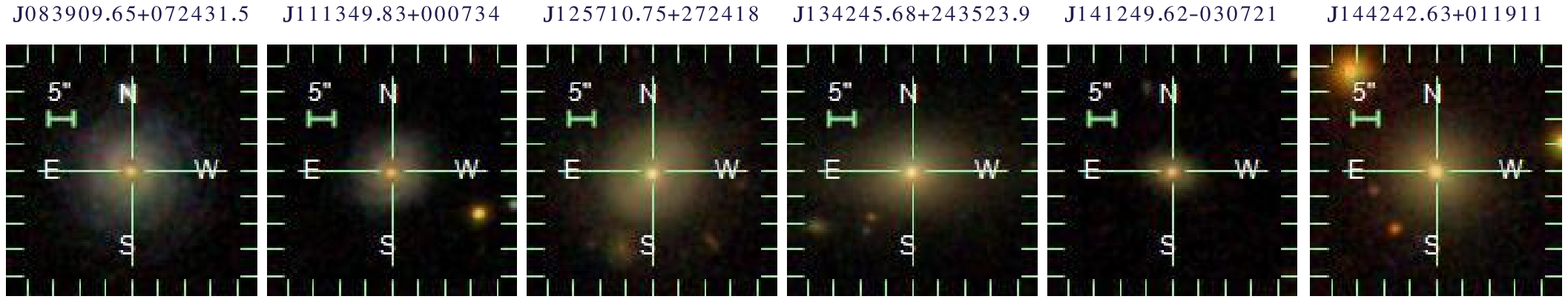}}
	  
	  \subfigure[SDSS $g,r,i$ composite images of the low-$z$ CL AGNs. \label{fig:sdssimage_clagn}]{
	  \includegraphics[width=0.7\textwidth]{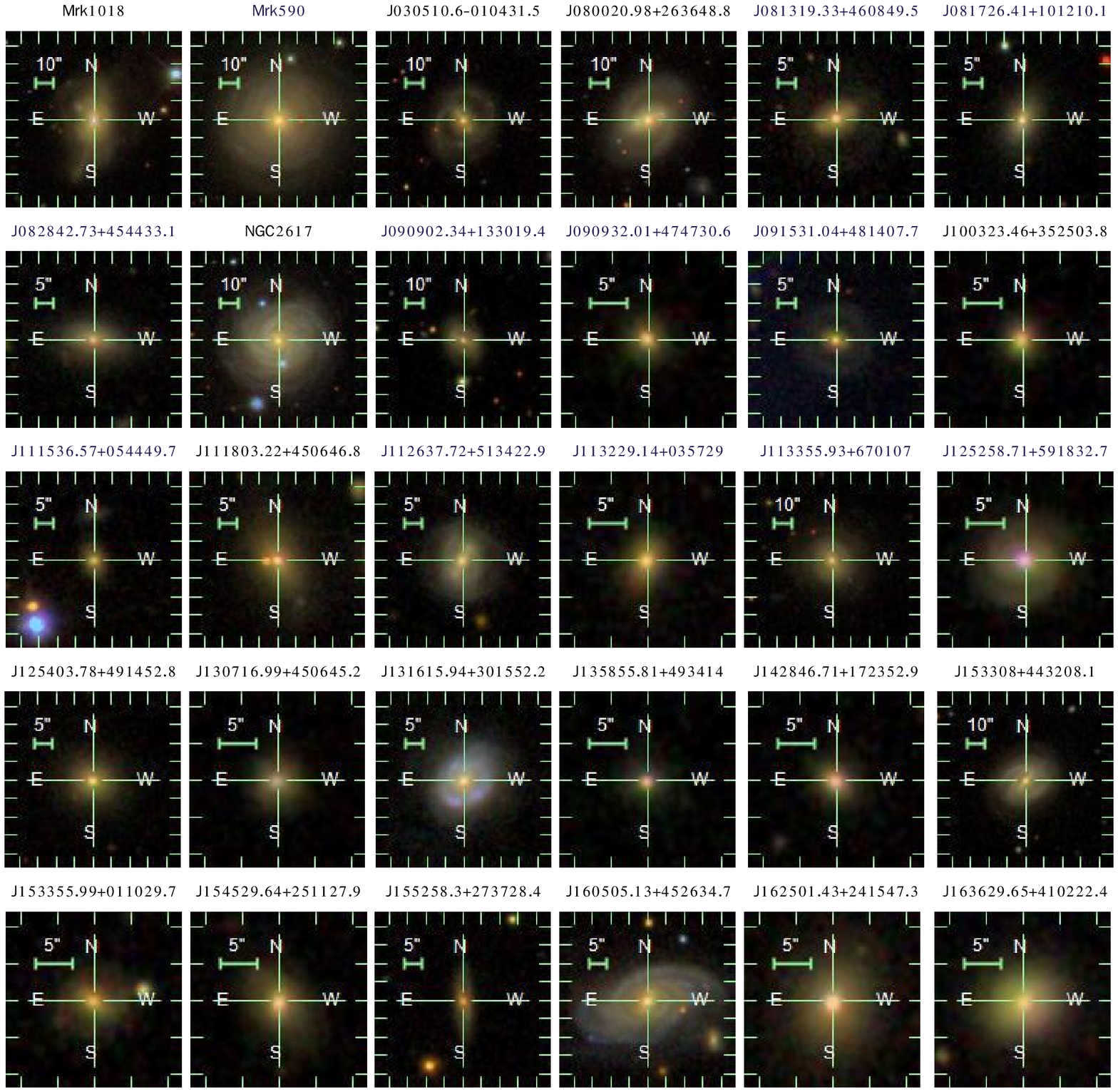}}
	
	  \subfigure[SDSS $g,r,i$ composite images of the large-variability reverberation mapped AGNs. \label{fig:sdssimage_mapping}]{
	  \includegraphics[width=0.7\textwidth]{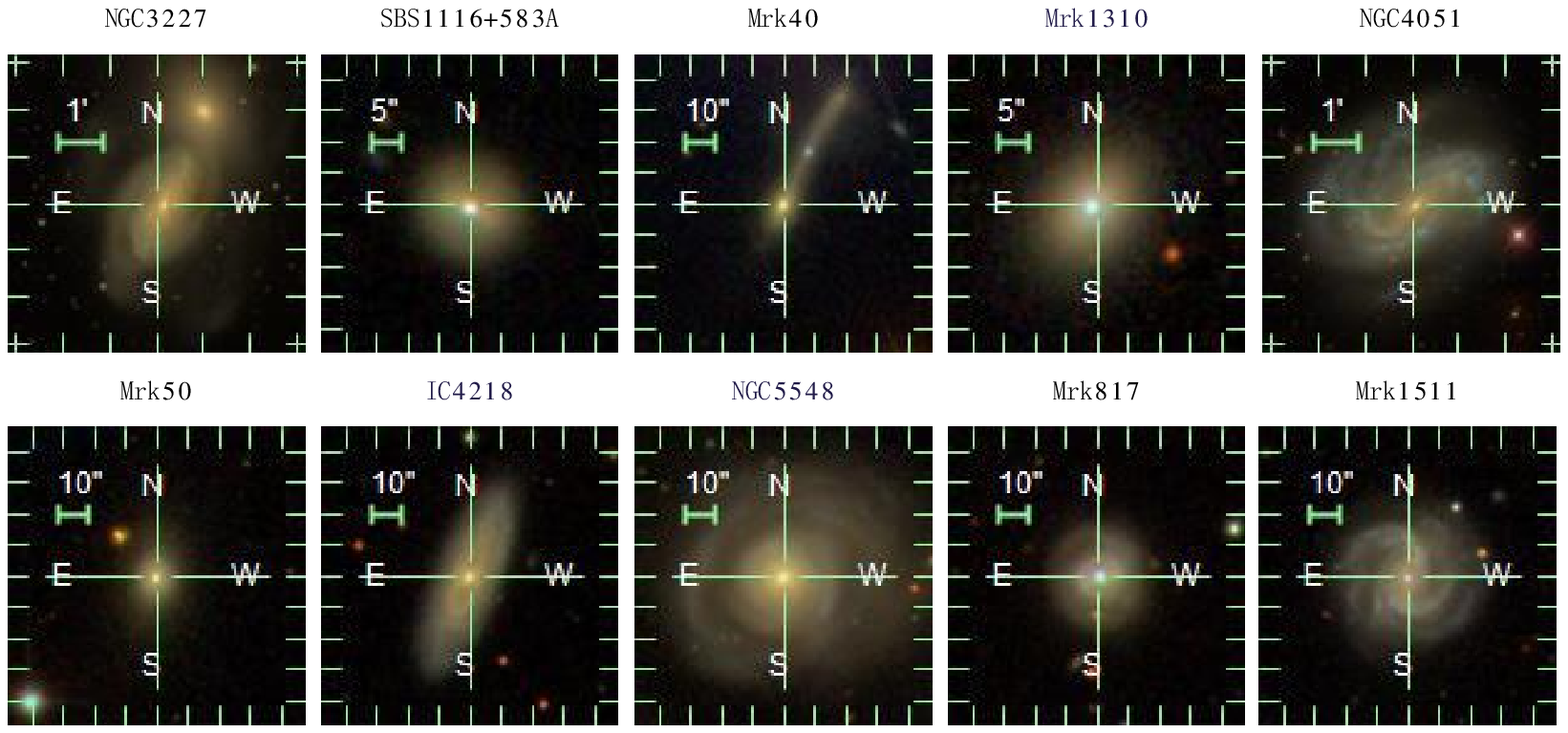}}
	  \caption{SDSS $g,r,i$ composite images. \label{fig:sdssimage}}
\end{figure}


\begin{figure}[htbp]
      \centering
      \includegraphics[width=6.8in]{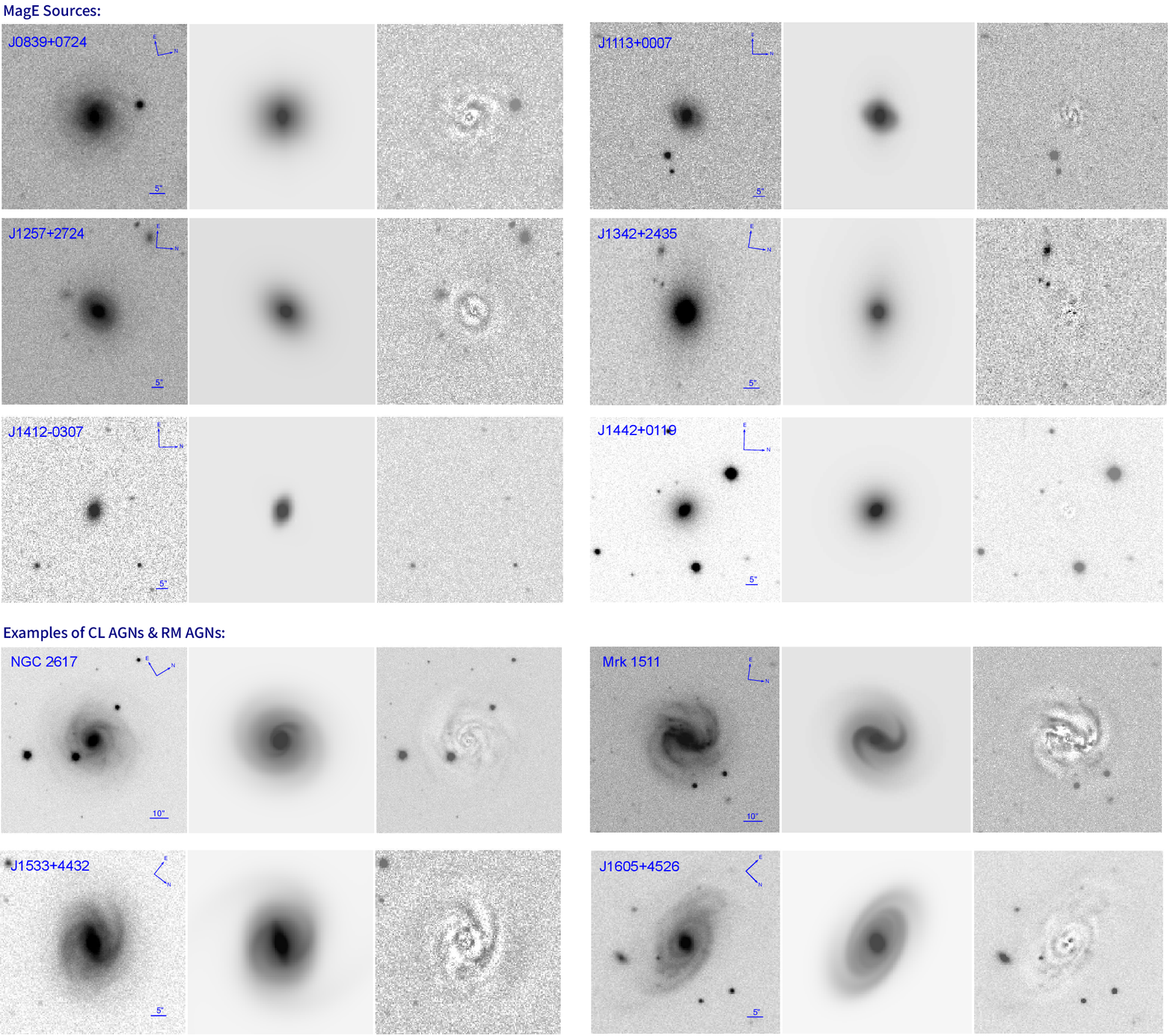}
      \caption{\footnotesize Examples of the two-dimensional imaging decomposition by GALFIT.
      	Among them, four MagE sources are fitted with the traditional approach, whereas
      	the other two MagE sources (J1113$+$0007 and J1343$+$2435) and the demonstrated four CL or RM sources
      	are fitted with the NG approach.
      	For every source, the left image is the original SDSS $r$-band image,
      	the middle the best-fit model, and the right the residual. \label{fig:galfit} }
  \end{figure}

 \begin{figure}[htbp]
	\centering
	\includegraphics[width=0.95\textwidth]{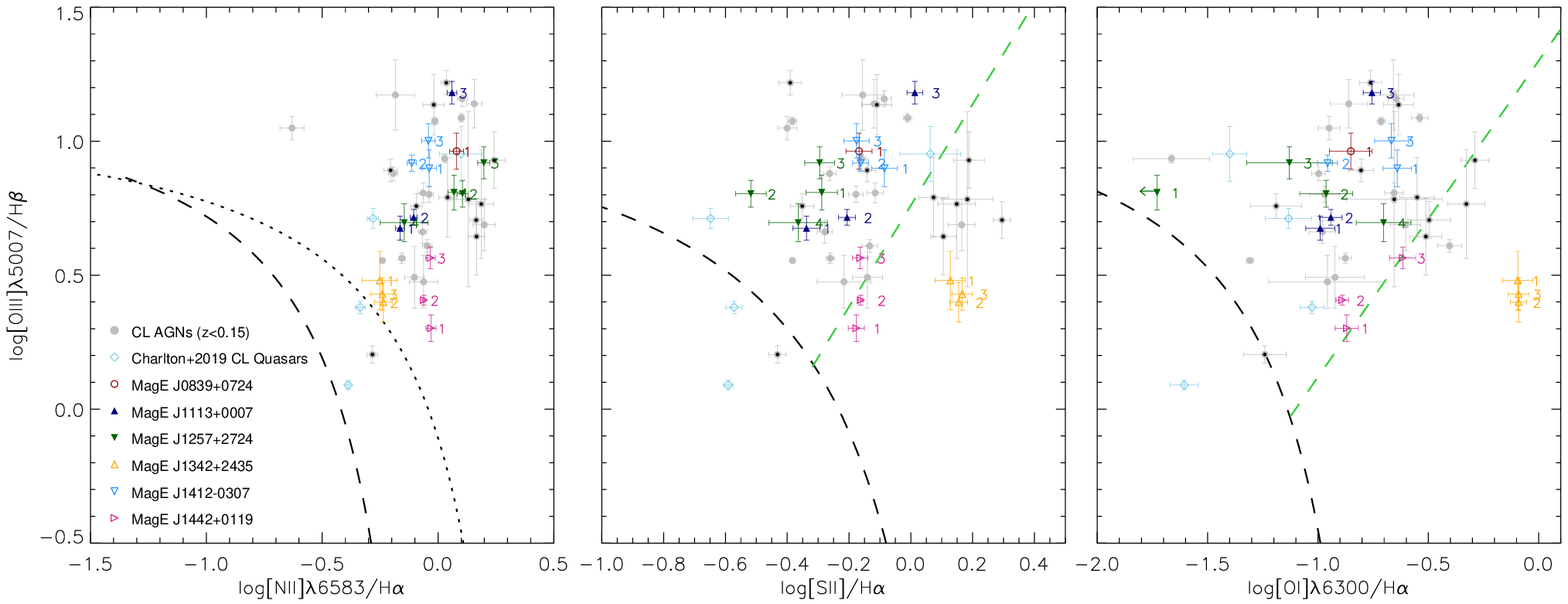}
	\caption{\label{fig:bpt}
		Narrow-line diagnostic diagrams for the six large-variability MagE sources, the 31 low-$z$ CL AGNs,
		and the four CL quasars at $z \gtrsim 0.2$\,.
		The six MagE AGNs are denoted by an asterisk, with different colors for their multiepoch spectra
		(labeled in chronological order of observation).
		Gray filled circles represent the low-$z$ CL AGNs,
		including all of the 11 $z<0.15$ CL AGNs of \citet[][each marked with an additional black dot at the center]{2021ApJ...907L..21D}.
		The cyan filled circles represent the four CL quasars in \citet{2019ApJ...876...75C}.
		The dashed lines separating H\,{\scriptsize II} regions, Seyfert galaxies, and LINERs are
	taken from \citet{2001ApJ...556..121K,Kau-bpt_2003MNRAS.346.1055K} and \citet{2006MNRAS.372..961K}, respectively. }
\end{figure}

 \begin{figure}[htbp]
	\centering
	\includegraphics[width=0.8\textwidth]{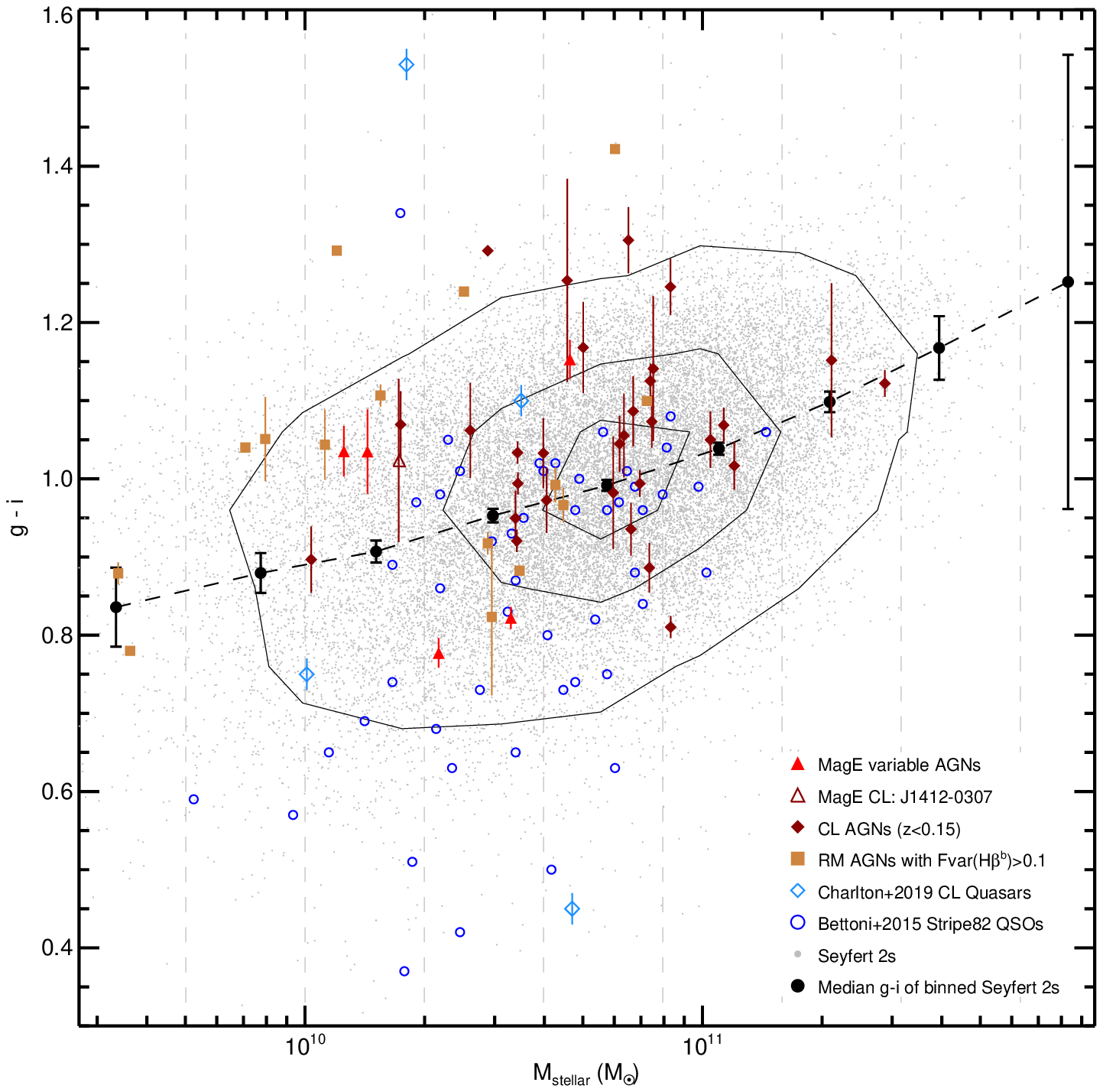}
	\caption{\label{fig:C-M-diagram}
	  Distribution of the 6 MagE AGNs (filled and open triangles, Table~\ref{table:galaxy}),
	  30 CL AGNs at $z < 0.15$ (dark red filled diamonds; see Table~\ref{table:clagn}),
	  and the 15 large-variability RM AGNs (brown filled squares; see \ref{table:rmAGN})
	  on the diagram of the global $g-i$ color versus \mgal.
	  Also shown are the 4 CL quasars with $z \gtrsim 0.2$ of \citet[][light blue open diamonds]{2019ApJ...876...75C},
	  and the host galaxies of 52 QSOs from SDSS Stripe 82 analyzed by \citet[][blue open circles]{2015MNRAS.454.4103B}.
	  As a comparison, we plot (by gray dots) about 26,000 low-$z$ Seyfert 2 galaxies
      selected in terms of BPT diagram by \citet{Dong-2010ApJ...721L.143D}.
	  The contours represent 90\% (the outermost), 50\%, and 20\% of the Seyfert 2 galaxies enclosed, respectively.
	  The black filled dots and dashed line denote the median $g-i$ of Seyfert 2 galaxies
	  in the \mgal\ bins (grouped by vertical gray dashed lines);
	  see the text in \S\ref{subsubsec:whole-color-Mgal} for the detail.}
	
\end{figure}

\begin{figure}[htbp]
	\centering
	\includegraphics[width=0.8\textwidth]{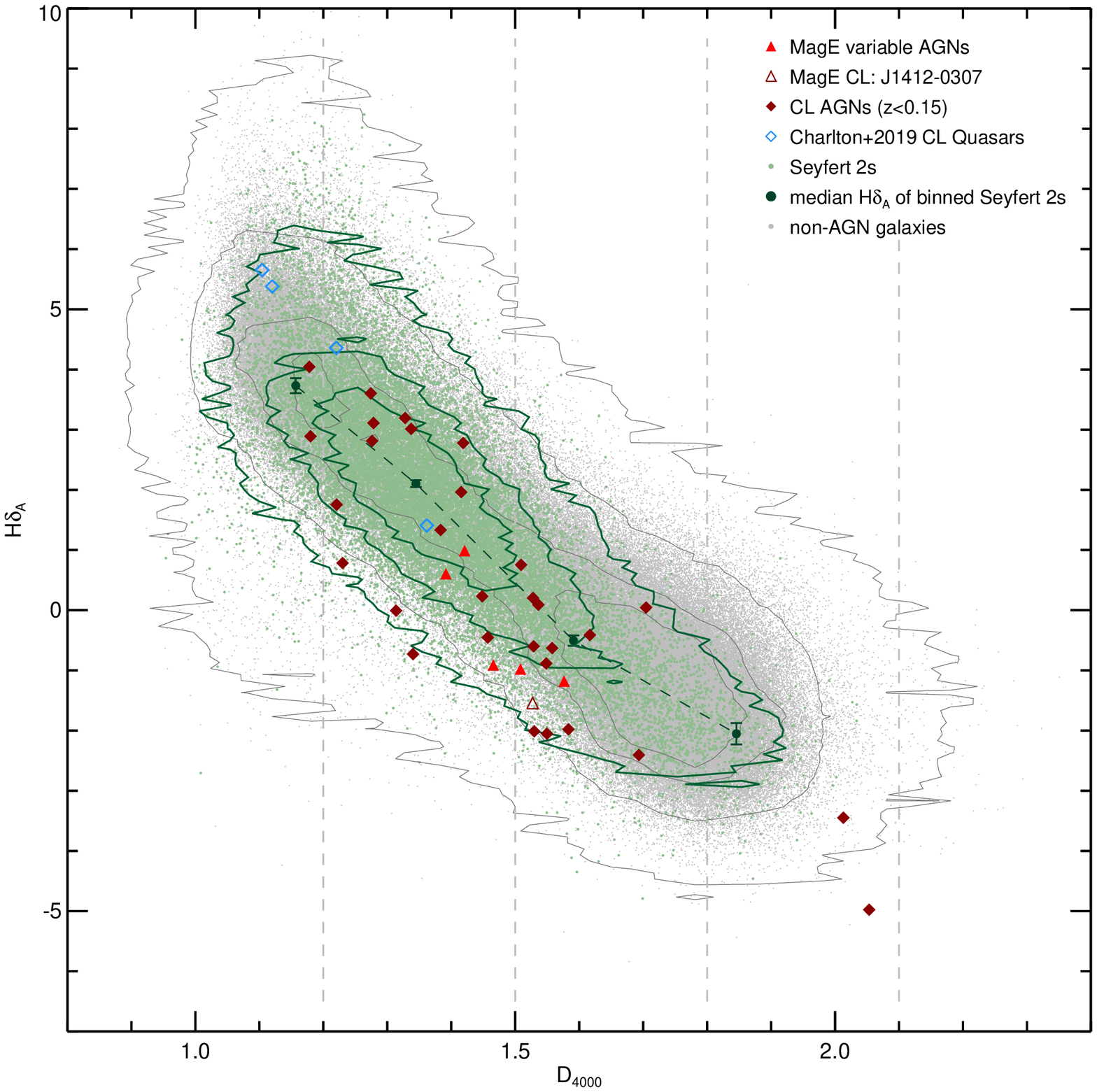}
	\caption{\label{fig:d4000-hda}
	  Distribution of the 6 MagE AGNs (Table~\ref{table:galaxy}),
	  29 low-$z$ CL AGNs (Table~\ref{table:clagn}),
	  as well as the 4 CL quasars at $z \gtrsim 0.2$ (Table~\ref{table:clagn}),
 	  on the diagram of \dbreak\ and \hda. Those AGNs are denoted in the same way as in Figure~\ref{fig:C-M-diagram}.
	  As comparison, we also plot the aforementioned $\approx$26,000 Seyfert 2 galaxies (green dots),
	  as well as the $\approx$32,000 low-$z$ normal galaxies selected by \citet[][gray dots]{Dong-2012ApJ...755..167D}.
	  The gray contours represent 99.9\% (the outermost), 95\%, 68\%, and 40\% of the normal galaxies enclosed, respectively;
      the green contours represent 95\%, 68\%, and 40\% of the Seyfert 2 galaxies enclosed, respectively.
      All of the \dbreak\ and \hda\ are measured from the SDSS spectra.}
\end{figure}

\begin{figure}[htbp]
	\centering
	\includegraphics[width=0.8\textwidth]{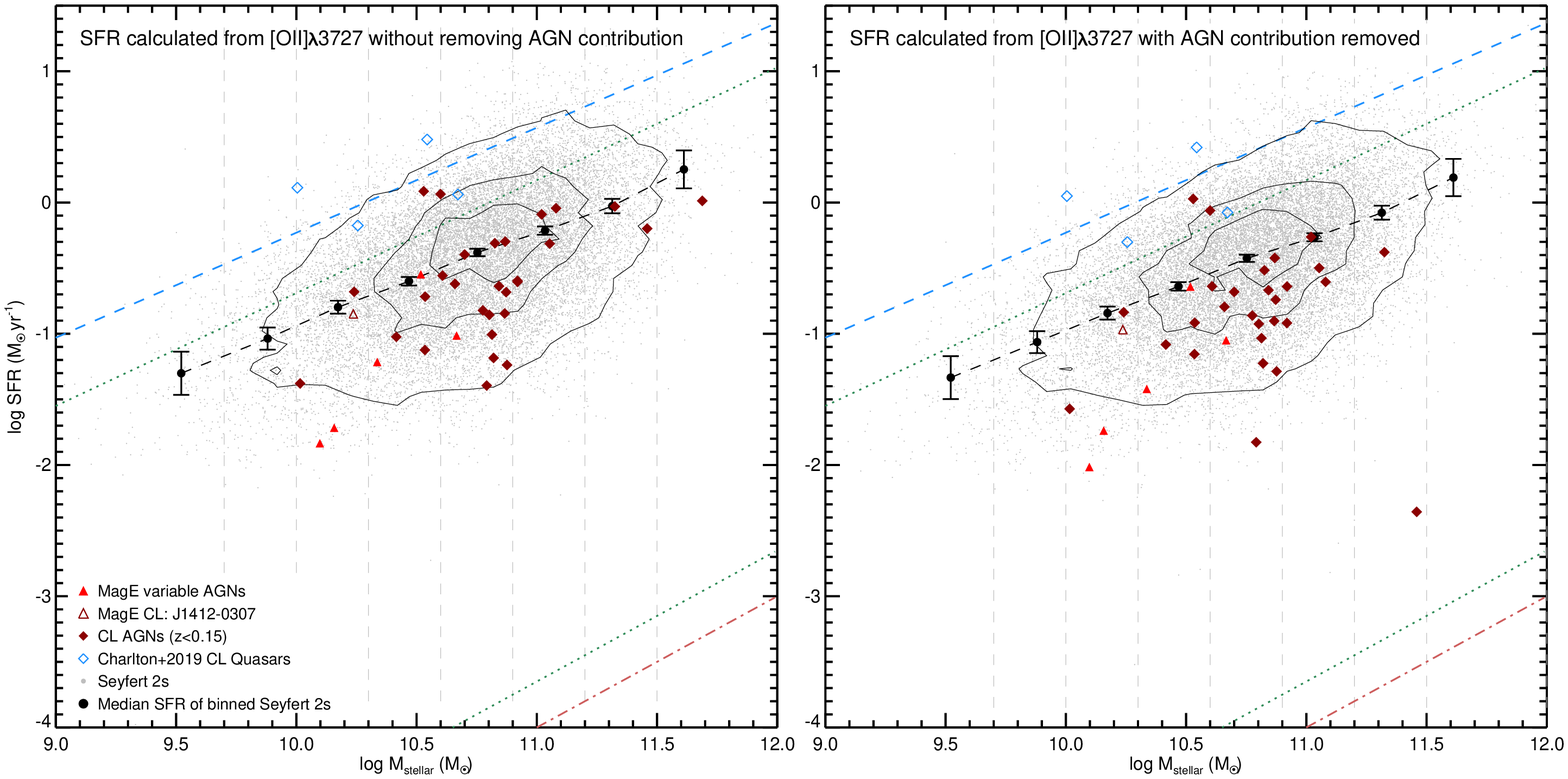}
	\caption{\label{fig:sfr-mgal}
		Distribution of the 6 MagE AGNs (Table~\ref{table:galaxy}),
		29 low-$z$ CL AGNs (Table~\ref{table:clagn}),
		as well as the 4 CL quasars at $z \gtrsim 0.2$ (Table~\ref{table:clagn}),
		on the diagram of SFR versus \mgal. Those AGNs are denoted in the same way as in Figure~\ref{fig:C-M-diagram}.
		As comparison, the aforementioned $\approx$26,000 Seyfert 2 galaxies (gray dots) are also plotted.
		The SFR values are derived from the [O\,{\scriptsize II}]$\lambda3727$ lines in the SDSS spectra,
		without (left panel) and with (right panel) correcting for the AGN contamination.
		The contours represent 90\% (the outermost), 50\%, and 20\% of Seyfert 2 galaxies enclosed.
		THe blue dashed line represents the star forming main sequence,
		the red dashed-dot line represents the so-called quiescent sequence of old red galaxies,
		and the region between the two green dotted lines is the green valley;
		see the text in \S\ref{subsubsec:inner-region-diagrams} for details.}
\end{figure}

\begin{figure}[htbp]
	\centering
	\includegraphics[width=0.8\textwidth]{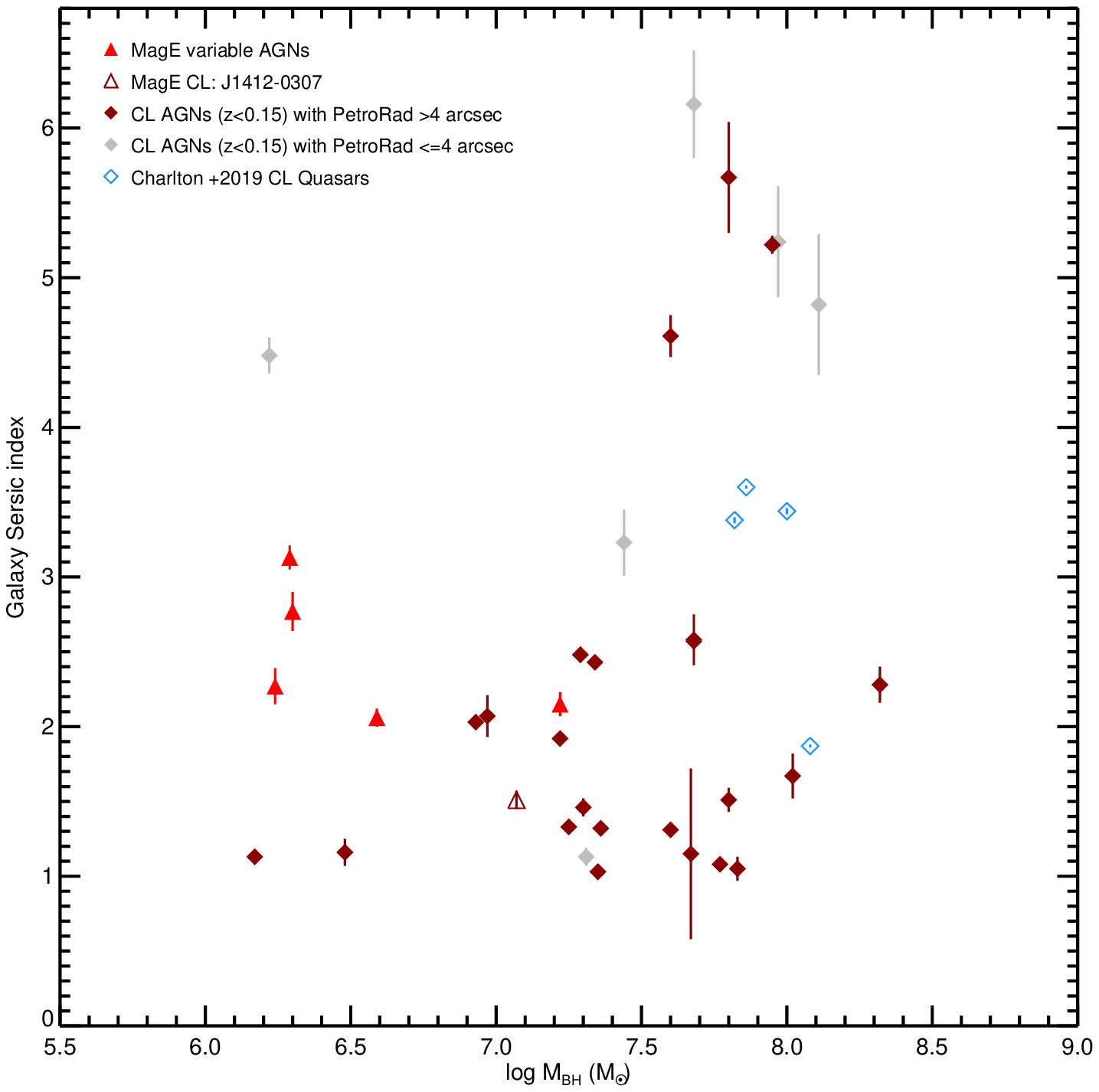}
	\caption{\label{fig:sersic_mbh}
	  Distribution of the 6 MagE AGNs (Table~\ref{table:galaxy}), 29 low-$z$ CL AGNs (Table~\ref{table:clagn}),
	  as well as the 4 CL quasars at $z \gtrsim 0.2$ (Table~\ref{table:clagn})
		on the diagram of inner \sersic\ index versus \mbh.
	        Among low-$z$ CL AGNs, those with $r$-band Petrosian radii $<4$\arcsec\
	    (i.e., resulting in unreliable inner-\sersic\ indexes) are marked in gray; otherwise, they are in dark red.		
	}
\end{figure}

\renewcommand{\thefigure}{A\arabic{figure}}
\setcounter{figure}{0}
\begin{figure}[htbp]
      \centering
      \includegraphics[width=6.8in]{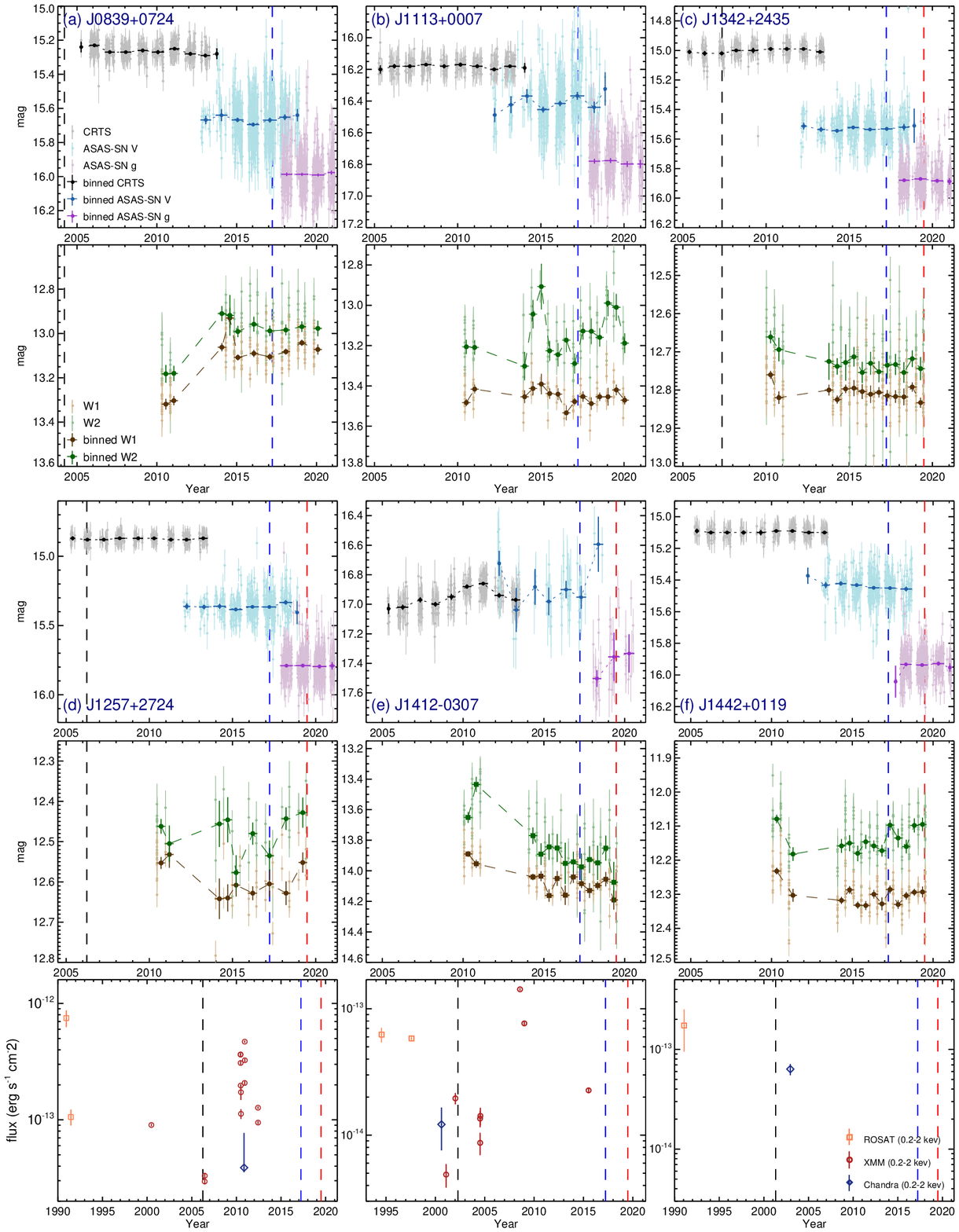}
      \caption{\footnotesize Multi-band (optical, MIR and X-ray) light curves
      	of the six MagE sources.
      	For every source, the optical and MIR light curves are displayed;
      	for the three sources having X-ray observations, light curves are also displayed
      	with the fluxes in the uniform rest-frame energy range (0.2--2 keV).
      	See the text in \S\ref{subsec:LCdata} for the detail.
        \label{fig:light_curve} }
  \end{figure}

\begin{figure}[htbp]
	\centering
	\includegraphics[width=0.95\textwidth]{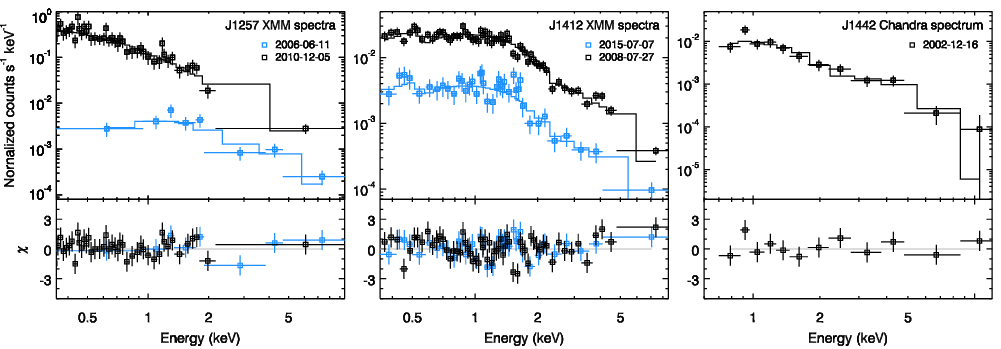}
	\caption{X-ray spectra and their best fits (upper panels) and the residuals (bottom panels).
	  For J1257$+$2724 and J1412$-$0307, the high- and low-flux states are denoted with black and blue, respectively.
	\label{fig:xrayspec}}
\end{figure}

\end{document}